\documentclass{aastex63}
\newcommand{\um }{$\mu $m}

\newcommand{\kms}{km~s{$^{-1}$}}

\newcommand{\Msol}{M{$_{\odot}$}}

\newcommand{\Lsol}{L{$_{\odot}$}}
\newcommand{\Vlsr}{{V$_{LSR}$}}
\newcommand{\Cplus}{{C$^+$}}

\newcommand{\Cii}{[C{\sc ii}]}

\newcommand{\tco}{{$^{13}$CO}}

\newcommand{\Feii}{Fe~{\sc ii}}
\newcommand{\Feiii}{Fe~{\sc iii}}
\newcommand{\Hii}{H{\sc ii}}
\newcommand{\Hi}{H{\sc i}}
\newcommand{\Nii}{N{\sc ii}}
\newcommand{\Ni}{N{\sc i}}
\newcommand{\Sii}{S{\sc ii}}
\newcommand{\Oi}{O{\sc i}}
\newcommand{\Oii}{O{\sc ii}}
\newcommand{\Oiii}{O{\sc iii}}
\newcommand{\Ha}{H$\alpha$}
\newcommand{\BrG}{Br-$\gamma$}

\newcommand{\Hei}{He{\sc i}}
\newcommand{\Htwo}{H{$_2$}}
\newcommand{\HII}{H{\sc ii}}
\newcommand{\per}{$^{-1}$}
\newcommand{\masyr}{mas~yr$^{-1}$}

\received{June 7, 2021}
\revised{September 9, 2021}
\accepted{October 5, 2021}

\submitjournal{ApJ}

\shorttitle{Supersonic Expansion of S106}
\shortauthors{Bally et al.}

\begin{document}

\title[S106 Outflows]{Supersonic Expansion of the Bipolar \Hii\ Region Sh2-106: A 3,500 Year-Old Explosion?}

\correspondingauthor{John Bally}
\email{john.bally@colorado.edu}

% The list of authors, and the short list which is used in the headers.
% If you need two or more lines of authors, add an extra line using \newauthor
\author[0000-0001-8135-6612]{John Bally}

% List of institutions
\affiliation{Center for Astrophysics and Space Astronomy, 
     Department of Astrophysical and Planetary Sciences \\
     University of Colorado, Boulder, CO 80389, USA} 
     
\author{Zen Chia} 
\affiliation{Center for Astrophysics and Space Astronomy,
     Department of Astrophysical and Planetary Sciences \\
     University of Colorado,  Boulder, CO 80389, USA} 
     
\author{Adam Ginsburg}
\affiliation{Department of Astronomy, 
     University of Florida, 
     PO Box 112055, USA}
    
\author{Bo Reipurth}
\affiliation{Institute for Astronomy, University of Hawaii at Manoa, 
      640 North Aohoku Place, Hilo, HI 96720, USA}
     
\author{Kei E.I Tanaka}
\affiliation{Center for Astrophysics and Space Astronomy, 
     Department of Astrophysical and Planetary Sciences \\
     University of Colorado, Boulder, CO 80389, USA} 
\affiliation{
     ALMA Project, National Astronomical Observatory of Japan, Mitaka,  
     Tokyo 181-8588, Japan} 

\author{Hans Zinnecker}
\affiliation{Universidad Autonoma de Chile, Av Pedro de Valdivia 425, 
      Providencia, Santiago de Chile, Chile} 
      
\author{John Faulhaber}
\affiliation{Center for Astrophysics and Space Astronomy,
     Department of Astrophysical and Planetary Sciences \\
     University of Colorado, Boulder, CO 80389, USA}

\begin{abstract}
Multi-epoch narrow-band HST images of the bipolar \Hii\ region Sh2-106 reveal highly supersonic nebular proper motions which increase with projected distance from the massive young stellar object S106~IR, reaching over $\sim$30 mas yr$^{-1}$ ($\sim$150 \kms\ at D=1.09 kpc) at a projected separation of $\sim$1.4\arcmin\ (0.44 pc) from S106~IR. 
We propose that S106~IR experienced a $\sim10^{47}$ erg explosion $\sim$3,500 years ago.  The  explosion may be the result of a major accretion burst, a recent encounter with another star, or a consequence of the interaction of a companion with the bloated photosphere of S106~IR as it grew from $\sim$10 through $\sim15$ \Msol\  at a high accretion rate. 
Near-IR images reveal fingers of \Htwo\ emission pointing away from S106~IR and an asymmetric photon-dominated region surrounding the ionized nebula. Radio continuum and \BrG\ emission reveal a C-shaped bend in the plasma, either indicating motion of S106~IR toward the east, or deflection of plasma toward the west by the surrounding cloud.  The \Hii\ region bends around a $\sim$1\arcmin\ diameter dark bay west of S106~IR that may be shielded from direct illumination by a dense molecular clump.   Herbig-Haro (HH) and Molecular Hydrogen Objects (MHOs) tracing outflows powered by
stars in the  Sh2-106 proto-cluster such as the Class 0 source S106 FIR are discussed.

\end{abstract}

\keywords{
stars: pre-main-sequence
stars: massive
stars: mass-loss \\
ISM: bubbles, \Hii\ regions, Sh2-106}

%%%%%%%%%%%%%%%%%%%%%%%%%%%%%%%%%%%%%%%%%%%%%%%%%%

%%%%%%%%%%%%%%%%% BODY OF PAPER %%%%%%%%%%%%%%%%%%

% Section 1  sssssssssss

\section{Introduction}

The birth and early evolution of massive stars remain one of the least 
understood aspects of star formation. Massive stars play essential roles 
in the feedback and self-regulation of star formation and profoundly 
impact the environments in which lower mass sibling stars and their 
planetary systems form
\citep{Krumholz2011,Krumholz2012,Krumholz2014,Dale2012,Dale2014,Federrath2014}.
The nearest region of on-going massive star formation, Orion OMC1, experienced 
a powerful explosion about 550 years ago 
\citep{BallyZinnecker2005,Zapata2009,Bally2015,Bally2017,Bally2020}.  
Several other massive star forming regions contain explosive protostellar outflows 
\citep{Zapata2017,Zapata2020}.  Here we show that the exciting 
star of the \Hii\ region Sh2-106 (S106 for short) likely experienced a powerful 
explosion several thousands of years ago.
 
The gravitational collapse of molecular clouds drives the formation of 
clumps and protostellar cores, which undergo inside-out collapse to 
form  young stellar objects (YSOs).  The infalling flow's angular momentum 
leads to the formation of circumstellar disks in which viscous dissipation 
fuels further accretion onto the YSO.  Shear amplification of entrained magnetic
fields and convection-powered magneto-hydrodynamic (MHD) dynamos can drive winds
and collimated jets, which entrain matter from the surrounding cloud to produce 
bipolar molecular outflows  \citep{PudritzRay2019}.  Thus, most stars 
produce bipolar outflows as they accrete from their parent clouds and 
grow in mass \citep{Bally2016}.   

These protostellar outflows are a potent source of feedback in the 
self-regulation of star formation because they inject momentum and 
kinetic energy into their host clouds efficiently.  Outflows create 
turbulence, dissociate molecules, and can disrupt the star-formation environment.  
Outflow momentum and kinetic energy injection rates increase with protostellar 
luminosity and mass.   Thus, as forming 
massive stars accrete from their parent clouds and cores, they usually drive
the most powerful molecular outflows in their environment.    Hence, in a 
forming cluster of stars, the most massive young stellar object (MYSO) tends 
to have the greatest feedback impact on the parent cloud
\citep{Maud2015a,Maud2015b,Bally2016}. 

Most MYSOs form as parts of multiple star systems inside clusters, typically
containing hundreds of lower mass YSOs \citep{ZinneckerYorke2007}.   
Multiplicity and surrounding cluster members can dramatically alter the 
evolution of MYSOs \citep{Peters2010a,Peters2010b,Peters2010c}.    
Multi-body interactions with sibling stars can 
re-orient accretion disks and alter  outflow orientations  
\citep{Cunningham2009,Bally2016}.    Compact, non-hierarchical  
multiple systems, binary-binary, and N-body interactions can eject cluster 
members \citep{ReipurthMikkola2010,ReipurthMikkola2012,ReipurthMikkola2015}.
Such interactions can lead to protostellar mergers and explosive 
outflows such as in Orion OMC1 located behind the Orion
Nebula  \citep{Bally2015,Bally2017,Bally2020}.  Although several other
explosive outflows have been identified such as DR21 \citep{Zapata2013}
and G5.89-0.39 \citep{Zapata2020}, the event rate of protostellar 
explosions is not yet known.   Thus, it is important to identify 
additional examples to constrain the event rate.

MYSO cores rapidly reach the main-sequence \citep{ZinneckerYorke2007}.  However, 
their envelopes become bloated in the presence of high-accretion rates.   
Photospheric radii can grow to 1 AU or more as their masses reach 
$\sim$10 \Msol .   Thus, rapidly accreting MYSOs 
($\dot M > 10^{-4}$ \Msol yr \per ) have low effective temperatures 
resembling post-main-sequence supergiant stars \citep{HosokawaOmukai2009}.   
As  MYSOs grow beyond 15 \Msol\ and accretion subsides, they
eventually shrink in radius and become hot O-stars, emitting
hydrogen-ionizing Lyman continuum (LyC) radiation.   This 
radiation dissociates, heats, and ionizes a bubble, including the bipolar 
molecular outflow generated during its main accretion phase.   
Such feedback can halt star-formation 
by blowing out the gas supply fueling the growth of the MYSO and that of 
other stars.    In areas where the sound speed in the 
photo-ionized gas, typically around 10 \kms , is larger than the gravitational 
escape speed,  the plasma can be expelled from the parent cloud.
Such feedback can lead to the cloud's destruction 
(e.g. Bressert et al. 2012).

Massive, main-sequence stars power stellar winds with speeds up to 
several thousand  kilometers per second and mass-loss rates of the order 
$\dot M \sim 10^{-8}$ to over $10^{-6}$ \Msol ~yr$^{-1}$
\citep{Puls2008}.    
Such winds 
create hot, X-ray-bright bubbles of million-Kelvin plasma surrounded by 
swept-up shells of cooler material.   During the early phases of stellar-wind 
bubble evolution, the ram pressure of the wind is converted to 
thermal pressure in a reverse-shock.  Expansion of the hot bubble 
sweeps-up the surrounding \Hii\ region into a shell in an energy-conserving 
interaction.  At later times, 
when radiative and conductive cooling of the hot bubble becomes dominant, 
this ram-pressure continues to drive the shell in a momentum 
conserving interaction  \citep{Castor1975,Weaver1977,Geen2020a,Geen2020b}.    

Some forming MYSOs such as S106~IR drive winds with much slower speeds but 
higher mass-loss rates than main-sequence stars 
\citep{Simon1983,Jaffe1999}.   The dense plasma near the base of the wind produces 
strong near- and mid-IR hydrogen recombination lines;  free-free 
continuum emission at radio frequencies originates farther from the star.  
Such winds are seen as  compact radio sources at the location of the MYSO.   
For a constant velocity and
constant mass-loss-rate wind or jet that spreads with a constant opening 
angle, the electron density decreases as $r^{-2}$.
Because the wind photospheric radius (where the free-free optical depth $\sim$1) 
shrinks with  increasing frequency, the flux-density of such winds 
increase roughly as $\nu ^{0.7}$  \citep{SimonFischer1982,Simon1983,Bally1983}.

S106  provides a unique opportunity 
to study the short-lived transition from massive protostar to a main-sequence star 
surrounded by an emerging \HII\  region.   Of particular interest in this transition  
is the nature of the feedback mechanisms that halt accretion and destroys the 
parent cloud.   S106 provides a unique 
nearby laboratory in which to study the transition from outflow driven 
feedback to UV and wind-powered feedback in the destruction of the parent 
molecular cloud and emergence of a young cluster surrounding an O star. 

In this paper, we present an analysis of nebular proper motions based on 
multi-epoch Hubble Space Telescope (HST) images taken with an interval of 16 years.    
The images were registered using field stars whose positions were determined 
during the epoch of each HST observation using proper motions measured by 
the Gaia satellite and presented in Gaia EDR3.   The nebular
proper motions increase linearly with projected distance from S106~IR, reaching
speeds of order 150 \kms\ at the outer edge of the ionized nebula.
The nature of the supersonic expansion and what it implies for the evolutionary 
stage of S106 is discussed.

We present narrow-band images of S106 in the 2.12 $\mu$m S(1) emission line
of \Htwo\ and the 2.16 $\mu$m Brackett-$\gamma$ hydrogen recombination emission line.  
The  \Htwo\ emission shows the photon-dominated region's structure (PDR) 
surrounding the \Hii\ region.  These images reveal 
fingers of \Htwo\ emission pointing directly away from S106~IR and knots of
\Htwo\ emission beyond the PDR that may trace debris ejected by S106~IR and/or
shocks powered by protostellar outflows from lower-mass protostars in the S106 cluster.
The \BrG\ emission line shows that the plasma in S106 exhibits C-shaped symmetry
with the northern and southern lobes deflected towards the west.  The ends of each 
nebular lobe are capped by `bright-bars' of emission about 80\arcsec\ from S106~IR.

Section 2 presents an  overview of S106.
Section 3 describes the data sets presented here. 
Section 4 describes the reduction and registration of images and the nebular proper motion measurements.
Section 5 presents new near-IR images.
Section 6 combines the proper motion analysis, IR-images, with results from the
literature to interpret the physics of S106. 
Section 7 presents a summary.
Additional images and figures showing features discussed in the text are presented 
in Appendices.

\begin{figure}
    \includegraphics[width=\columnwidth]{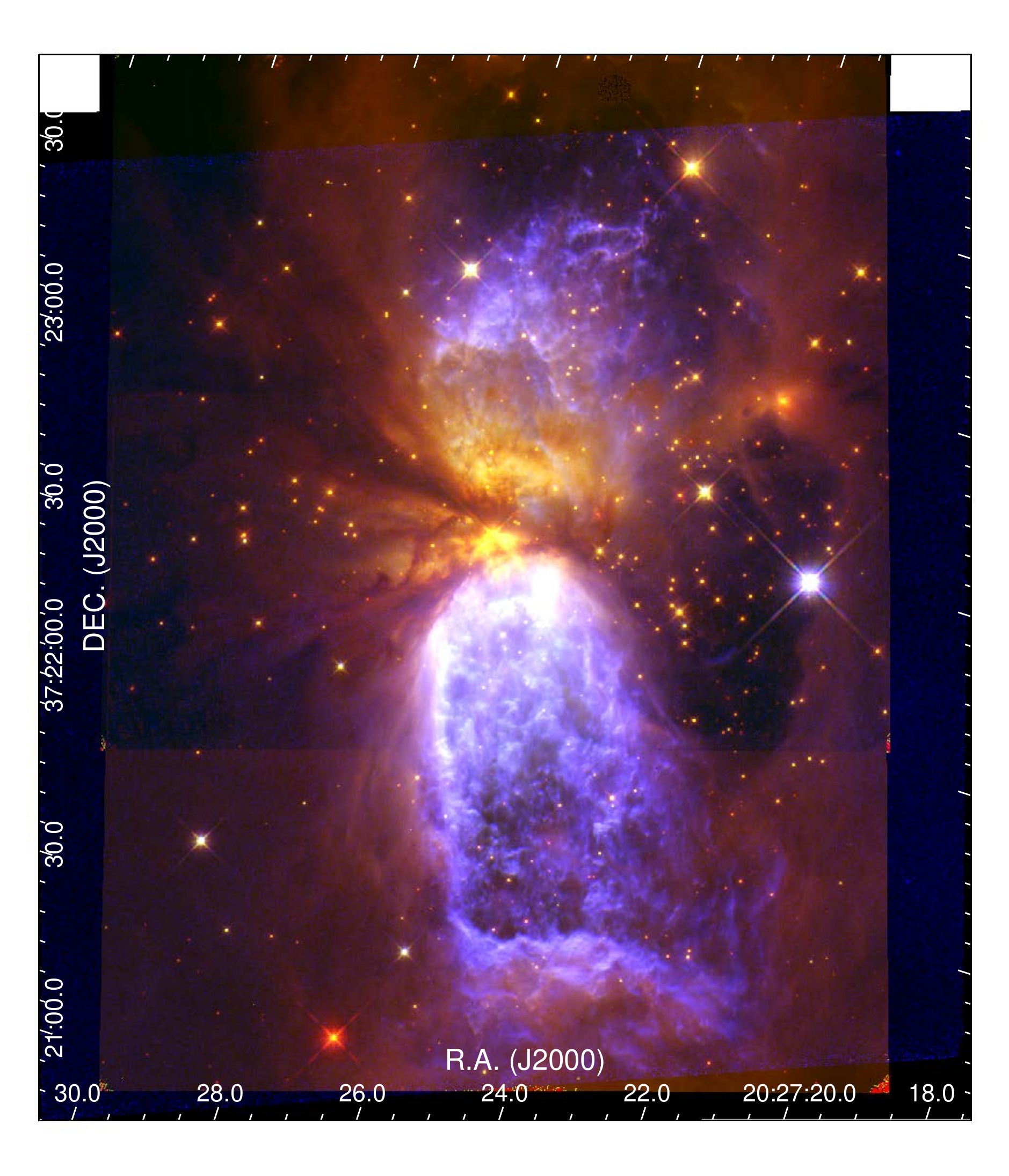}
    \caption{An HST image showing S106  in the  H-band  F160W filter at 1.6 \um\ (red), the 
    J-band F110W filter at 1.1 \um\ (green), and  a narrow-band F658N filter transmitting
    [\Nii ] (blue) using data obtained in 2011.  The image was rotated so north is 18$^o$
    to the right of vertical.  The vertical extent of this image is 186\arcsec\ (0.98 pc). }
    \label{fig1}
\end{figure}

%  Section 2 sssssssssss

\section{Overview of S106}

\begin{figure}
	\includegraphics[width=\columnwidth]{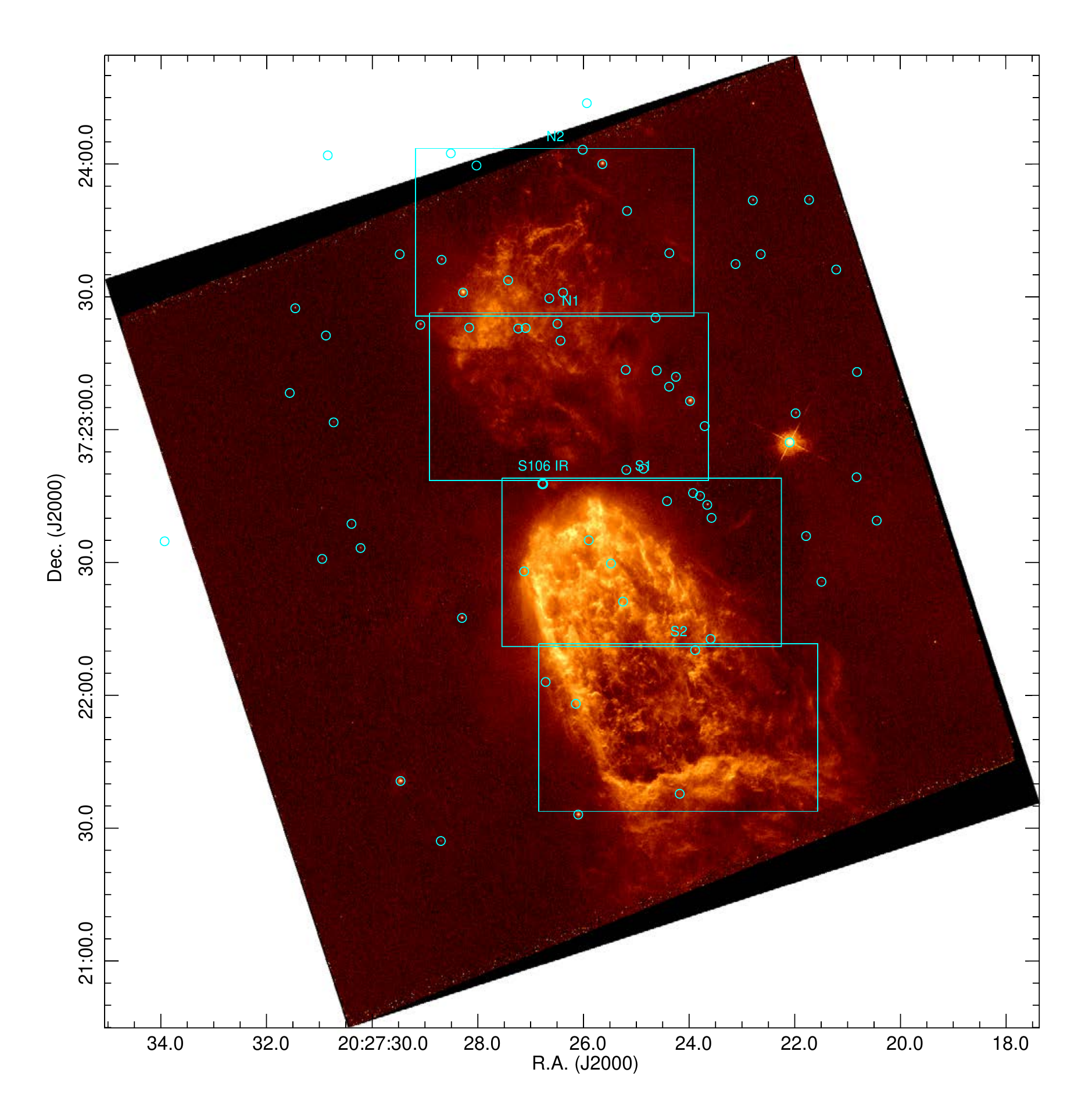}
    \caption{An HST WFC3 image showing S106 in the  narrow-band [\Nii ] filter
    F658N that excludes the \Ha\ emission line.  This image was obtained in 
    2011.  Boxes mark the locations of the four  sub-frames shown in 
    Appendices A and B.  Cyan circles mark the locations of stars for which
    Gaia EDR3 provides proper motions.    The stellar proper motions were
    traced back from the 2015.5 reference date to the dates on which the 
    1995 and 2011 images were acquired. The computed positions of these
    stars at the time the images were taken were used to register the images.
    }
    \label{fig_HST_Nii}
\end{figure}

S106 is the nearest bipolar \Hii\ region, a subclass of very young
photo-ionized nebulae, in which a nearly edge-on disk or belt of dense material 
splits the ionized nebula into a pair of lobes \citep{Eiroa1979,Neckel1982}. 
Long-slit spectroscopy demonstrated that \Ha\ exhibits supersonic 
expansion away from the central ionizing source \citep{Solf1980}.
S106 is ionized by a highly obscured late-O star known as S106~IR embedded 
in the dark lane \citep{HodappSchneider2008}.   Figure \ref{fig1} shows an HST 
image of S106 obtained in the broad-band 1.1~$\mu$m and 1.6~$\mu$m filters 
F110W and F160W with WFC3 on HST.  Figure \ref{fig_HST_Nii} shows a 
narrow-band 6584\AA\ [\Nii ] image acquired with the F658N filter in WFC3.

S106 is located in the Cygnus-X region at Galactic coordinates 
$l$=76.4$^o$ $b$=$-$0.6$^o$  where radial velocities provide an 
unreliable measure of distance.   Thus, distance 
estimates have ranged from $\sim$500 pc to $\sim$5.7 kpc 
(for a review, see  Hodapp \& Schneider 2008).   
Parallax measurements at radio wavelengths towards the 22 GHz H$_2$O maser 
from the Class 0 source S106~FIR \citep{Furuya1999,Furuya2000} give a distance of 
$\sim$1.3$\pm 0.1$ kpc \citep{Xu2013}.   The most reliable estimate uses the Gaia DR2 
distances to dozens of stars toward the S106 molecular cloud that lie either in front 
or behind the cloud.  The parallax at which the extinction and reddening increase abruptly 
(in a so-called `Wolf' plot named after Max Wolf, who pioneered the method about a 
century ago)  gives a distance of 1,091$\pm$54 pc \citep{Zucker2020}.   In this paper, 
we adopt a distance of 1.09 kpc.

S106 is a young \Hii\ region whose central star is in the late stages of formation 
and still embedded in a cometary molecular cloud \citep{BallyScoville1982,Schneider2002}. 
\citet{Schneider2002} found a total mass of 
$\sim$2,000 \Msol\ for the cloud using \tco\ and assuming a distance of 600 pc.  
Scaling this to our adopted distance of 1.09 kpc implies a total mass of $\sim$6,600 \Msol .  
S106 contains a cluster of $\sim$600 young stellar object, including a substantial number
of sub-stellar objects with an age less than 1 Myr  \citep{Oasa2006}.

\citet{IsraelFelli1978} measured the total radio continuum flux from S106 at 
1.4 and 5 GHz finding an approximately flat spectral index with a total flux 
density of S$_{1.4}$ = 11.1 Jy and S$_{5}$ = 12.3 Jy indicating optically 
thin free-free emission.   For their assumed distance of 3.6 kpc, 
\citet{IsraelFelli1978}  derived a total \Hii\ region mass of 35 \Msol.  
Scaling this to 1.09 kpc implies a total plasma mass of 3.1 \Msol.

\begin{figure}
	\center
	\includegraphics[width=4in]{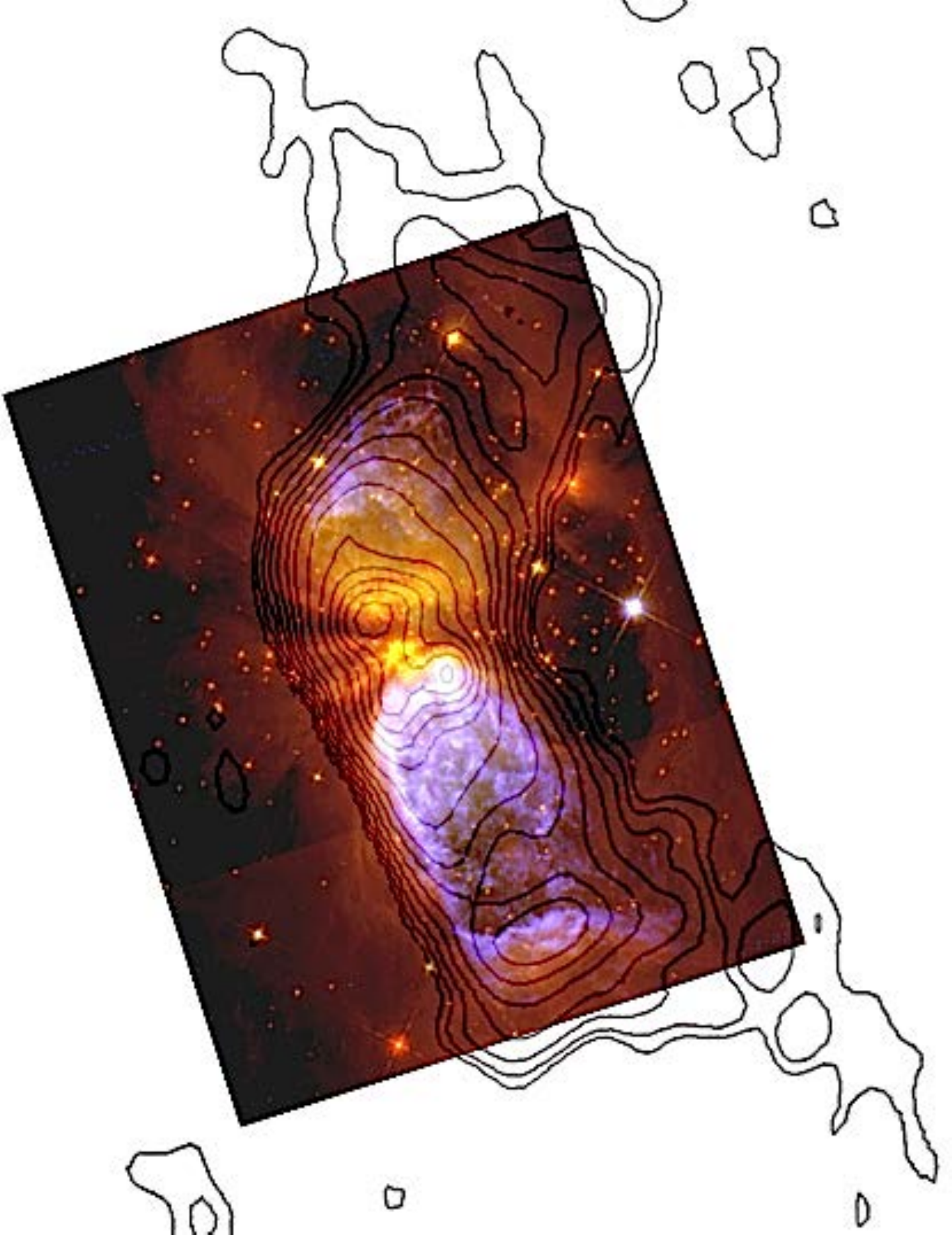}
	\endcenter
    \caption{Contours of 4.8 GHz radio continuum emission superimposed
    on a color image of S106 showing 1.6 $\mu$m (red), 1.1 $\mu$m (green),
    and 0.6584 $\mu$m  [\Nii ] emission.  Note the C-shaped bend and the
    presence of bars of enhanced free-free emission located at south end of
    the southern lobe and the north end of the northern lobe (for a
    clearer view of the South Bar and Northwest Bar, 
    see Figure \ref{fig_BrGgrey}.   The beam diameter
    is 10\arcsec .  Contour levels are at 
    13.4, 26.7, 40.1,  53.5, 66.9, 80.2, 93.6, 107.0, 120, 134, 
    267,  401, 535, 669, 802, 936, 1070, 1204, and 1337 mJy/beam.  
    The peak flux is 1.337 Jy/beam.  The figure is aligned with north along 
    the vertical axis.}
    \label{fig_5GHz}
\end{figure}

The southern lobe of S106 is much brighter than the northern lobe
at visual-wavelengths.  However, at centimeter wavelengths, the northern lobe 
has similar intensity and size to the southern lobe, indicating
that the northern lobe is hidden by more extinction than the southern lobe
(Figure \ref{fig_5GHz}).  The radio data shows
the bright features evident in the mid-infrared within 30\arcsec\ of S106~IR.  
The bright bar seen at visual and near-IR wavelengths located 1.2\arcmin\ 
south of S106~IR is also a prominent radio continuum emission feature. 

The dark equatorial band in Figure \ref{fig1} is also present in high-resolution
radio  images \citep{Bally1983}.  
Thus, this feature is not merely caused by foreground obscuration.  
Rather, it traces a lack of plasma due to the presence of either dense atomic or
molecular gas.   This region may be  shadowed by a compact disk.  
\citet{Barsony1989}  presented interferometric observations of \tco , CS, and HCN 
but failed to find evidence of a massive disk.  Dense gas was found to be concentrated
in two peaks along the eastern and western walls of the  \Hii\ region with a bridge of   
emission connecting the peaks at the location of the equatorial dark region.

From near-IR to sub-mm wavelengths, S106 has a dust luminosity 
$L_{dust} > 9 \times 10^4$~\Lsol\ \citep{Adams2015}.    The dust
temperature decreases with increasing distance from S106~IR, indicating
that this star is the primary heating source.  Mid-IR 
($\lambda \sim$3.6 to 12 $\mu$m) images show several fingers of dust in absorption, 
converging on the position of S106~IR, which is a bright point source
at these wavelengths.   In the Herschel 70 and 160 $\mu$m images, these dust 
streamers are seen in emission and may trace streamers of dense gas falling 
into the equatorial region of S106 
\citep{Adams2015,Schneider2018}. 

High-resolution visual-wavelength spectroscopy  of \Ha , [\Nii ], and [\Sii ]  
\citep{SolfCarsenty1982} shows that the northern lobe
of S106 is redshifted with a mean velocity of \Vlsr $\approx$22 \kms . 
Within a half-arcminute of S106~IR,  the line-widths are $\sim$45~\kms\ 
at half-maximum and up to $\sim$120 \kms\ at zero intensity.    The \Ha\ 
profile towards the southern lobe shows  line-splitting with two components 
separated by up to  100 \kms\  within  30\arcsec\ of S106~IR;   most of 
this emission is blue-shifted.  Beyond 30\arcsec\ from S106~IR, the brighter 
southern lobe has a slight blueshift of about \Vlsr $\approx-$12 \kms\ along 
with some gas close to \Vlsr = 0 \kms . \citet{Noel2005}  presented near-IR 
Fourier Transform Spectroscopy of  \Htwo, \Hei, Br-$\gamma$, and [\Feiii ] 
of the inner region of S106 within 30\arcsec\ of S106~IR, finding that Br-$\gamma$
exhibits emission over a velocity range of $-$45 to +80 \kms.

The \Hii\ region sits inside
a roughly cylindrical, $\sim$6\arcmin\  (1.9 pc) long cavity bounded by 
warm dust and PAH emission at 3.6 to 8.0 $\mu$m  (see Appendix C for 
more discussion).     This cavity is more than a factor of 
two longer and wider than the \Hii\ region at visual, near-IR, and radio wavelengths.   
S106~IR is displaced from the center of this cavity and located near its eastern edge.

\citet{Schneider2003} presented an extensive FIR sub-mm study of S106, finding
strong \Cplus ,  [\Oi ], and high-J emission from CO and a variety of dense 
gas tracers commonly associated with PDRs in 
intense UV radiation fields.   \citet{Schneider2007} found blue-shifted 
CO emission from the southern lobe of 
S106, red-shifted emission towards the northern lobe,  and interpreted 
these features as tracing part of a bipolar flow from S106~IR.   
The S106 molecular cloud is cometary with the S106 \Hii\ region
located in the dense head  of the cloud at its northern
edge. This indicates that feedback from several Cygnus-X OB  clusters, 
especially  NGC~6913, located north of S106 have shaped this cloud  \citep{Schneider2007}.

\citet{Simon2012}  mapped the 158 $\mu$m \Cplus , CO J=11$-$10, and 350 $\mu$m
dust continuum emission with high spectral resolution and $\sim$6\arcsec\ angular 
resolution, finding complex morphology and kinematics.    
These tracers identify a belt of warm, high-velocity gas extending from the 
eastern rim of the \Hii\ region to S106~FIR, closely following the northern 
portion of the dark lane (Figures \ref{fig1} and \ref{fig_HST_Nii}).  
Remarkably, the \Cplus\ emission towards the southern lobe close to S106~IR 
is red-shifted \citep{Simon2012}, opposite to the radial velocity of the \Ha\ 
emission. It appears that the front-side of the  southern cavity has been eroded, 
and most of the 
visual-wavelength emission likely originates from the rear wall of the 
cavity, with the  \Cplus\ emission originating in a PDR 
behind the ionization front at a somewhat higher radial 
velocity than the \Ha\ emission, especially near S106~IR  \citep{Schneider2018}.
\citet{Schneider2018} show a cartoon of the suspected geometry.

\subsection{The young O-star, S106~IR}

S106~IR is a binary with an orbital period of $\sim$5 days, a semimajor axis of 
0.17 AU, and eccentricity of $\sim$0.2.   The primary is a $\sim$20 \Msol\ O8 or O9 star;   
the secondary is a $\sim$3 \Msol\ B8 star \citep{Comeron2018}.    The binary 
exhibits short-term (hours time-scale) photometric variability whose intensity 
peaks around the time of periastron, possibly indicating accretion bursts.  

\citet{Beuther2018} used the NOEMA interferometer to show that S106~IR is 
surrounded by a massive disk or core with a mass of about 1 \Msol\ and an outer
radius of about 800 AU using a distance of 1.3 kpc.  This corresponds to a
mass of $\sim$0.7 \Msol\ and a radius $\sim$650 AU at a distance of 1.09 kpc. 
\citet{Beuther2018} also found a $\sim$0.03 \Msol\ core (scaled to D$\sim$1.09 kpc)
4\arcsec\ northeast of S106~IR in the dust lane connecting the east side of S106
to S106~IR.

\citet{Bally1983}
found  that S106~IR is an unresolved, point radio source at centimeter wavelengths
in high-resolution VLA observations.  It has a spectral index $S_{\nu}$ 
rising as $\nu^{0.7}$  between 5 and 22 GHz, indicating that the radio emission
is produced by an ionized stellar wind or an outflow with a density profile 
decreasing with  distance as $r^{-2}$. \citet{GibbHoare2007} 
found that the radio continuum emission at 22 GHz with 0.03\arcsec\
resolution takes the shape of a torus surrounding S106 IR.  It is elongated and 
measures $\sim$20 by 60 AU in extent, with its major axis aligned along the 
equatorial dark band seen in the radio, infrared, and visual wavelength images. 

\citet{GibbHoare2007} interpreted this feature as a dense, equatorial wind which
may be responsible for shielding the dense gas in the dark lane from ionizing 
radiation.   This wind must be sufficiently enhanced in the equatorial direction
to absorb the Lyman continuum.  However, it must be optically thin to Lyman 
continuum along its polar axis in order to ionize the S106 
\Hii\ region. \citet{Lumsden2012}  found spectroscopic evidence for such a 
wind which may be responsible for shadowing the dark lane. 
Such an equatorially enhanced wind may trace the ionized surface
of a dense accretion disk.  

The full-width-half-maxima (FWHM)  of the Brackett 12, 
Brackett-$\gamma$ , and [\Feii ] line profiles  range from 185 to 280
\kms .  The wind-velocity of $\sim$200 \kms\ \citep{SimonFischer1982,
Lumsden2012} combined with the radio spectral index of $\nu^{0.7}$ implies a 
mass-loss rate from S106~IR of about $10^{-6}$ \Msol\  yr$^{-1}$. 
More precise wind velocity measurements by 
\citet{Drew1993} found a mass-loss rate of at least 
$\dot M > 2.7 \times 10^{-6}$ \Msol\ yr$^{-1}$ and 
a  wind velocity at infinity of at least  340 \kms .    The
slow wind velocity, and large mass-loss rate from S106~IR is unusual for 
a main sequence late-O star.  Most O stars power winds with mass-loss rates
around $10^{-8}$ to $10^{-7}$ \Msol\  yr$^{-1}$ and terminal velocities 
$\sim 10^{3}$ \kms\ \citep{Puls2008} .  However, winds with parameters similar to the wind 
from S106~IR are found to be produced by some other massive young stellar 
objects \citep{Simon1983}.  

The presence of a 0.7 \Msol\ compact ($\sim$ 700 AU) core inferred from the 
NOEMA 1.3 mm observations of S106~IR combined with the high
resolution 22 GHz image suggest that a dense neutral disk surrounds S106~IR 
whose surface is ionized within 30 AU of S106~IR.   The broad line
profiles of the near-IR emission lines from S106~IR, such as \BrG\, may be produced 
by the Keplerian rotation of the photo-ionized disk surface.  If the central mass is
23 \Msol , the Kepler speed at 30 AU is 26 \kms ; The Kepler speed at 1 AU is
143 \kms . The rising radio  spectral index seen at centimeter-wavelengths 
indicates a dense wind.  Such winds  could be driven either by the central O-star, 
or by magneto-centrifugal processes  at the disk surface, or a combination of 
the two mechanisms.

\citet{Murakawa2013} used adaptive-optics-assisted integral-field spectroscopy
and spectro-astrometry to study S106~IR.  They found evidence for a rotating, 
wide-angle,  disk-wind  emerging from the inner  $\sim$0.43 AU portion of an edge-on  
disk (scaled to our assumed 1.09 kpc distance from their assumed 1.7 kpc distance)  
in  Br$\gamma$ and CO overtone emission at 2.3 \um.   
The major axis of the disk has an orientation  $\approx$100$^o$ to 119$^o$, 
approximately at right-angles to the major axis of the bipolar \Hii\ region.  Modeling the
$\pm$200 \kms\ velocity difference on opposite sides of S106~IR as Keplerian
rotation at a radius of $\sim$0.43 AU around S106~IR in a disk inclined by 83$^o$ 
implies an enclosed mass of 19$\pm$4 \Msol\ \citep{Murakawa2013} .

%  Section 3  ssssssssss
\section{The Data Sets}

\subsection{Archival Data}

The proper motion analysis presented here is based on narrow-band 
HST images acquired in 1995 and 2011.
S106 was observed by HST in 1995 under GO program 5963 \citep{Bally1998} 
targeting \Ha\  with  WFPC2 using filter F656N.      S106 was observed again 
by HST in 2011 targeting the [\Nii ] 6584\AA\ emission line with 
WFC3 using filter F658N (GO program 12326:  PI Keith Noll).  
In this program, images were also obtained in HeII 
4686\AA, [\Oiii ] 5007\AA\ emission lines and the 1.1 and 1.6 \um\  
continuum using the wide-band filters F110W and F160W.  
The observations used in the present analysis are summarized in Table~1.   
The time interval between the observations of the northern lobe of S106 
in 1995 and the 2011 images was  $\sim$15.578 years.
For the southern lobe, the time interval was 15.124 years.   

The far-infrared images presented here were obtained with the Spitzer Space Telescope
and downloaded from the public archive at IPAC.  The Sub-mm images were obtained with
the Herschel Space Observatory under the Hi-GAL program. 
\citep{Molinari2010a,Molinari2010b,Molinari2016}.  

\subsection{H$_2$, Br-$\gamma$, H$\alpha$, and 4.8 GHz Radio Observations}

Narrow-band near-infrared images presented here were obtained using the Apache Point 
Observatory (APO) 3.5 meter telescope with the NICFPS camera on the dates indicated
in Table~2. NICFPS uses a 1024 $\times$ 1024 
pixel Rockwell Hawaii 1-RG HgCdTe detector.   The pixel  scale of this instrument is 
0.273\arcsec\ per pixel with a field of view  4.58\arcmin\ on each side. 
Images with 180
second exposures were obtained in the 2.122 $\mu$m S(1) line of H$_2$ and in 
the 2.16 $\mu$m  \BrG\ hydrogen recombination lines.   The narrow-band filters 
have band-passes of $\sim$0.4\% of the central wavelength.   Narrow-band
filters centered off-line were used to obtain an off-line continuum frame
to remove the effects of reflection nebulosity.  The central-wavelengths and 
band-passes  are listed in Table~2.
Separate off-source sky frames in each filter were interspersed with on-source 
images using the same exposure time at a location 600\arcsec\ east.   

During each observation, a set of 5 dithered images were obtained both on-source 
and on the sky position. A median-combined set of unregistered, 
mode-subtracted sky frames were used to 
form a master sky-frame that was subtracted from each individual image.   The reduced 
images were corrected for optical distortions.   Field stars were used to align 
the frames, which were median-combined to produce the final images.  Atmospheric 
seeing produced  $\sim$0.9\arcsec\ FWHM stellar images.

A continuum subtracted image showing only \BrG\ emission was made by subtracting 
the reduced and registered image acquired with the 2.17 $\mu$m off-line narrow-band 
filter from the reduced image obtained with the 2.16 $\mu$m \BrG\ filter.  
Because the seeing deteriorated during the acquisition of the images with the 
2.17 $\mu$m filter, the resulting difference image contains a negative bowl 
surrounding a spike at the stars' positions.  However, extended reflection 
nebulosity is removed to reveal the pure recombination-line structure of S106. 

A continuum subtracted image showing only \Htwo\ emission was formed by subtracting
the 2.13 $\mu$m image from the 2.12 $\mu$m image.  As with the \BrG\ difference 
image, seeing variations resulted in slightly mismatched PSFs which generated 
residuals at the locations of stars.

\Ha\ and [\Nii ] images were obtained using the APO 3.5 meter telescope on 
21 June 2020 and  21 October 2020 using the 2048 by 2048 pixel  
ARCTIC CCD camera using narrow-band filters with 30\AA\ band-passes 
centered at 6570\AA\ and 6590\AA.   Exposure times were 60, 300, and 900 
seconds. Three frames were acquired at each exposure time and median 
combined to remove cosmic rays.   Standard procedures were used for 
Bias and Dark current removal, and flat-fielding was done using twilight flats. 

The previously unpublished radio continuum map used here was 
obtained at a frequency of 4.8 GHz with the Very Large Array's D-configuration (VLA) 
radio telescope on 14 June 1983 under VLA program AB~0206.  The continuum image was 
obtained as part of a study of the polarization of the formaldehyde (H$_2$CO) absorption 
toward S106~IR.  
The total integration time was  14,000 seconds.    With a maximum baseline of 1.3 km, 
the synthesized beam has a full-width-half-maximum diameter of about 10\arcsec. 
The beam is nearly circular since S106 transits close to the zenith at the VLA.
The flux calibrator was 3C286.  A nearby bright, compact source, 2005+403, was used as a phase calibrator.   The 1 $\sigma$ rms noise was $\sim$10 mJy/beam.

% Data is located on FREESPACE/DATA/MAST_S106/
% Table 1
\begin{table}
	\centering
	\caption{HST Observations Used in the Proper Motion Analysis}
	\label{tab:HST}
	\begin{tabular}{llcccl} % four columns, alignment for each
		\hline
		Field	& Date          & MJD       & Instrument  & Filter      & Exposure   \\
		\hline
		S106N2 	& 30 Dec 1995   & 50081    & WFC2       & F656N \Ha     & 1200s    \\
		S106N   &  30 Dec 1995  & 50081    &   "        &     "         & 1600s  \\
		S106S 	& 17 Jul 1995   & 49915    &   "        &     "         & 1600s    \\
  		S106	& 12 Feb 2011   & 55604    & WFC3       & F658N [\Nii ] & 2400s  \\
	    S106	& 13 Feb 2011   & 55605    &    "       & F110W         & 1198s  \\
	    S106    & 13 Feb 2011   & 55605    &   "        & F160W         & 1198s \\
		\hline
	\end{tabular}
\end{table}

% Table 2
\begin{table}
	\centering
	\caption{Near-Infrared \Htwo\ and \BrG\ Observations}
	\label{tab:NIR obs}
	\begin{tabular}{lllll} % four columns, alignment for each
		\hline
	    Date    & Filter    & $\lambda _c$ (nm)  & $\Delta \lambda$ (nm) & Exposure   \\
		\hline
  14 Sept 2020 &  H2-2.12  H$_2$     & 2121.63   & 6.93    &  5 $\times$ 180s \\
  14 Sept 2020 &  H2r-2.13  off-line & 2129.64   & 7.40    &  5 $\times$ 180s \\
  21 Oct 2020  &  H2-2.12  H$_2$     & 2121.63   & 6.93    & 10 $\times$ 180s \\
  21 Oct 2020  &  H2r-2.13  off-line & 2129.64   & 7.40    & 10 $\times$ 180s \\
  26 Dec 2020  &  BrG-2.16  \BrG\    & 2166.35   & 6.90    &  5 $\times$ 180s  \\
  26 Dec 2020  &  BrG-2.17  off-line & 2173.91   & 7.20    &  5 $\times$ 180s  \\	
    \hline
	\end{tabular}
\end{table}

% Section 4  ssssssssssss

\section{Nebular Proper Motions}

The analysis of the nebular proper motions is based on the comparison of the 1995
\Ha\ images with the 2011 [\Nii ] image.    This comparison assumes that the WFPC2
18\AA\ wide F656N filter, which transmits both the 6563\AA\ \Ha\ 
emission line, and the 23.6\AA\ WFC3 F658N filter, which only transmits the 
6584\AA\ [\Nii ] line, trace the same plasma.   
To check the validity of this assumption, we identified all public-domain HST WFC3 
images which used both the WFC3/F656N filter ($\Delta \lambda$=13.9\AA ) that 
transmits only the \Ha\ line and the WFC3/F658N filter ($\Delta \lambda$=23.6\AA ) 
on the same target. 
% HST observed the Bubble Nebula (NGC 7635; D $\sim$2.7 kpc) using these
% two narrow-band filters in WFC3 (HST Proposal 14471, PI.  Z. Levay).   
% The raw \Ha\ images were downloaded from MAST and were subsequently registered and combined using DrizzlePac's AstroDrizzle package. The raw [\Nii\] images were also downloaded from MAST, but combination was done using basic IRAF routines. Both images ran through WCS correction using Gaia's EDR3 Database.
Although the fluxes of the \Ha\ and [\Nii ]  emission lines vary, there are no 
detectable displacements between the structures traced by these two emission lines 
at the sub-pixel level. 
% Despite the near factor of three greater distance, any spatial offset between 
% the \Ha\ and [\Nii ] emission comparable to that seen in S106 would have been apparent.

Although the 1995 and 2011 observations used different filters transmitting H$\alpha$ 
and [\Nii ] , the emission in these two lines traces the same nebular plasma.
The ionization potential of atomic hydrogen and nitrogen are similar; 13.6 eV 
versus 14.6 eV.  As recombining hydrogen in the \Hii\  region interior is 
re-ionized,  mostly by photons with energies just greater than 13.6 eV
(because the cross-section to \Hi\ ionization from the ground state is given by
$\sigma_{\rm H} \approx 6.3\times10^{-18} (h\nu/13.6{\rm~eV})^{-3} {\rm~cm^{2}}$),
the Lyman continuum flux impinging on the transition layer from 
\Hii\ to \Hi\ (the I-front) becomes slightly harder than that 
emitted by the source star.   Thus, nitrogen in the \Hii\ region will be mostly 
singly ionized.  On the other hand, in the neutral hydrogen outside the
\Hii\ region beyond the I-front, nitrogen is expected to be neutral.  
The transition from  \Nii\ to \Ni\  occurs mostly {\it within} the I-front.
The thickness of the I-front is given by  
$\Delta x \sim 1/(\sigma_{\rm H} n_{\rm H}) \sim 10(n_{\rm H}/10^3{\rm~cm^{-3}})^{-1}{\rm~AU}$, i.e., a scale unresolved by HST (here $n_{\rm H}$ is the atomic hydrogen density).
We note that, while the mean plasma density of S106 is about $10^3{\rm~cm^{-3}}$,
the compact, arcsecond-scale knots used for proper motion measurements must be 
much denser. Thus, the thickness of their I-fronts must be even less than $10{\rm~AU}$.
Moreover, the [\Nii ] features in the 2011 image are downstream 
(e.g., farther from S106 IR) compared to the H$\alpha$ features in the 1995 image.
This is opposite of what might be expected if the compact knots in S106 were 
not moving and if the [\Nii ] emission originated from a region between 
the hydrogen I-front and the ionizing source S106 IR.
Furthermore, inspection of H$\alpha$ and [\Nii ]  HST images of the 
Orion Nebula and several other \Hii\  regions shows that,
although there are variations in the relative intensities of these 
emission lines, the spatial structures revealed by these two species are 
coincident at the resolution of HST.  Thus, as discussed below, the differences 
in the positions of nebular features are well interpreted as proper motions.

The measurement of proper motions requires that images obtained at different times 
with different instruments be processed to remove optical distortions and registered 
using field stars.   The assembly of the individual 1995 WFPC2 images into a single 
mosaic covering the full extent of S106 was described by \citet{Bally1998}.

To check the accuracy of the original mosaic published by \citet{Bally1998}, the 1995 
data was re-processed with the Python-based {\sf DizzlePac} software package from STScI.
This analysis de-distorts the images using the latest distortion coefficients, and assembles 
all data in the S106 field into a single image using a pixel scale given by the
PC chip in WFPC2,  The astrometry of the final drizzled image was checked against
Gaia DR2 and EDR3 stellar positions as described below.   The astrometry on the 
drizzled image was found to be better than in the mosaic image generated for publication 
in \citet{Bally1998}.   Thus, we used the newer, drizzled version of the 1995 data
for this analysis.  The residual astrometric errors are discussed below.

The {\sf AstroDrizzle} and {\sf TweakReg} routines in the {\sf DrizzlePac} package, 
made available in a Jupyter Notebook, takes each CCD frame in the S106 data set 
and stitches them together into a mosaic.  The 1995 WFPC2 images are processed by
{\sf tweakReg} which accesses STScI databases to determine the correct WCS using 
stars in the field.   The images are the passed through {\sf AstroDrizzle} where 
the image scale and cosmic ray removal parameters can be adjusted. The output \Ha\ 
image WCS is adjusted using Gaia EDR3 to correct for stellar proper motions. 

The 2011 WFC3 images used in this analysis  were downloaded from the 
Barbara A. Mikulski Archive for Space Telescopes.   These data have been de-distorted 
by the  Hubble Legacy Archive image processing pipeline, including corrections for 
alignment shifts between exposures.  The images are astrometrically corrected and 
aligned using the Hubble Source Catalog version 2 and drizzled onto a common pixel grid.   

Comparison of the 1995 and 2011 images reveals that many field stars 
have proper motions larger than the point-spread-function.  
We used the positions and proper motions of field stars
from Gaia EDR3 \citep{GAIACollaboration2016,GAIACollaboration2018b} to 
improve both the distortion corrections and astrometric registration of the 1995
and 2011 images. Gaia EDR3 proper motions and proper motion errors were
used to estimate the positions of 57 stars in the field when 
the 1995 and 2011 images were obtained.   The pixel coordinates of 
these stars in each HST image were matched to the R.A. and DEC. positions 
estimated by back-tracing the Gaia EDR3 proper motions.  We used IRAF routines 
{\sf ccmap} and {\sf ccsetwcs}  to determine the mapping of the pixel 
coordinates into the celestial coordinates.    For the southern lobe of S106, stellar
positions for the 1995 image were back-traced by 20.0 years (7289 days)
corresponding to the interval between the Gaia EDR3 reference epoch and
the observation date of the 1995 images;   for the northern lobe, stellar 
positions were back-traced by 19.5 years (7123 days).   For the 2011 image, 
stellar positions were back-traced by 4.4 years (1599 days).

Nebular proper motions need to be referenced to a frame at rest with respect to S106. 
Inspection of the 57 stars in the S106 field reveals a net streaming motion towards 
the southwest.  Gaia provides proper motions referenced to the 
Solar System barycenter, which has a $\sim$20 \kms\ motion with respect to the local 
standard of rest (LSR).   To jump to the S106 reference frame, we identified all 
stars in Gaia EDR3 within a 5\arcmin\ radius of S106~IR that have a parallax
range within $\pm$0.2 milli-arcseconds (mas) of the parallax of S106, $\pi$=0.917 mas.  
We determined the mean proper motions of all stars within $\pi$=0.917$\pm$0.2 mas, 
which corresponds to a distance of 894 to 1393 pc.    
The mean proper motion of $\sim$119 stars in Gaia EDR3 in 
this region of phase-space is [$\mu _{\alpha}, \mu _{\delta}$] = [-1.05, -5.4] mas~yr$^{-1}$.   
We checked the mean proper motion's sensitivity to the accepted parallax 
range by varying this parameter from $\pm$0.05 mas to $\pm$0.2 mas.  The $\pm$0.05 
mas bin contained 33 stars with 
[$\mu _{\alpha}, \mu _{\delta}]$ = [-1.0, -4.7] mas~yr$^{-1}$.   
We determined that the S106 reference frame has a mean proper motion of 
[-1.05$\pm$1.0, -5.4$\pm$1.0] mas~yr$^{-1}$.   This implies that the mean proper 
motion of S106 is 28 \kms\ towards PA = 191\arcdeg\ with respect to the Solar 
System barycenter.   Proper motions reported here are given in the S106 reference frame.

Inspection of 57 stars in Figure \ref{fig_HST_Nii} for which we have Gaia DR2 or EDR3 
proper motions on the registered images shows that registration has a 1$\sigma$ 
error of about 2 to 3  \masyr.   Unfortunately, in the southern part of S106, 
where the bright South Bar is located in Figures \ref{fig1} and \ref{fig_HST_Nii} 
has few stars. The registration in this part of the image may have a 
factor of two larger error because this portion of the image is close to the 
edge of the field imaged in 1995 and 2011. 

\subsection{Manual Measurement of Proper Motions}

Blinking of the registered images reveals a systematic expansion of the nebular 
lobes of S106.   These motions are also clearly seen in images formed by subtraction 
of the 1995 image from the 2011 image (see Appendix B).  Proper motions were 
initially measured by identifying local intensity maxima on each image.  
In each region, most of the nebular emission moves coherently with respect 
to the traced-back positions of the stars.   Local emission peaks in compact 
knots, bow-shaped features, and filaments were marked with DS9 regions on the more 
sensitive 2011 image.  These regions were then displayed on the 1995 image. 
Vectors were drawn between the intensity peaks on the 1995 image and the 
regions measured on the 2011 image.  Independent measurements of the same 
peaks by three of the co-authors were used to estimate measurement uncertainties. 
Typical errors were about 15 \kms .   In the final analysis, manual measurement 
of nebular proper motions were also used to check the automated measurements.

\begin{figure*}
    \begin{center}
	\includegraphics[width=6.5in]{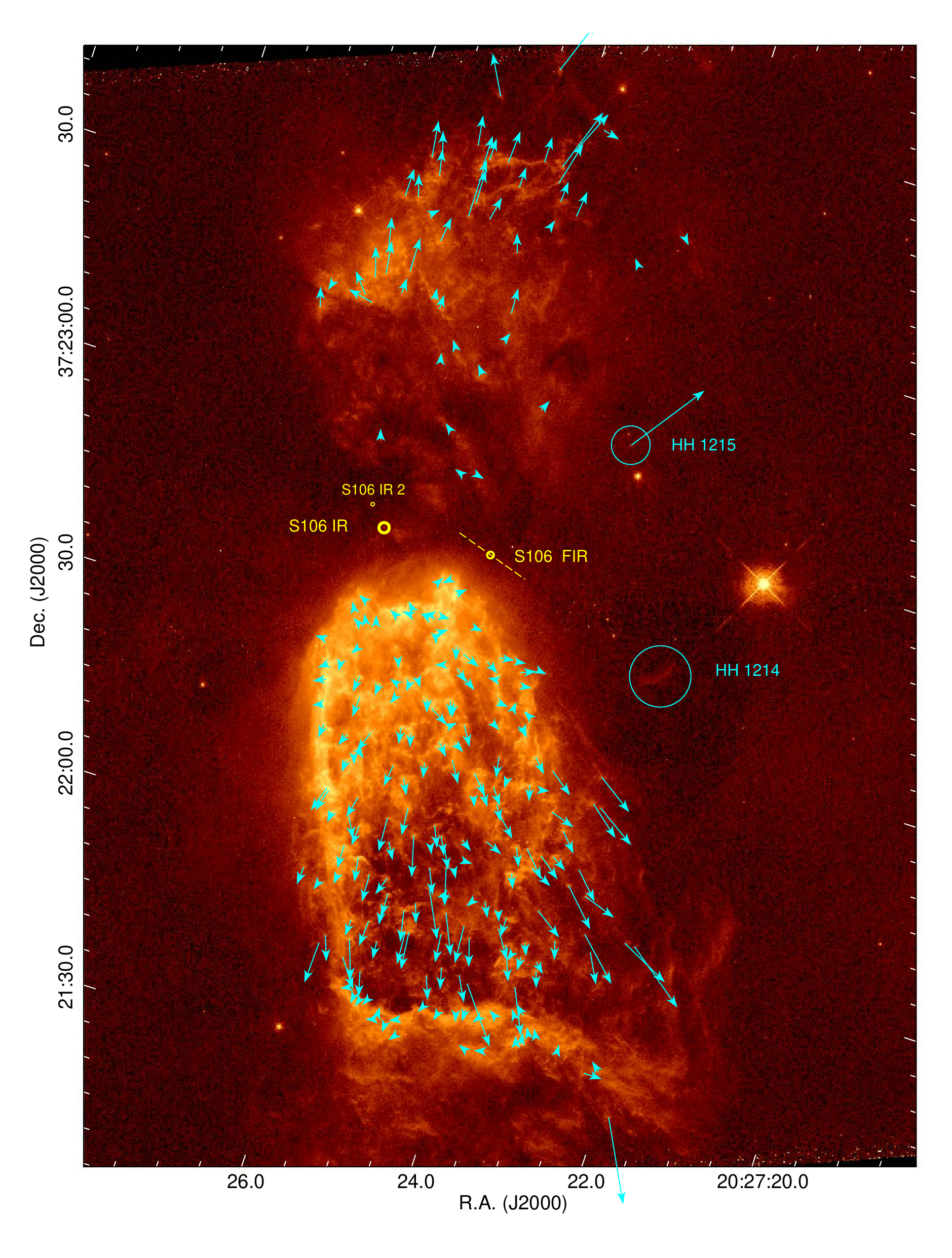}
    \caption{Proper motions of selected compact 
    features superimposed on the 2011 [\Nii ] image in the S106 
    reference frame, measured using the cross-correlation
    methods described in the text.  The vector lengths correspond to the
    motions over the next 400 years.  
    The positions of S106~IR, S106~IR 2 [the small secondary core found by 
    \citet{Beuther2018}], and S106~FIR are indicated by yellow circles. 
    The dashed line indicates the orientation of the maser jet from S106~FIR.
    The isolated, fast moving knot to the upper-right of S106~FIR is HH~1215.
    This image is rotated by 18$^o$ with respect to north.}
    \label{fig_CCPMs}
    \end{center}
\end{figure*}

\begin{figure*}
    \begin{center}
    
    \gridline{\fig {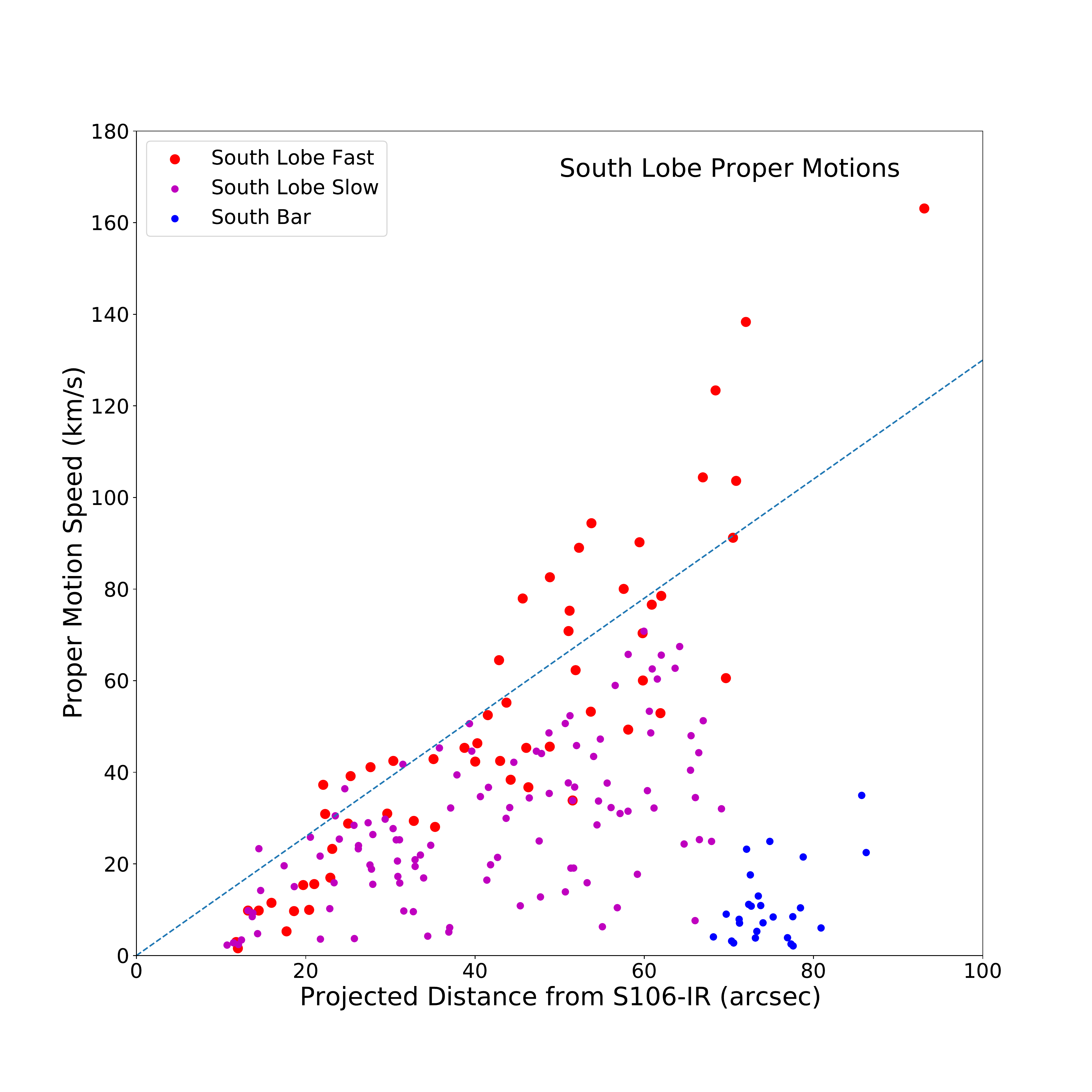} {0.5\textwidth} {(a)}
              \fig {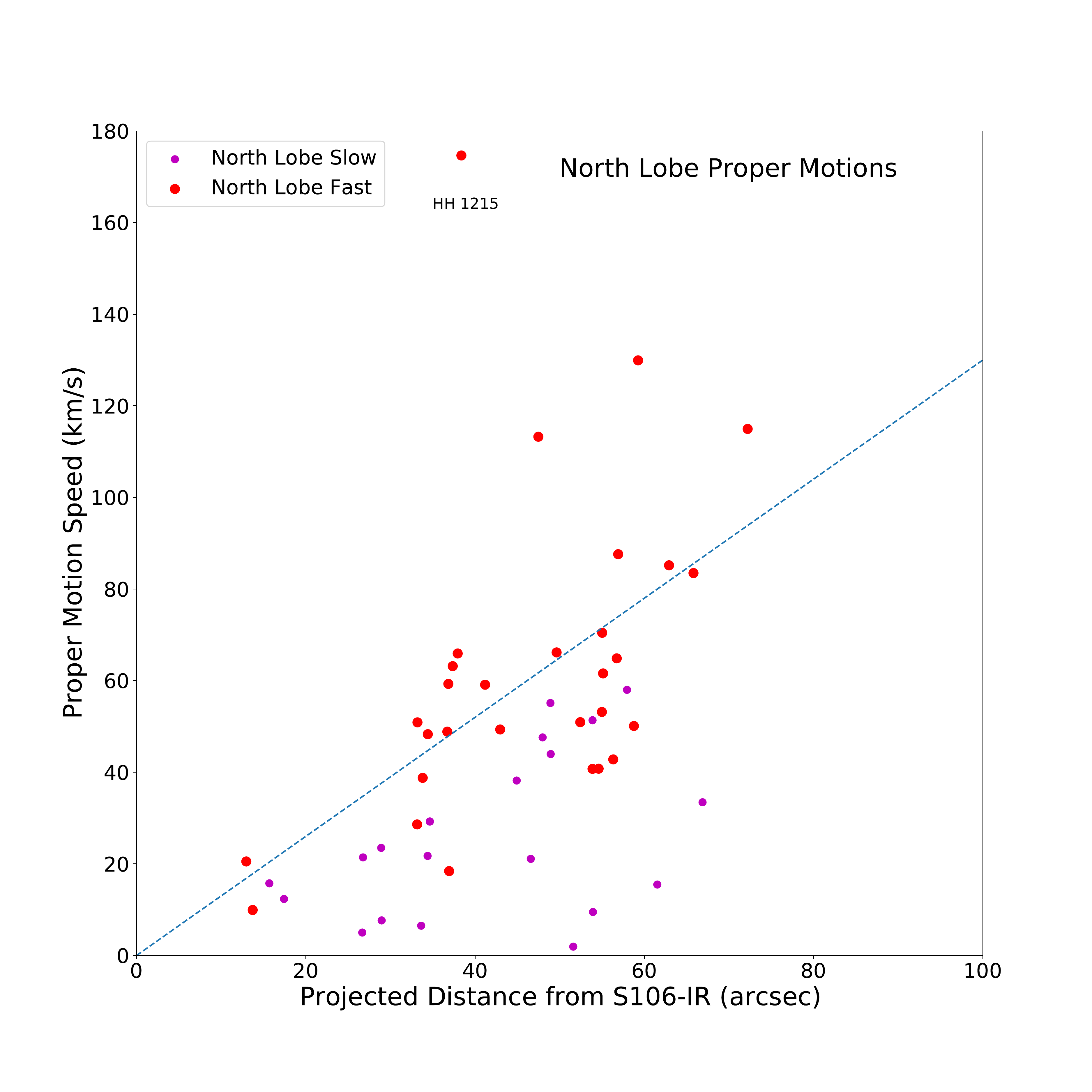} {0.5\textwidth} {(b)}
             }
    \caption{
    {\bf (a):} A plot of the proper motion amplitude for the  vectors
    shown in Figure \ref{fig_CCPMs} in the south lobe of S106 as a function of
    the projected distance from S106~IR.  The dashed black line shows a
    projected linear proper motion of 1.3 \kms\ per arc-second, corresponding to
    a linear velocity gradient of 246 \kms /pc.  The red dots show the motions
    designated as {\it fast} at a given projected distance from S106~IR and
    so listed in the Tables in the Appendix.   The magenta dots represent 
    mostly slower motions and listed as {\it slow} in the Appendix.   The 
    blue dots correspond to the slow-moving features in the South Bar.
    {\bf (b):} 
    A plot of the proper motion amplitude for the  vectors
    shown in Figure \ref{fig_CCPMs} in the North lobe of S106 as a function of
    the projected distance from S106~IR.  The color scheme is identical 
    to part (a).   The isolated fast knot near the legend is the motion of the
    compact knot, HH 1215.
    }
    \label{fig_PMs_All_VvsD}
    \end{center}
\end{figure*}

\subsection{Automated Measurement of Proper motions}

Proper motions were measured using a Python code that cross-correlates
marked regions in a pair of images.    Because the 1995 and 2011 images
were taken with different cameras on-board HST that have different image 
scales and orientations of its pixels,  the analysis was conducted on 
aligned and re-interpolated sub-frames extracted from the registered images.   
Four pairs of sub-frames shown in Figure \ref{fig_HST_Nii} were 
extracted from the full-field 1995 and 2011 mosaiced images shifted into the
S106 reference frame using {\sf SAOImage ds9}.  The images 
were displayed in {\sf ds9} with the 2011 epoch image displayed at full 
resolution using the {\sf ds9 align} function to so that the x- and y-pixels
are aligned east-west and north-south, respectively.  Matching the 1995 
epoch image to the 2011 image using the WCS resulted in a magnification 
of the drizzled 1995 epoch image by a factor 
of 1.14944.  This sub-frame was interpolated onto a pixel grid identical 
to the aligned 2011 epoch image.  For each of the four sub-fields, the 
resulting image pairs were saved  as {\sf fits} files to be used as input 
for proper motion measurements.  In a final step, the intensity scales on each
image pair were normalized to have similar peak intensities.

The Jupyter Notebook 
{\sf CrossCorrelate.ipynb} uses the Python 3.0 package {\sf SciPy.signal} 
and the {\sf image$\_$registration} package from 
{\sf https://pypi.org/project/agpy/}.  
{\sf CrossCorrelate.ipynb} ingests the sub-field image pairs along with a 
user-provided {\sf ds9 Region} file containing the pixel coordinates 
({\sf Points}) of emission peaks on one of the sub-frames.  Because
the 2011 epoch image has better signal to noise and smaller original 
pixel scale, it was designated the reference image.  The user specifies 
the number of pixels along each side of a measurement box  
centered on the features marked by {\sf Points}.  
{\sf CrossCorrelate.ipynb} finds the 
actual intensity maximum in each measurement box on the reference image 
(the 2011 image) and re-centers the box on this peak.   Measurement 
box-sizes used in the analysis range from 20 by 20 pixels for 
compact features to over 200 by 200 pixels for large features. For each 
marked point, the data inside the measurement box on the 1995 and 2011 images 
are cross-correlated.  The offset of the peak of the resulting cross-correlation 
image from the center of the box is used as an estimator of the proper motion.  
When the  more recent 2011 epoch images is used as the reference image,  the
sign of the motion is reversed to give the proper motion.   

% {\sf CrossCorrelate.ipynb}  
% generates a cross-correlation image and finds the peak intensity using 
% two methods, {\sf correlate2d} and {\sf cross$\_$correlation$\_$shifts} in 
% the {\sf scipy} and {\sf image$\_$registration} packages, calculating 
% the best fit for the movement between the two epochs. The shift of the 
% peak from the intensity maximum on the reference image to the peak of the 
% maximum in the cross-correlation image is a measure of the intensity-weighted 
% position shift of the feature between 1995 to 2011. Typically, the 
% two outputs of the two routines agree to about one pixel.  However, visual 
% inspection of the difference images or blinked image pairs shows that the 
% routine {\sf cross$\_$correlation$\_$shifts} provides a more precise 
% measurement of the proper motions.  

The output of {\sf CrossCorrelate.ipynb} consists of an {\sf SAO Image DS9} 
region file containing the measured displacement vectors along with the 
measurement boxes and a formatted (LaTeX) table of positions and proper motions.   
The region file shows the measurement boxes and the proper motion vectors 
scaled to represent motions over the next 400 years.  
Figure \ref{fig_CCPMs} shows the results on the 2011 epoch image.    
Tables in the Appendix (\ref{tab:table_south_lobe} to \ref{tab:table_north_lobe}) 
list the  peak positions of features, the proper motions in mas~yr$^{-1}$, the
speed in \kms , and the direction (position angle) of the motion in 
each of the four sub-fields.  
Each entry in the Tables is given a sequential identification 
number ranging from 1 to 194 for the southern lobe and from 1
to 51 for the northern lobe.  The numbers start in the southwest 
corner of sub-fields S2 and N1 and increase towards the east and north.
These numbers are also indicated in
the four figures in the Appendix, Figures \ref{fig_S2_PMs} to
\ref{fig_N2_PMs}.    

{\sf CrossCorrelate.ipynb} can be used to automatically identify all 
intensity maxima above a chosen intensity value and minimum separation 
criterion. However, this approach resulted in a large number of `faulty'
proper  motions  shown to be incorrect by visual inspection.  The
likely causes for the misbehavior of the code are discussed below.  
For the analysis presented here, we marked 245 locations on the 2011 
image for analysis which are deemed to be representative of the 
overall motions in each portion of S106 by visual inspection.  
As can be seen by from the movies shown in the electronic version 
of this paper (or by blinking the provided {\sf fits} files), 
these selected points are a good representation of the overall 
proper motion vector field.

Measurements made with {\sf CrossCorrelate.ipynb} are subject to
several limitations and types of error.   The code works best for bright, 
compact, and isolated knots or stars.   But, most of the emission 
in  S106 originates from extended structure, crowded 
clusters of clumps, forward and backward facing bow-shapes, and filaments. 
Features with low signal-to-noise ratios have larger uncertainties 
(in general, the 1995 image has a lower signal-to-noise ratio than the 2011). 
The fastest feature (\#1 in Table \ref{tab:table_south_lobe} 
is the tip of a bow-shaped feature in the lower-right corner of the 
S2 field.   It has a proper motion of $\sim$167 \kms , making it the
second fastest object in S106 (the Herbig-Haro object HH 1215 - 
entry 1 in Table \ref{tab:table_north_lobe} in the 
S1 field is faster).  But the object is very dim on the 1995 image.  
Visual inspection shows that  its motion is likely to have an  
uncertainty of tens of \kms .

In elongated structures such as filaments with relatively constant 
intensity, only the component of the motion parallel to minor dimension 
(orthogonal to the filament)  can be trusted.  The derived
position-angle of the proper motion vector can be very uncertain. 
For extended or complex, clumpy structure, the entry of a bright clump
into the measurement box on the 2011 image (or the exit of a clump)
which was not in the measurement box on the 1995 image can result in
a nonsensical proper motion vector.  The entry and exit of dimmer
features into the 2-nd epoch image was found to bias the cross-correlation 
results towards lower speeds.  The selection of peaks and the measurement 
box size was based on our desire to avoid the entry of new features and 
the departure of others on the 2011 image in the measurement box.   
Measurement box sizes of 41 by 41 pixels (on the 2011) image were 
used for the results presented here. In summary, the  results of
automatic measurements require  confirmation by visual inspection.

\subsection{Results: Supersonic Nebular Expansion}

The velocity on the plane of the sky is given by 
$V = 4.7405 ~PM_{\rm mas/yr} ~ D_{\rm kpc}$ km~s$^{-1}$ where 
$PM_{\rm mas/yr}$ is in mas~yr$^{-1}$ and $D_{\rm kpc}$ is in units 
of 1 kpc.  Thus at 1.09  kpc, $V = 5.167 ~PM_{mas/yr}$ (km~s$^{-1}$). 
The approximately 15-year interval between the two epochs 
reveals that most of the narrow-band emission in the nebular
interior traced by \Ha\ and  [\Nii] is expanding away from S106~IR 
with a mean proper motion of $\sim$14 \masyr\ or $\sim$70 \kms\ at 
the 1.09 kpc distance of S106,  $\sim$7 times the sound speed in 
photo-ionized plasma and 5 times the  typical expansion speed 
in \Hii\ regions \citep{ODell2017,ODell2018}.

Figure \ref{fig_CCPMs} shows proper motion vectors of selected regions 
superimposed on the 2011 epoch image.  In the southern lobe of S106, 
compact nebular features move towards the south and southwest 
(PA$\sim$190$^o$ to $\sim$220$^o$).  In the northern lobe of S106, 
compact nebular  features move towards the north and northwest 
(PA$\sim$330$^o$ to $\sim$360$^o$).  
In the northern lobe, only a few proper motions 
could  be measured within  $\sim30$\arcsec\ of S106~IR because 
of the high extinction towards this region.  The proper motions  
are closely aligned with the orientations of the nebular lobes
with motions generally pointing away from S106~IR.  Figure 
\ref{fig_CCPMs} shows that the fastest motions are far 
from S106~IR and that the motions show a systematic deflection
towards the west (e.g. exhibit C-shaped symmetry with respect
to S106~IR).  As discussed below,  the two nebular lobes 
together also exhibit a C-shaped bend towards the west.  

Figure \ref{fig_PMs_All_VvsD} shows the amplitudes of all vectors 
plotted in Figure \ref{fig_CCPMs} as a function of the projected 
distance from S106~IR.  Several general trends are apparent.  
The proper motions fill-in a region between zero speed 
and a line indicating a linear increase in speed 
with increasing projected distance from S106~IR. The dashed 
line in Figure \ref{fig_PMs_All_VvsD} indicates a velocity 
gradient of 246 ${\rm km~s^{-1}~pc^{-1}}$.  Thus, the upper 
bound  on the speeds increases with increasing distance from 
S106~IR.  While within $\sim$30\arcsec\  of S106~IR,  motions 
are typically slower than $\sim$40 \kms ,  motions at larger 
projected distance are faster.  The fastest motions 
are southwest and northwest of S106~IR.  Here,  some features 
have proper motions up to  about 30 \masyr, corresponding 
to a speed  of $\sim$150 \kms .  

The fastest motions tend to avoid the bright projected walls 
of the \Hii\ region.  Along the projected eastern edge of the 
southern lobe and in the South Bar located about 1\arcmin\ 
south-southwest of S106~IR,  proper motions are either 
absent or slower than about 30 \kms . Because of the high extinction 
towards the northern lobe, it is not clear if the pattern of slower 
motions along the lobe walls also holds.  There are some features 
in the lobe interior farther from S106~IR than 30\arcsec\ that 
are also moving slowly.  Differential motion is evident when 
the images are blinked in DS9 or viewed as the movies (shown
in the electronic version of this paper).    These slow-moving 
features may be close to the foreground or background walls of 
the nebular lobes along lines of sight through the lobe interior. 
Consequently, as shown in Figure  \ref{fig_PMs_All_VvsD}, 
at any particular projected  separation from S106~IR, 
a range of proper motions are seen from low values to a maximum 
value which increases with projected separation from S106~IR.
Blinking of the  images shows filaments and knots just north 
of the South Bar approaching  this quasi-stationary structure. 
This behavior provides confirmation that, despite the lack 
of stars near the South Bar, the image registration is good. 

Figure \ref{fig_PMs_All_VvsD} shows the amplitudes of fastest 
nebular motions in the north and south lobe regions tend to
avoid the eastern lobe edges and the South Bar.  
Figure \ref{fig_PMs_All_VvsD} shows that for the motions 
in the nebular lobe interior and excluding the South Bar,  
the expansion pattern shows increasing proper motions with 
increasing  projected distance from S106~IR (a `Hubble flow').  
The proper motions are shown in each of the four sub-regions
in greater detail in Appendix A as vectors superimposed on the
2011 sub-frames.   

Appendix B shows difference images obtained by matching the mean 
flux of the nebular emission in the  1995 \Ha\ image with the mean 
flux of the 2011 [\Nii ] image and taking the difference.  In 
these figures, the 1995 epoch image is shown in black, while 
the 2011  image is shown in white.  In the electronic version 
of this manuscript, we present MP4 movies showing the 
changes and motions in the four sub-fields between 1995 and 2011.

We compared the positions and proper motions of field stars 
measured on our images with the motions measured by Gaia for 
over two dozen stars in the same field.  For the majority of 
stars, the motions agree to about 2 mas~yr$^{-1}$,  and for 
motions as large as 6 mas~yr$^{-1}$ the stellar proper motion 
directions on our images agree with those measured by Gaia 
EDR3 to better than $\sim$15$^o$. 

% Section 5  ssssssssssss

\section{Radio and Near-Infrared Images   }

\subsection{A C-Shaped Bend in the Ionized Plasma}

Figure \ref{fig_5GHz} shows a previously unpublished, low-resolution but deep 
$\sim$4.8 GHz (6 cm) radio continuum contour map of S106 superimposed on the 
image shown in  Figure \ref{fig1}. 
The radio contours show that the northern lobe has similar 
intensity and size to the southern lobe.  However, the northern lobe is 
deflected towards the west by about 45$^o$ with respect to the axis of 
symmetry of the southern lobe and the axis of the cavity seen 
in the far-infrared images shown in Appendix C. 

The bright bar located 1.2\arcmin\ south of S106~IR (South Bar)
is a prominent radio continuum emission feature. There is a second bright bar 
at the northwest end of the northern lobe, also about 1.2\arcmin\ 
from S106~IR. 
This `Northwest Bar' is dimmer at 4.8 GHz than the South Bar by about 
a factor of two. This feature is not seen in visual wavelength images due 
to foreground extinction but apparent in the \BrG\ image discussed below.
Motivated by the radio images, we obtained deep near-IR images of
S106 in the 2.16 $\mu$m \BrG\ hydrogen-recombination line which is much
less impacted by extinction than \Ha .

\begin{figure*}
	\includegraphics[width=7in]{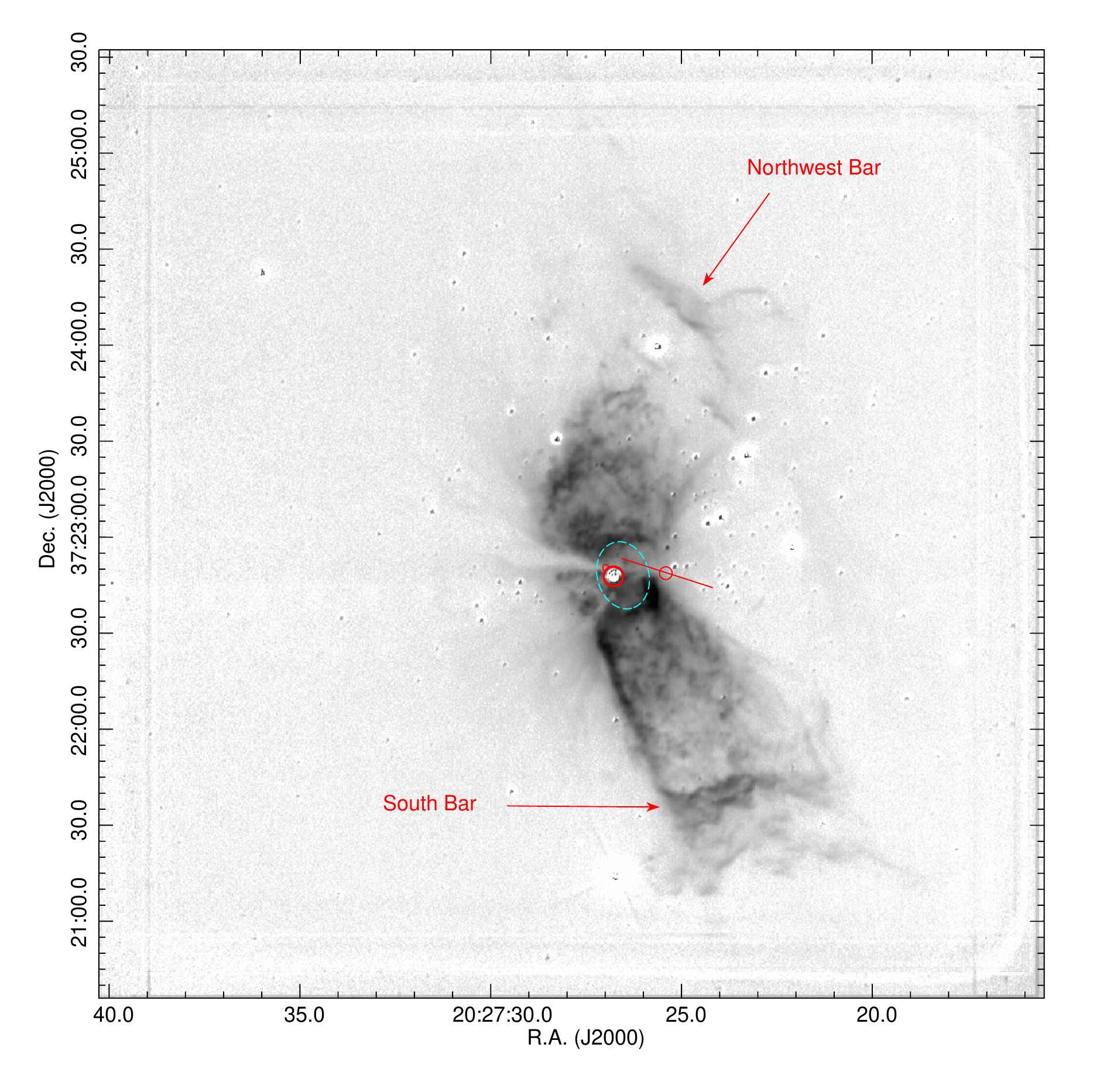}
    \caption{The continuum subtracted \BrG\ image shown with 
    an inverted logarithmic grey scale.  The South Bar and Northwest Bar
    are indicated.  The thick red circle in the middle of the figure marks
    the location of S106~IR.  The small red circle at the upper-left edge 
    of the circle marking S106~IR indicated the location of the faint secondary 
    continuum source found by \citet{Beuther2018} northeast of S106~IR.
    The circle with a line marks the location
    of S106~FIR;  The line indicates the direction of the outflow from
    S106~FIR \citep{Furuya1999,Furuya2000}. The dashed cyan oval shows the 
    8\arcsec\ by 11\arcsec\ cavity discussed in the text.  A version of
    this image emphasizing the faint emission is shown in the Appendix. }   
    \label{fig_BrGgrey}
\end{figure*}

Figure \ref{fig_BrGgrey} shows a continuum-subtracted \BrG\ image of S106 
with 0.9\arcsec\ angular resolution.   This image shows the C-symmetric 
bend and both the South and Northwest Bars at the ends of the southern 
and northern lobes of S106. The locations of S106~IR, S106~FIR, and the 
0.2 \Msol\ secondary core (scaled to a distance of 1.09 kpc) 
found 4\arcsec\ northwest of S106~IR 
by \citet{Beuther2018} are marked.   The orientation of the protostellar 
outflow from S106~FIR traced by H$_2$O masers \citep{Furuya1999,Furuya2000} 
is indicated by a red line.

\subsection{Cavities Surrounding S106~IR and the S106 Ionized Nebula}

In the high-resolution radio images of \citet{Bally1983}, 
there is a roughy 8\arcsec\ by 11\arcsec\ elliptical cavity flanked by bright 
free-free emission centered about 4\arcsec\ west of S106~IR.  
This feature is also clearly seen in the \BrG\ images.
The major axis of this  structure is at PA$\sim$15$^o$.  
The southern and western parts of
this elliptical feature are also seen in the HST images where the brightest 
\Ha, [\Nii ], \BrG, and radio  continuum emission is located.   This inner
cavity is bounded by the brightest \Cii\ and \Oi\ emission in S106 
\citep{Simon2012,Schneider2018}.  The cavity walls have very low
proper motions.  The relatively low free-free, \BrG , and \Ha\ emission
in the cavity interior compared to the S106 nebular lobes suggests
that the cavity has low density.   It may have been evacuated by
the slow stellar wind powered by S106~IR.

On larger scales of several arcminutes, the S106 \Hii\ region 
is located in the interior of a roughly cylindrical 
cavity with  limb-brightened walls at mid- to far-IR wavelengths.   
The cavity walls are seen clearly in mid to far-IR images \citep{Adams2015} and
in molecules \citep{Schneider2002}. In Spitzer 3.6 to 8 $\mu$m images, 
the bright part of the cavity containing the \Hii\ region
is surrounded by straight and  nearly parallel walls,
is about $\sim$ 90\arcsec\ ($\sim$ 0.48 pc) wide 
and $\sim$400\arcsec\ (2.1 pc) long,  with its axis of symmetry oriented 
towards PA$\sim$ 15$^o$ to 20$^o$ (see Appendix C). 
S106~IR is displaced from the axis of 
symmetry toward the east by about 25\arcsec\ (0.13 pc).  The Spitzer/IRAC 
3.6 and 4.5 $\mu$m images show a concentration of stars in S106 with the 
centroid of the distribution centered within the cylindrical cavity 
and 15 to 30\arcsec\ west of S106~IR.   It is possible that
this cylindrical cavity is the fossil remnant of a bipolar outflow
powered by S106~IR produced during its main accretion phase as it grew from
a sub-stellar mass object to its current mass.   Various color combinations
of the mid- to far-IR data are presented in Appendix C.

\subsection{Molecular Hydrogen Images}

Figure \ref{fig_H2_subtracted} presents a continuum-subtracted \Htwo\ 
image of S106.  The \Htwo\ morphology is different from both the radio and hydrogen 
recombination line emission.  The brightest \Htwo\ emission is within 
$\sim$30\arcsec\ of S106~IR.  To the south, west, and north, the \Htwo\ 
emission closely follows the \BrG\ and radio continuum which reveals a 
limb-brightened cavity with a radius ranging from  8 to  11\arcsec .  
The brightest  \Htwo\ emission occurs 1\arcsec\ to 3\arcsec\ outside 
this cavity. Such a separation between the hydrogen ionization front (I-front) 
and the peak of the \Htwo\ emission is consistent with PDR models.  Assuming 
that the penetration column density (the column density between the I-front 
and the peak of the fluorescent \Htwo\ emission) of non-ionizing far-ultraviolet 
(FUV) radiation is $N_{PDR} =2  \times 10^{21}$~cm$^{-2}$ ($A_V \sim$1), 
the volume density of the gas between the I-front and the \Htwo\ photo-center 
must be $n(H) \approx N_{PDR} / \Delta x_{PDR}$ = 2 to 6$\times 10^4$~cm$^{-3}$.

The \Hii\ region is surrounded by a scalloped, inverted C-shaped PDR with 
a radius of about 1 to 1.5\arcmin\ wrapping around  the \Hii\ region from the south, 
through the west, and to the north.   
To the south, the PDR forms a clumpy ridge.  To the east and northeast, the 
PDR consists of filaments pointing away from the nebular core and oval cavity.
These features appear to wrap around the dense molecular gas and dust adjacent to the
east wall of S106 \citep{Schneider2002,Schneider2018,Simon2012}.   They wrap around
the prominent finger of dust pointing at S106~IR seen in absorption in the 2 $\mu$m
images and emission in dense gas tracers and 350 $\mu$m dust continuum.  
\citet{Schneider2018} interpreted this structure as a streamer falling into the
core of S106 with an infall rate of $\sim2.5 \times 10^{-4}$~\Msol ~yr$^{-1}$.

Outside the inner 30\arcsec\ core of S106, the South Bar is the brightest 
part of the entire PDR 
structure surrounding S106. The angular separation between the I-front traced 
by \Ha\ and \BrG\, and the peak \Htwo\ emission ranges from 5\arcsec\ to 15\arcsec, 
implying nearly an order of magnitude lower density between the I-front and \Htwo\ 
peaks than in the PDR surrounding the central elliptical cavity. 
Figures \ref{fig_H2_BrG_color1}, \ref{fig_H2_BrG_color2}, and \ref{fig_H2_Ha_color} 
show color composites made from 
the continuum subtracted \Htwo, \BrG, and \Ha\ images. 

The CO maps of \citet{Schneider2002} show the presence of a clump of molecular gas 
just south of the South Bar.   In the \Ha , [\Nii ], and \BrG\ images, the South Bar 
may be the ionized  surface of a protrusion of dense gas being overrun by the 
expanding plasma flow in S106.   The small proper motions
seen in the ionized gas is consistent with this interpretation.  The corrugated and
filamentary \Htwo\ emission extending at least 1\arcmin\ father south indicates
that non-hydrogen ionizing FUV radiation penetrates farther into the large, cylindrical
cavity seen in the mid- to far-IR and in molecules, presumably either in front, 
or behind the clump creating the South Bar.  

\subsection{Candidate Molecular Hydrogen Objects (MHOs)}

There are several compact \Htwo\ emission knots beyond the PDR surrounding 
the \Hii\ region. These are listed as entries 1 through 6 in Table \ref{tab:MHOs} 
and given the formal designations MHO~4079 through 
MHO~4084\footnote{\url{http://astro.kent.ac.uk/~df/MHCat/}}.  
MHOs~4079, 4080, and 4081 are located west of S106~IR, beyond the western 
PDR traced by the \Htwo\ and \BrG\ images.  MHOs~4082, 4083, and 4084 are 
located east of the eastern PDR.   These Molecular Hydrogen Objects 
(MHOs) likely trace shocks in protostellar 
outflows from YSOs in the S106 cluster, or possibly from  S106~IR or S106~FIR.
A jet-like feature (entry 7 in Table \ref{tab:MHOs}) points to MHO~4083 
and is thus given the same designation.  MHO~4085 refers to the collection
of objects likely to originate from S106~IR and are listed as entries 9 to 12 in 
Table \ref{tab:MHOs}.   See \citet{Davis2010} for a description of the 
MHO catalog.   
Figure \ref{fig_H2_subtracted} shows the location of these MHOs along with 
their entry numbers in Table \ref{tab:MHOs}.   

MHO~4081 is located within 5$^o$ of the axis of the outflow 
from S106~FIR \citep{Furuya1999,Furuya2000}.     
A bow-shaped \Htwo\ protrusion or finger 
points to this knot and is therefore also designated MHO~4081.  
The eastern portion of this MHO extends for $\sim$ 30\arcsec\ towards 
the compact knot in the west which is located 93\arcsec\ (0.49 pc) 
from S106~FIR and 107\arcsec\ (0.57 pc) west of S106~IR.
Entry 8 (the main body of MHO~4081) is seen in projection toward 
the 60\arcsec\ `dark bay' located due west of S106~IR and S106~FIR.  
A bow-shaped HH object seen in \Ha\ and equally in [\Sii ], HH~1214 
discussed below, is located just beyond and below the
west tip of this streamer.

MHO~4084,  east of S106~IR,  is about 5$^o$ north of the axis defined by
S106~FIR and MHO~4081.  MHO~4084 is 120\arcsec\ from S106~IR (0.63 pc) 
and 135\arcsec\ (0.71 pc)  from S106~FIR. 
S106~IR is $\sim$8\arcsec\ south of a line connecting the knot in 
the western part of MHO~4081 and MHO~4084.

Entry 7, labeled MHO 4083 in Table \ref{tab:MHOs} and 
Figure \ref{fig_H2_subtracted}, is a linear  feature resembling a jet.  
It points within one degree of the compact, bow-shaped knot (entry 5) 
and is thus given the same designation, MHO~4083. 
At very low levels, there is faint \Htwo\ emission connecting the jet-like
feature to knot 5 in Figure \ref{fig_H2_subtracted} 
(see the deep-cut figure in the Appendix).   Thus, this
knot which is elongated in the direction of the jet-like-feature may 
be a terminal bow shock in a highly collimated, jet-like outflow.

A number of fingers of \Htwo\ emission within about 40\arcsec\ of S106~IR 
point directly away from this source and therefore are collectively given
the designation MHO~4085, centered at 20:27:26.8, +37:22:48
(MHO~4081 is excluded from this because of its
possible association with S106~FIR).
The arrows in Figures \ref{fig_H2_subtracted} and \ref{fig_H2_BrG_color2} show 
the locations and orientations of several bow-shock-like \Htwo\ streamers 
originating from the vicinity of S106~IR.   Entries 9 and 10 in 
Figure \ref{fig_H2_subtracted} mark bow-shaped streamers whose
axes of symmetry point directly away from S106~IR.  Several other,
unmarked linear features between entries 9 and 10 also point away from S106~IR.
Entry 11 marks one of the brightest \Htwo\ knots
in S106.  This object lies well outside the ionized zone in S106 but
in the interior of the PDR.
A line connecting S106~IR to entry 11 contains entry 12, located at the 
eastern tip of a collection of compact \Htwo\ knots which together outline a
bow-shaped structure pointing away from S106~IR.
Table \ref{tab:MHOs} lists the coordinates of the features marked in Figure
\ref{fig_H2_subtracted}.  

% Table 2
\begin{table}
	\centering
	\caption{Molecular Hydrogen Objects (MHOs) in and Near S106 }
	\label{tab:MHOs}
	\begin{tabular}{llll} % four columns, alignment for each
		\hline
\#  &	    R.A.    & Dec.     & Comments   \\
		\hline
	
1   &   20:27:16.7  &   37:22:49    &  MHO 4079: Compact knot west of S106~IR \\
2   &   20:27:16.9  &   37:23:07    &  MHO 4080: Compact knot north of MHO 1  \\  
3   &   20:27:17.9  &   37:22:36    &  MHO 4081: Compact knot south of MHO 1 \\  
4   &   20:27:32.7  &   37:20:53    &  MHO 4082: Compact knot south-southeast of S106~IR \\
5   &   20:27:34.9  &   37:21:24    &  MHO 4083: Southeast of S106~IR.  Jet (\#7) and terminal shock? \\
6   &   20:27:36.5  &   37:23:18    &  MHO 4084: East of S106~IR.  Counterflow from S106~FIR ?\\
7   &   20:27:28.9  &   37:22:50    &  MHO 4083: Jet-like filament in PDR   \\
8   &   20:27:26.0  &   37:22:49    &  MHO 4081: West-facing bow from S106~FIR? \\
9   &   20:27:26.4  &   37:22:54    &  MHO 4085: Northwest-facing bow from S106~IR\\
10  &   20:27:26.2  &   37:22:47    &  MHO 4085: Southwest-facing bow from S106~IR \\
11  &   20:27:28.1  &   37:22:40    &  MHO 4085: Bright knot, part of southeast-facing bow? \\
12  &   20:27:29.1  &   37:22:35    &  MHO 4085: Tip of southeast-facing bow?\\
    \hline
	\end{tabular}
\end{table}

\subsection{The `Dark Bay' west of S106~IR and S106~FIR}

\begin{figure*}
     \includegraphics[width=7in]{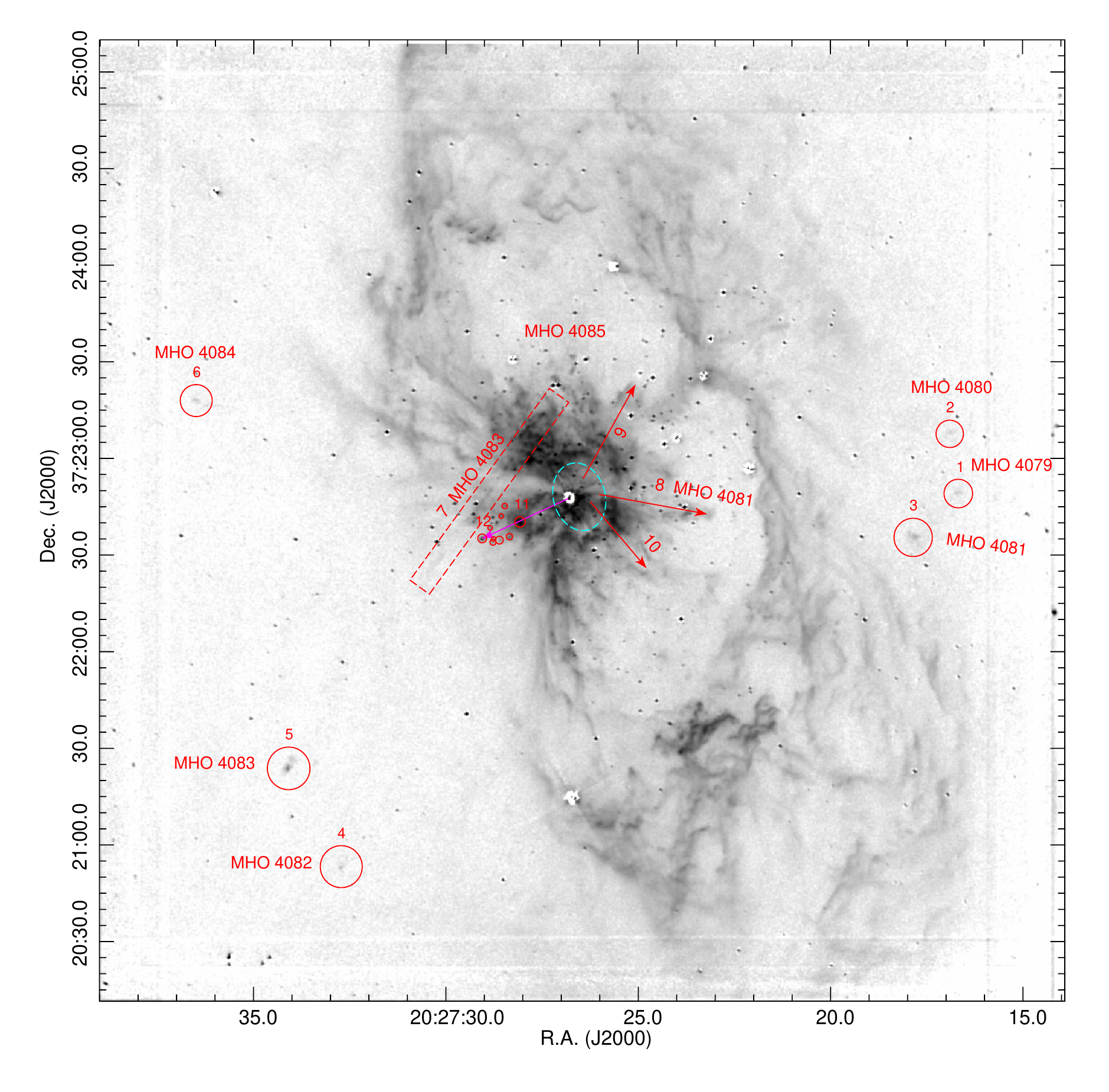}
    \caption{A continuum-subtracted near-infrared image showing S106  
    the 2.12 \um\ \Htwo\ emission line.
    Large red circles mark \Htwo\ knots located outside the main
    PDR.  
    Objects labeled 1 through 6 are molecular hydrogen objects (MHOs), 
    likely shocks powered by protostellar outflows.  
    Objects \#7 through \#12 are seen towards
    the projected interior of the PDR surrounding S106.
    The dashed rectangular box shows
    a jet-like feature which is aimed at object
    \#5 in the lower left of the figure.  
    This is MHO~4083.  MHOs 4081 (\#3) and MHO~4084 (\#6)
    are located close
    to the axis of the H$_2$O outflow from S106~FIR.
    The three red arrows mark \Htwo\ fingers that point away from the
    vicinity of S106~IR.   The dashed cyan oval shows the 
    8\arcsec\ by 11\arcsec\ cavity discussed in the text and shown in
    Figure \ref{fig_BrGgrey}.   
    The small, unnumbered circles mark a candidate bow-shock
    whose axis point back to S106~IR (magenta arrow).  The brightest 
    knot,  \#11, is located in the interior of this bow-shaped feature
    pointing away from S106~IR.  The tip of this structure is object
    \#12.  The \Htwo\ features pointing radially away from S106~IR
    are collectively designated MHO~4085 (except for MHO~4081 which
    may be associated with S106~FIR).
    A version of
    this image emphasizing the faint emission is shown in the Apppendix.
    }
        \label{fig_H2_subtracted}
\end{figure*}

\begin{figure*}
	\includegraphics[width=7in]{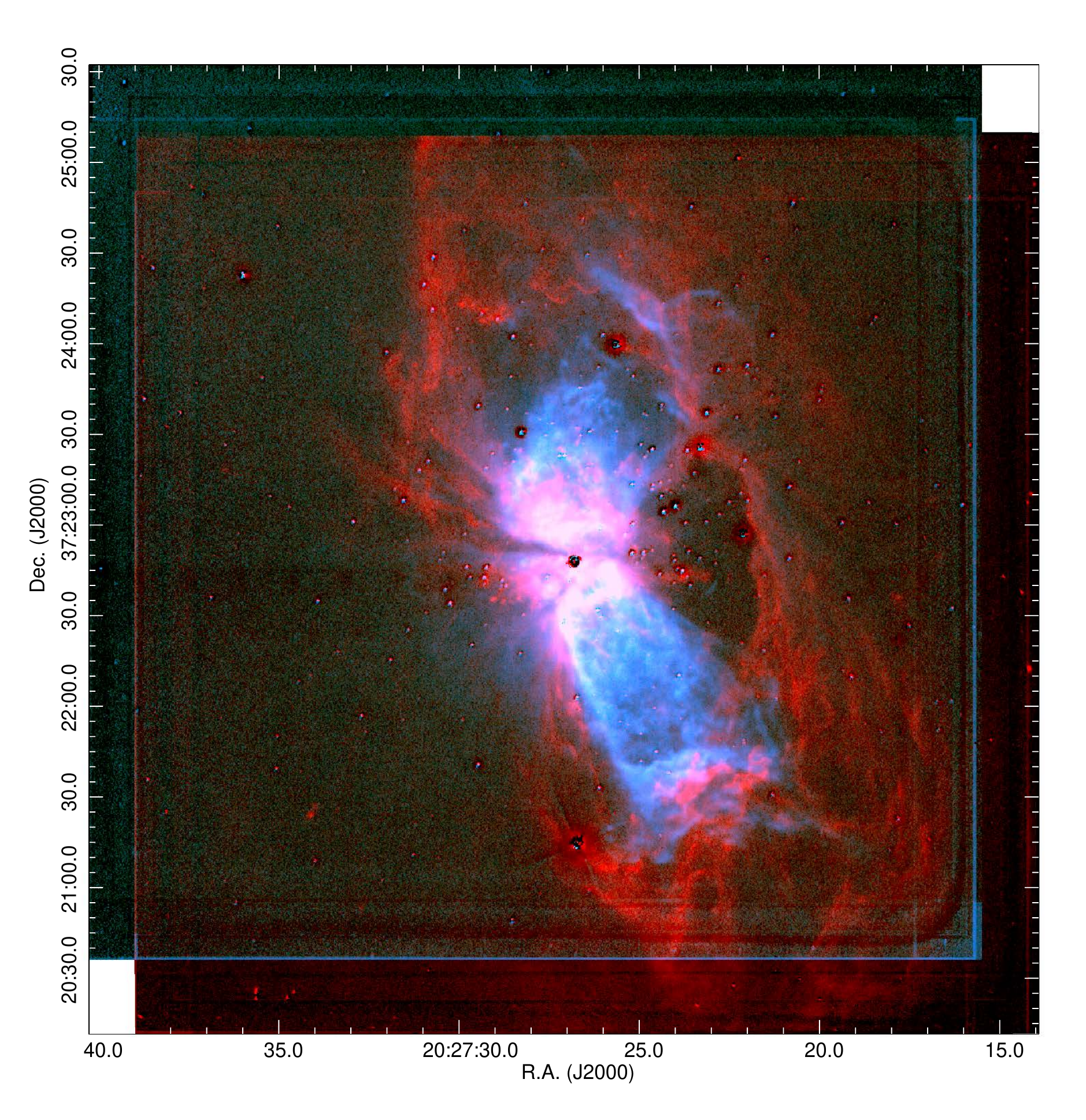}
    \caption{A color composite image showing the continuum 
    subtracted \BrG\ image (blue and green) combined with the
    continuum subtracted \Htwo\ image (red).   Stars are still
    visible due to the variable seeing and slight registration 
    errors.  }
    \label{fig_H2_BrG_color1}
\end{figure*}

\begin{figure*}
	\includegraphics[width=7in]{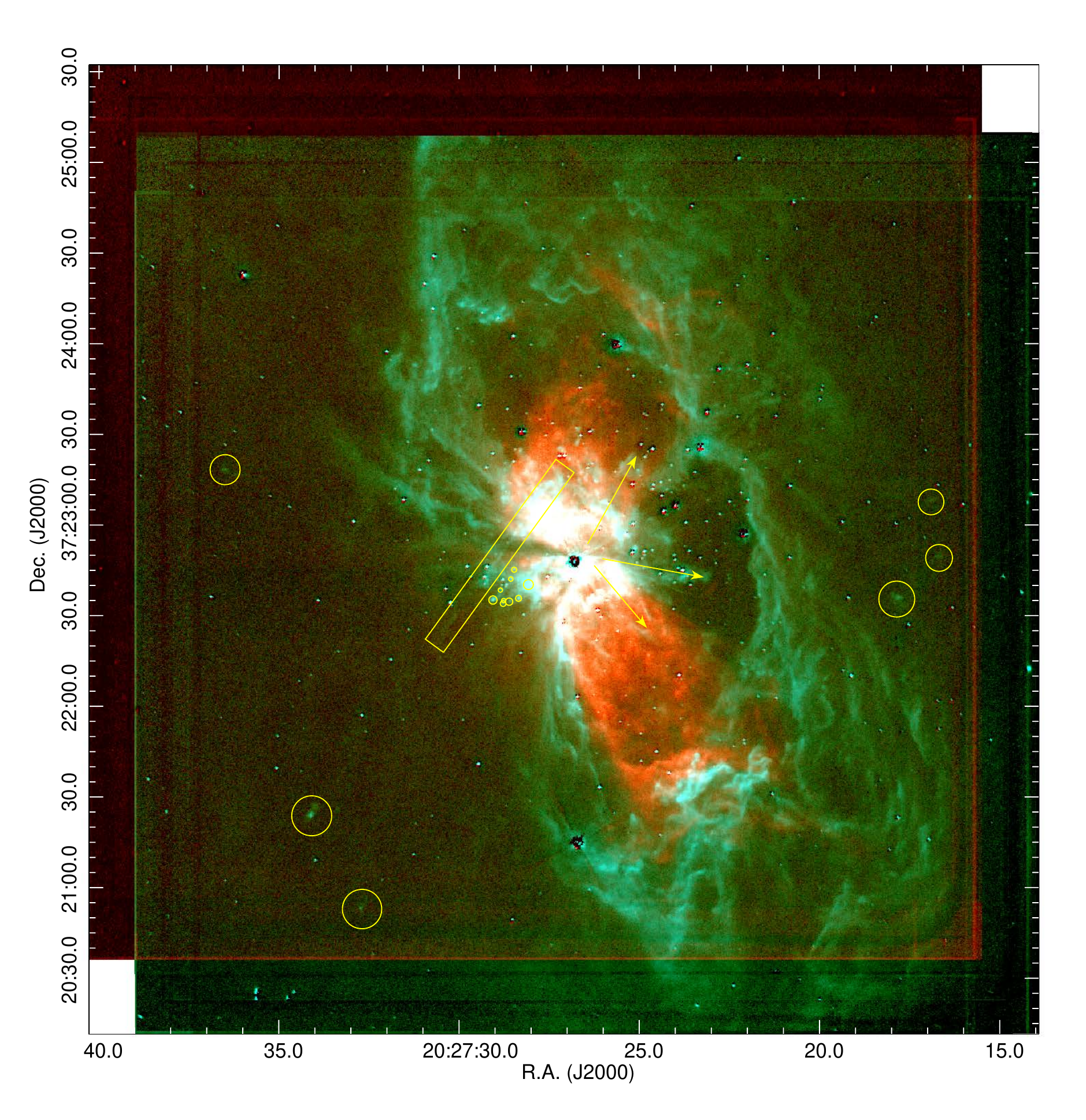}
    \caption{A color composite image showing the continuum 
    subtracted \BrG\ image (red)  combined with the
    continuum subtracted \Htwo\ image (blue and green).
    The yellow circles mark compact 2.12 $\mu$m \Htwo\ 
    features that may trace shocks.  The three yellow
    arrows mark \Htwo\ fingers that point away from the
    vicinity of S106~IR.   The yellow box marks the
    linear feature suspected to trace an \Htwo jet.
      }
    \label{fig_H2_BrG_color2}
\end{figure*}

\begin{figure*}
    \includegraphics[width=7in]{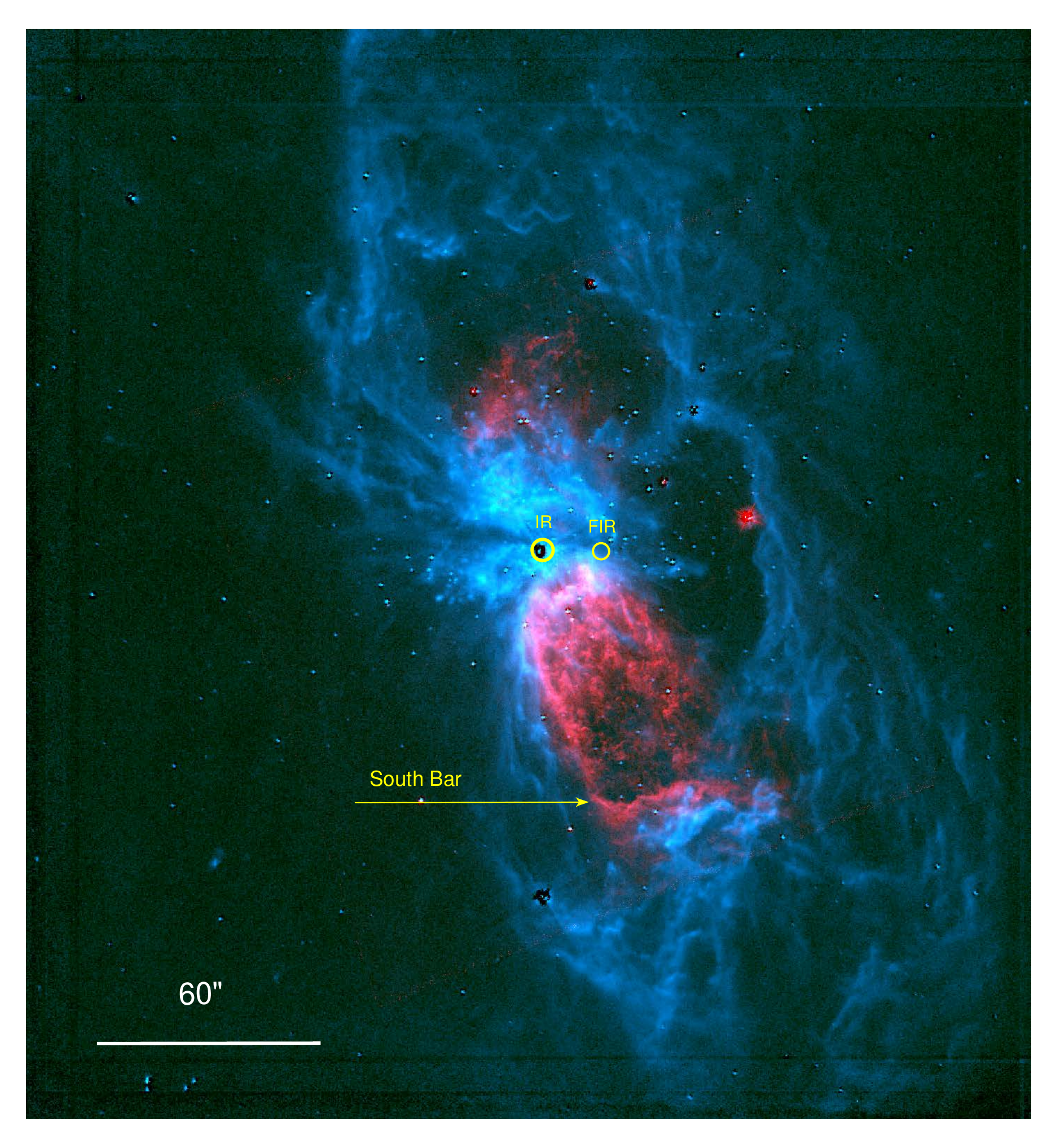}
    \caption{A color image comparing the APO \Htwo\ image (blue),
    and the HST \Ha\ image (red). }
    \label{fig_H2_Ha_color}
\end{figure*}

The near-IR images reveal a large, roughly 60\arcsec\ diameter `dark bay' 
west of S106~IR opening towards the west.  The west rim of the region is 
bounded by the \Htwo\ PDR.   To the north and south,  the cavity is bounded 
by the ionized plasma in S106,  contributing to its C-shaped symmetry
(See Figures \ref{fig_BrGgrey} to \ref{fig_H2_Ha_color}).   This cavity 
may be shielded from direct UV illumination by S106~IR by a combination of the
nearly edge-on disk around S106~IR, the cloud core harboring S106~FIR 
located 16\arcsec\ to the west of S106~IR, and possibly a streamer of 
gas and dust flowing in from the east that overshoots S106~IR.
The cloud core is seen as a compact, bright  
350 $\mu$m peak in the maps presented by  \citet{Simon2012}.  The peak 
emission in the core is located between S106~FIR and S106~IR.

The \Htwo\ emission in the PDR at the west end of the 60\arcsec\ cavity 
indicates illumination
by FUV radiation.  If the cavity is primarily created by the shadow of 
S106~FIR source, then FUV may propagate 
either in-front of or behind  the shadowed region to produce the 
PDR on the west side of the cylindrical far-IR cavity.  Alternatively, 
the PDR may be illuminated by FUV radiation produced in the \Hii\ 
region lobes by recombining hydrogen to produce Lyman-alpha light, and/or
FUV from S106~IR scattered by dust.  

The deep-cut continuum-subtracted \BrG\ and \Htwo\ images in the Appendix 
show that there is faint \BrG\ emission associated with the PDR traced
by \Htwo .  In PDRs, the ionized gas associated with ionization fronts
is located between the source of Lyman continuum and the peak \Htwo\ emission.
In the South Bar, there is a  several arcsecond offset between the peak of 
the \Ha\ and \BrG\ emission and the peak in \Htwo\ with these 
hydrogen recombination lines peaking closer to S106~IR than \Htwo .  In
contrast, the images in the Appendix show that the \BrG\ emission associated with 
PDR $\sim$60\arcsec\ west of S106~IR is either coincident with, or 
{\it slightly farther} from S106~IR than the \Htwo\  emission.  This may 
indicate that the faint \BrG\ emission is not an indication of local ionization
but a \BrG\ reflection nebula.   \BrG\ emission from the S106 lobes may be 
scattered by  dust in the PDR.

The dark bay may contain lower density neutral gas than the CO cloud
east and west of the S106 far-IR cavity.  The density has to be sufficiently high
to exclude indirect ionization by recombination-generated Lyman continuum
shining on this region from the bipolar lobes of S106.  Yet, the density
has to be low enough to allow non-ionizing FUV photons to produce the PDR.
Such gas may emit in the 157 $\mu$m \Cii\ line.  \citet{Simon2012} presented
evidence for such emission.

\subsection{Is S106~IR Moving with Respect to the S106 Cluster?}

The Gaia EDR3 proper motion of S106~IR is 
$\mu _\alpha  cos(\delta )$ = -2.183$\pm$0.13  mas~yr$^{-1}$ and 
$\mu _\delta$ = -5.861$\pm$0.15 mas~yr$^{-1}$, implying a motion of 6.25 mas~yr$^{-1}$ 
towards PA = 200$^o$ in coordinates referenced to the Solar system barycenter.
However, it is unclear how reliable the Gaia proper motion measurement is given the
extended nebulosity surrounding this star.  
S106~IR has phot-g mean magnitude' is 17.87 and the formal proper motion errors may
be an underestimate.

On the registered HST images in the S106 reference frame described 
above, S106~IR moved 
0.067\arcsec\ in the 15.6-year interval between the first and second epoch
images.  This implies 
a proper motion of 4.3 mas~yr$^{-1}$ or $\sim$21 \kms\ towards position angle of
PA = 60\arcdeg .   At this speed the star would have moved about 30\arcsec , 
roughly the distance to the axis or center of the infrared cavity surrounding S106,
in about 7,400 years.  It would take about 4,000 years to cover the current 16\arcsec\ 
distance from the embedded class-0 source S106~FIR located due west of S106~IR. 
Such a motion might explain the C-shaped symmetry seen in 
both the nebular plasma and in the nebular proper motions.   
However, the error on the proper motion of S106~IR is $\sim$2 \masyr . 
Therefore, it is unclear if the apparent proper motion of S106~IR 
is real or an artifact of errors in the image de-distortion combined with 
uncertainties in determining the S106 reference frame.  New observations 
of the position of S106~IR using near-IR, visual, or radio wavelength are 
needed. New radio observations with the JVLA could be compared to the 1980s 
data taken with VLA and Merlin.  New HST or JWST observations could be compared 
to the 1995 and 2011 epoch images. 

\subsection{Candidate Herbig-Haro (HH)} 

We detected several candidate Herbig-Haro (HH) objects, shock-excited nebulae 
powered by outflows from forming and young stars, located outside the photo-ionized 
body of S106.  HH~1214 is an arc of emission located about 45\arcsec\ 
west of S106~IR at  $\alpha$=20:27:22.96,  $\delta$=+37:22:39.0  that 
resembles a bow shock. This 5\arcsec -wide feature lies within a few degrees 
of the orientation of the water maser micro-jet and compact CO outflow emerging 
from the Class 0 source S106 FIR \citep{Furuya1999,Furuya2000}.  Its bow-shape 
is consistent with being powered by this YSO.  
This feature is also seen in the \BrG\ image shown in Figure \ref{fig_BrGgrey}.
The candidate bow shock is $\sim$31\arcsec\ from S106~FIR.  The proper motion
is difficult to measure because of its diffuse morphology.  Proper motions are
less than $\sim$20 \kms .  

\begin{figure*}
	\includegraphics[width=6in]{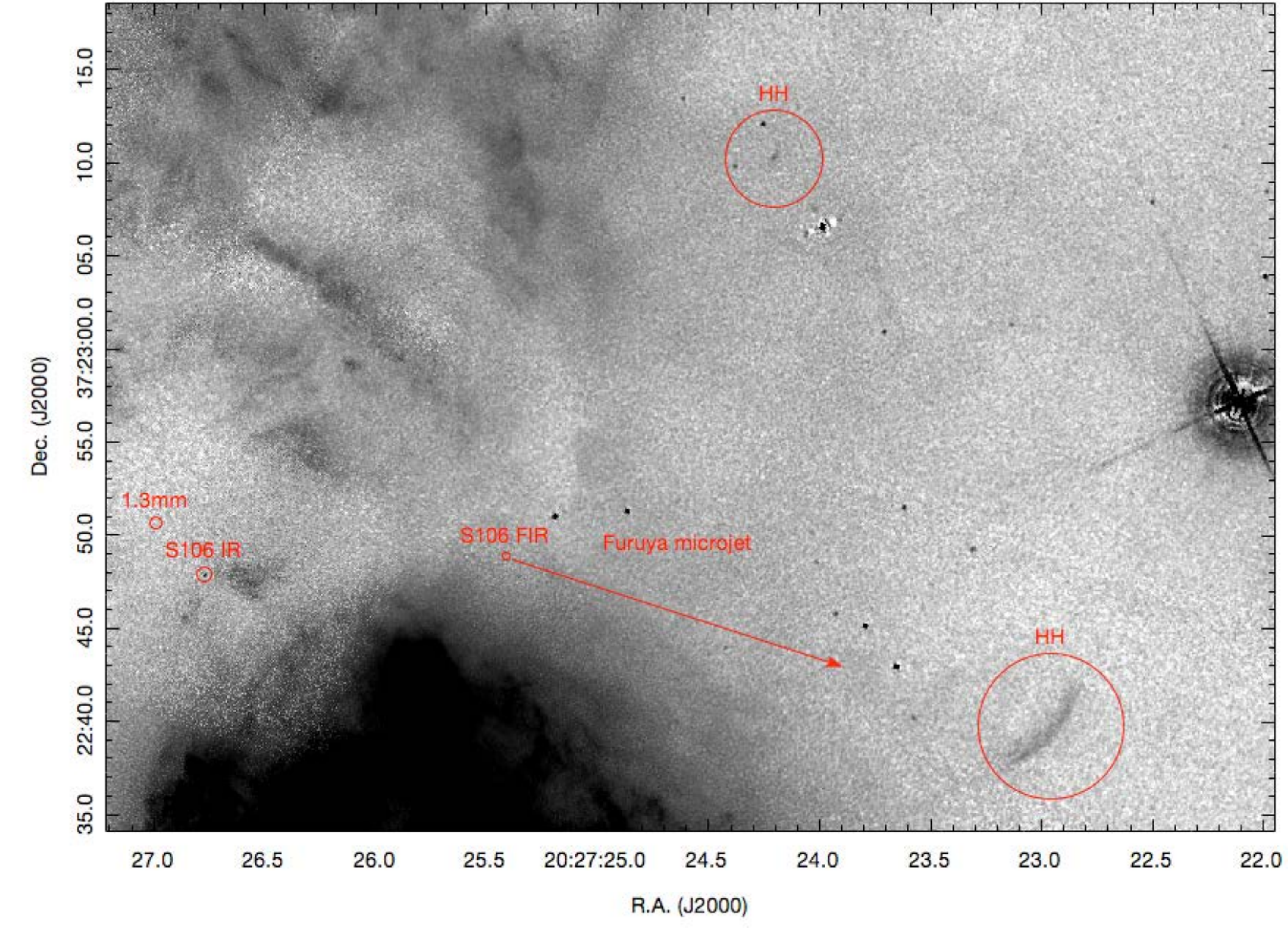}
    \caption{A portion of the 2011 epoch HST [\Nii ] image showing the candidate HH objects,  HH~1214 (lower-right) and HH~1215 (top-center),
    the locations of S106~IR, the 1.3 mm source found by \citet{Beuther2018}, and
    S106~FIR.   The orientation of the H$_2$O maser and CO micro-jet is indicated \citep{Furuya2000}.}
    \label{fig_HHs}
\end{figure*}

A candidate HH object, HH~1215 (\# 1 in Table~8 and Figure \ref{fig_HHs}),
is located northwest of S106 IR at $\alpha$=20:27:24.20,
$\delta$=+37:23:10.5.  This object consists of a small group of unresolved knots 
in a 0.5\arcsec\ diameter region which exhibits a proper motion of $\sim$176 \kms\ 
towards PA = -36\arcdeg\ (northwest), making this knot the highest proper motion
object in S106. It is shown in Figure \ref{fig_CCPMs}. % Fig 4. 
However, its not clear if it is related to S106~IR. 
These HH objects are dimly visible in the F110W 
and F160W filter images, which may indicate the presence of shock-excited [\Feii ] 
emission.

\newpage
% Section 6  sssssssssssss

\section{Discussion and Interpretation}

\subsection{The Nature of the Supersonic Motions}

\Hii\ regions are expected to expand with a velocity comparable to the sound speed 
in photo-ionized plasma.  At near-solar metallicity, 
forbidden transitions from trace metals and metal ions with $\sim$2~eV transitions 
such as the [\Oi ], [\Oii ], [\Sii ],  [\Nii ] and [\Oiii ] 
in the near-UV and visual bands, tend to set the temperature 
of the plasma to around 6,000 to 10,000 K. The resulting sound 
speed is  $C_s = (k T / \mu m_H)^{1/2} \approx$~10 \kms.  Density and pressure 
gradients can increase the expansion velocity by up to a factor of two. Doppler 
shifts of expanding \Hii\ regions confirm that typical blister \Hii\ regions such 
as the Orion Nebula exhibit motions of around 5 to 20 \kms\ away from their 
parent molecular clouds with faster motions occurring in
a few percent of the emitting plasma \citep{Odell2001,Pabst2019,Pabst2020}.  
Thus, the highly supersonic expansion shown in Figure~\ref{fig_CCPMs} 
is remarkable.

Explosions are not the only mechanism that can produce `Hubble flows'.
\citet{MacLow1994} argued that the motions of the W49N 22 GHz water masers
were produced by shocked molecular gas accelerated by an an expanding 
cocoon at the head of a high-speed protostallar jet.
\citet{McCaughrean_MacLow1997} used numerical modeling to argue that
the Orion fingers may have been produced by a fast wind interacting with 
a previously launched slower wind.  

We explore several possible explanations for the 
supersonic expansion of S106: 
 
(1)  The fast motion could be an artifact. Large residual errors 
in de-distortion and  registration  may remain despite 
our best efforts to correct them.   However, the 
stellar positions on the images match their expected locations 
on the de-distorted images to about 0.02\arcsec .   
Nevertheless, the lack of stars in the extreme southwest corner 
of S106 may make measurements much less reliable. 
The small proper motions in the South Bar gives us
confidence in the registration of the images.  New HST 
images would provide confirmation of the de-distortion 
and registration accuracy and result in a more precise 
characterization of the proper motion field.

(2)  There could be a systematic offset between the \Ha\ and [\Nii ]. 
To explain the observed proper motions, 
such a model must predict a systematic variation of the gap size 
between the \Ha\ and [\Nii ] emission with the intensity of the emission 
and distance from S106~IR.   As discussed above,  
the ionization potentials of \Hi\  and \Ni\  are
similar, with \Ni\ being higher by a fraction of an eV.
Thus, in the absence of any true motion, the [\Nii ] emission, would be 
expected to peak closer to the ionizing source than the \Ha\ emission,
not farther away. 
Because the [\Nii ] image was acquired {\it after} the \Ha\ image, any
offset between \Nii\ and \Ha\ would result in an underestimate of the
proper motions.

(3)  In the presence of a steeply decreasing density or pressure gradient, 
the leading edge of a freely expanding plasma cloud undergoing 
thermal expansion can reach speeds of several times the sound speed 
in the plasma ($\sim$10 \kms ).  In models of the free-expansion of
an isothermal cloud,  the mass involved in supersonic expansion 
decreases  exponentially with increasing speed.  
But, the S106 expansion is seen in most of the emitting plasma, making 
gradients an unlikely explanation of the proper motions.   

(4) Acceleration of gas along the walls of a photo-ionized cavity by 
stellar wind from S106~IR.  The radio, infrared, and visual images of 
the southern lobe show that there is an evacuated region surrounding 
S106~IR with a radius of about 11\arcsec\ (see Figures \ref{fig1} and 
\ref{fig_H2_subtracted}).  The brightest emission is seen from cometary structures 
lying just outside this region.  This bright region may mark the interface 
between the low-density, wind-dominated zone around S106, and the denser
nebular plasma which dominates the emission-line fluxes.  However, if the wind
reached farther into the nebular lobes,  this model predicts that the 
proper motions should decrease with increasing distance from the source 
because the ram-pressure of a steady, constant velocity wind decreases 
as an inverse square law. This is contrary to the observed proper motion field.

(5) Projection effects, combined with the presence of a 
quasi-spherical, constant velocity, swept-up shell 
in the foreground driven by a stellar wind can also produce proper motions 
increasing with projected distance from the wind source. 
Consider a quasi-spherical shell swept up by the forward shock of a 
stellar wind from S106~IR.  Along the lines of sight close to the source, 
the shell-expansion velocity vector is aligned close to the line of sight 
and will exhibit a large blueshift from the foreground portion of the 
shell (and comparable redshift from the background portion).  The proper 
motion is proportional to the sine of the vector's angle with respect to 
the line of sight (or the cosine of the angle of the vector with respect 
to the plane of the sky).  The maximum proper motion will be seen at
the projected edge of the shell.  This model requires that most of the 
emitting plasma be located in the shell which surrounds a low-density region.  
The main problem with this model is that the high radial velocity gas,
suspected to trace the zone dominated by a dense wind, is 
confined to the inner 30\arcsec\ of S106.   Additionally, the nebula 
morphology is far from that of a
quasi-spherical shell.  Although the south and Northwest Bars do appear to 
be limb-brightened, the nebula is better described as consisting of a 
pair of bipolar, cylindrical cavities. 

(6)  The bipolar cavities of S106 could represent recently photo-ionized 
parts of a bipolar molecular outflow launched by S106~IR before it started 
to emit Lyman continuum radiation.  As most stars grow by accretion from
their parent molecular cloud cores they power supersonic jets and bipolar
winds \citep{Bally2016} with speeds ranging from tens to hundreds of \kms .  
As these primary flows interact with the molecular gas in the parent cloud, 
they produce bipolar molecular outflows that can reach parsec-scale 
dimensions.   The momentum injection rates of such protostellar outflows
increases with the source mass and luminosity \citep{Maud2015a,Maud2015b}.  
Thus, the most massive
stars in a region will generally drive the most powerful and largest outflows. 
Such a flow likely  produced the cylindrical cavity in which S106 is located.  

As a massive protostar reaches the main sequence, it starts to ionize 
its surroundings, which has been pre-processed by the star's protostellar 
outflows.  Thus, the large nebular proper motions in S106 could have 
been produced by a previous bipolar molecular outflow phase.  In this model, 
the ionized lobes of S106 trace the recently ionized walls of a fossil
bipolar molecular outflow cavity.  This model can accommodate a 
variety  of proper motion velocity fields, depending on the history of 
the mass-loss rate and ejection velocity of S106~IR prior to its
reaching the ZAMS, the geometry and opening angle of the flow, 
and the orientation of the outflow axis with respect to our line-of-sight.  
The main problems with this model is the very short
($\sim$3,500 year) dynamical age of the S106 nebular lobes.  
Additionally, this model would predict
that there still should be a remnant bipolar molecular flow beyond the 
\Hii\ region's ionization fronts in or along the walls
of the cylindrical cavity in which the \Hii\ region sits.  No such flow had been
seen in any species such as carbon monoxide.

(7) S106 may have experienced a `Hubble flow' explosion similar to the Orion 
fingers emerging from Orion Molecular Core 1 (OMC1) located $\sim$0.1 pc 
behind the Orion Nebula  \citep{Bally2020, Bally2017, Bally2015}. 
In S106, the ejecta are now being photo-ionized by the central 
OB star, S106~IR. This model predicts that proper motions lie between 
0 \kms\ and a line indicating increasing proper motions with 
projected distance from the source. Figure \ref{fig_CCPMs}, 
shows just such a pattern. The fastest motions are seen at the
extreme southwest corner of the southern lobe of S106 and along the northern 
lobe's extreme northern edge.   The images show the presence of multiple
bow-shaped protrusions at the southern end of the S106 \Hii\ region where
compact ejecta may be interacting with slower or stationary gas.

The morphology of the plasma in S106 is unique among \Hii\ regions
associated with massive young stars.  Most \Hii\ regions, such as the 
Orion Nebula, M16, M17, and others exhibit a relatively smooth plasma morphology. 
In contrast, as discussed by \citet{Bally1998}, the plasma
in S106 is highly structured and appears fragmented.  It consists of compact 
knots, along with forwarding and reverse-facing bow-shocks.    This morphology 
is reminiscent of some planetary nebulae, containing hundreds of compact 
knots and features (an example is the Helix Nebula).  The knotty, complex 
structure of such planetary nebulae is interpreted in terms of the impact 
of fast winds and outflows on earlier ejected shells of slower-moving but 
denser ejecta. A combination of Rayleigh-Taylor instabilities, cooling 
instabilities, and radiation hydrodynamics is thought to be responsible 
for sculpting these high-contrast structures. 

Recent observations of the nearest massive protostars, namely those
in Orion OMC1, have shown that a dynamical interaction $\sim$550 years
ago ejected two massive and one moderate mass star with speeds of
10 to 55 \kms\  \citep{Bally2020}.   The stellar ejection was accompanied 
by the launch of an explosive outflow with `Hubble flow' CO streamers 
whose proper motions and radial velocities increase with projected distance 
from the ejection center \citep{Zapata2009,Bally2017,Bally2020}. These streamers power the 
shock excited 'fingers' of molecular-hydrogen 
emission in this explosive outflow \citep{Bally2015}. 

Apparently, S106 exhibits similar behavior.   
In Orion, the  `Hubble flow' pattern with velocities increasing in proportion 
to projected distance is traced by CO, \Htwo,  and [\Feii ].   In S106, this 
pattern is revealed by proper motions of the photo-ionized plasma. The fastest 
proper motions occur in the bow-shaped southwest portion of the S106 
southern lobe located about $R \approx$ 96\arcsec\ from S106~IR.  The 
motions in this part of the nebula reach values of $V \approx$150 \kms. 
This implies a dynamic age for the S106 explosion of order 
$t_{dyn} = R / V$ $\sim 5 \times 10^3$ years.
 
Explosive outflows can be produced by major accretion events such as 
those that occurred in 2015 in  Sh2-255 IRS1, or NGC 6334 I.  These 
events apparently caused the accreting stars to increase their luminosity 
by $\sim 4 \times 10^4$ \Lsol\ for about 6 months 
\citep{Caratti_o_Garatti2017,Hunter2017}. 
Alternatively, as in Orion, the explosion in the gas was associated 
with the ejection
of stars by an N-body interaction. If S106~IR was ejected from its
parent core and accretion subsided, its photosphere could rapidly 
develop main-sequence properties such as the emission of intense 
Lyman continuum radiation that could start to ionize its surrounding. 
Having exited the core, its radiation field could ionize its previously 
ejected bipolar outflow or debris launched during the stellar ejection event.  

\subsection{Energetics}

Radio measurements of the mass of plasma in S106 imply a nebular mass of 
about 3.1 \Msol .
The  mean nebular expansion speed of about 75 \kms\ implies a nebular 
kinetic energy, E(\Hii)$\approx {\rm 1.7 \times 10^{47}}$ erg and a nebular 
momentum, P(\Hii )$ \approx$225~\Msol \kms . 
These rough estimates represent the current kinetic energy and momentum
content of the moving nebular plasma.   However,  when dissipation
due to shocks radiating away some of the original kinetic energy is
taken into consideration,  the energy requirement of the event that
set the nebular clumps into motions is larger than $1.7 \times 10^{47}$ ergs.

Assuming a Hubble flow explosion with no deceleration, the explosion would 
have occurred about 3,500 years ago. Assuming the current mass loss rate of 
S106~IR to be $10^{-6}$~\Msol year$^{-1}$ and a wind velocity of 200 \kms, 
the energy generated in 4,000 years in about ${\rm 1.6 \times 10^{45}}$ erg.  
Thus, the current wind fails by about than two orders of magnitude to explain 
the nebular energetics.

How much mass must be involved in an accretion event to produce the observed
kinetic energy in the S106 \Hii\ region in an energy-conserving interaction
of a stationary medium with the ejecta?
Given a 15 to 20 \Msol\ star with a radius of R$_{star} = 10^{12}$ cm,  the 
release of $2 \times 10^{47}$ ergs of gravitational potential energy requires the
accretion of $m > $0.03 to 0.05 \Msol\ onto the star.  A short-lived 
accretion accretion burst delivering this
amount of mass onto S106 IR could generate the kinetic energy  in the \Hii\ region.

How much mass must be involved in an accretion event to produce the observed 
momentum in the S106 \Hii\ region ($\sim$225 \Msol\ \kms ) in an momentum-conserving 
interaction with ejecta produced by an accretion event?
Assuming that a violent accretion event launches ejecta with speed (far from the
star) $V_{ejecta} = (G M / R_{star})^{1/2}$ $\sim$500 \kms , momentum conservation
would require accretion of $\sim$0.5 \Msol .
 
\subsection{Did S106~IR Have a Bloated Photosphere in the Recent Past?}

Massive protostars accreting at high rates tend to develop extended 
photospheres \citep{HosokawaOmukai2009}.   At accretion rates of 
$\dot M > 10^{-3}$ \Msol ~yr$^{-1}$, accreting massive stars 
develop AU-scale photospheres because they can not get rid of the
entropy developed by their accretion flows.  Currently, S106~IR has a mass
of order 22 \Msol\ and may still be accreting.   At an accretion rate of 
$\dot M > 10^{-3}$ \Msol ~yr$^{-1}$, S106~IR would only have a mass
of 10 \Msol\ 10$^{4}$ years ago.  As it grew past 10 \Msol\ it would have
developed an AU-scale photosphere which would have been cool, resembling
a red supergiant star.  The accretion disk feeding the star likely drove
a very powerful bipolar jet or wind.  Such a bipolar outflow could have 
produced the axi-symmetric cavities seen at mid- and far-infrared wavelengths. 

Models show that as a rapidly accreting, massive protostar grows past 
$\sim$15~\Msol , its AU-scale  photosphere is expected to shrink 
and heat, even if high accretion rates persist. 
As it grew past this mass, S106~IR's photoshpere would have heated and 
started to emit Lyman  continuum photons.   In this scenario,  
the ionization of the S106 \Hii\ region would have only started
within the last few-thousand years.   We hypothesize that during its
bloated phase, S106~IR either captured its current companion from the S106
cluster, or nearly circularized the orbit of a previously
acquired companion.  The interaction between a 15 to 20 \Msol\ star with
a bloated photosphere could have led to a violent interaction.  However,
the density and total mass in  such bloated stellar envelopes is low
and small.  The total mass ejected by the common-envelope evolution of 
a massive star and a $\sim$3 \Msol\ companion over $\sim 10^4$~years
would likely eject much less less than 1 \Msol .   At least one-half of
S106's plasma is moving supersonically.   To deliver the observed momentum
in the supersonic motion, debris from an ejected, bloated photosphere
the would have to have a much faster
speed than the fastest clumps currently seen in S106.    

The ejection speed of debris generated by a short-phase of common-envelope
interaction is likely to be comparable to the Kepler speed at the mean
orbit radius of the companion.   At a mean orbit radius of 0.3 AU, a 
companion orbiting a 20 \Msol\ MYSO the circular mean orbit speed is
$\sim$240 \kms .   Common envelope evolution of a $\sim$20 \Msol\ star
with a $\sim$3 \Msol\ companion would have resulted in a substantial increase
in the luminosity of the system as the companion moves through the 
envelope of the primary.  We speculate that the resulting radiation
pressure could have blown off a portion of a bloated  primary's
envelope.  Additionally, radiation pressure acting on the surrounding
disk and envelope, possibly aided by magnetic fields could have
launched additional mass into S106 lobes.  

As plausibility argument, we note
that the gravitational binding energy of the 3 \Msol\ companion as it orbits 
the $\sim 20$ \Msol\ primary in a circular orbit with a radius of 0.17 AU
is about $E_B \sim 4 \times 10^{48}$ ergs.
If the companion's orbit shrank to this value from one that is substantially larger,
it would have injected much of this energy into the primary's envelope. 
If the common-envelope phase lasted $\Delta t$ = 100 years,  and  
injected this much energy into the primary star's envelope, the 
resulting increase in the star's luminosity would be 
$L \sim E_B / \Delta t$ $\sim 3 \times 10^5$ \Lsol , an order-of-magnitude 
increase over the current luminosity of S106~IR.

An alternative scenario is that S106~IR was ejected towards the east
by a dynamic interaction with a protostar embedded within the cloud core 
located west of S106~IR where S106~FIR is currently located.  Such an
interaction may have been similar to the event that occurred in Orion OMC1
about 550 years ago \citep{Bally2020}.  S106~FIR is located about 16\arcsec\ 
from S106~IR.  

Either process could have ejected S106~IR towards the east as suggested
by our stellar proper motion measurements.    Such an ejection could explain
both the proper motions of S106~IR and the C-shaped symmetry of the 
plasma in S106.

Finally, it is possible that S106~IR is not moving in the reference frame
of the S106 cluster.   If so, the C-shaped symmetry of the \Hii\  region would
require continued mass inflow towards S106~IR from the east.  The orientation
of the dust continuum streamers seen at far-infrared and sub-millimeter 
wavelengths is consistent with the scenario.  The 
C-shaped orientation of the S106 nebular lobes 
could have been produced by feedback.

\subsection{Other Regions Similar to S106}

Bipolar \Hii\ regions represent a small subset of \Hii\ regions  \citep{Samal2018}.    
Bipolar structure can result from 
the formation of a massive star in a sheet of dense gas.  As the \Hii\ region
forms, ionizes surrounding gas,  and blows out of such a sheet, the plasma can
be channeled into a bipolar structure.  The bipolar geometry, however, may only
be seen if the confining sheet is seen nearly edge-on from our vantage point. 
The expansion rates of such regions will be dominated by thermal
pressure of photoionized plasma, and will therefore exhibit expansion with  
a speed comparable to the sound speed in the plasma, or about $\sim$10 \kms, or less.  
Faster expansion requires additional sources of energy and momentum input such
as can be provided by powerful stellar winds or explosive phenomena such as 
supernovae.  As far as we know, S106 is the first bipolar \Hii\ region
which exhibits highly supersonic motions.

The morphology and kinematics of S106 are remarkably similar to some proto-planetary 
nebulae and planetary nebulae such as NGC~6302, OH231.8+4.2, and IRAS~22036+5306.
These planetary nebulae exhibit bipolar morphologies, supersonic expansion, 
and exhibit nebular proper motions increasing with projected distance 
from the central star.  Bipolar planetary nebulae with such Hubble flows 
may indicate  an explosive origin \citep{SokerKashi2012}.   The kinetic energies
of these nebulae range from $10^{46}$ to $10^{49}$ ergs  \citep{SokerKashi2012,NGC6302}. 
They likely experienced explosions resulting form the merger of a compact binary
located at their centers.  The total energy released by these 
explosions is larger than those produced by novae but less than 
produced by supernovae.  They form a class known as
`Intermediate Luminosity Optical Transients' 
(ILOTs).  The SPitzer InfraRed Intensive Transients Survey (SPIRITS) 
has discovered many dozens of infrared transients having a similar luminosity 
as ILOTs in nearby galaxies \citep{Kasliwal2017}.

% Other MYSOs in transition has been observed.  However, most are associated with 
% compact or hyper-compact \Hii\ regions much smaller than S106. These bipolar \Hii\ 
% regions associated with young massive stars include MWC349A \citep{GvaramadzeMenten2012}.  

% G45.47+0.05 \citep{Zhang2019} is a 30 to 50 \Msol\ MYSO which is surrounded by 
% an ionized accretion disk that is losing mass due to photo-evaporation at a 
% rate $\dot M \sim 2$~to~$ 3.5 \times 10^{-5}$ \Msol ~yr$^{-1}$. 

% Section 7  sssssssssssss

\section{Conclusions}

S106 is the nearest bipolar \Hii\ region ionized by the $\sim$20 \Msol\ 
primary member of the binary star, S106~IR, which is transitioning from an embedded, 
protostellar phase  onto the main-sequence.   The secondary is a $\sim$3 \Msol\ 
object with a semi-major axis of $\sim$0.17 AU.   Archival HST images of 
S106 obtained in 1995 and 2011 are used to measure nebular proper motions.  
New near-infrared images of S106 obtained at Apache Point Observatory 
in the 2.16 $\mu$m \BrG\ and 2.12 $\mu$m \Htwo\ emission lines are presented.

The main result of this study are:

1)  Much of the \Ha\ and [\Nii ] emission in S106 is produced by compact 
knots whose proper motions are highly supersonic.  These motions point 
away from  S106~IR and their speeds increase with projected distance 
from S106~IR, 
reaching values larger than $\sim$150 \kms\, 1.5\arcmin\ ($\sim$0.48 pc) 
from S106~IR.  The proper motion vector field resembles an explosion
(a `Hubble flow') .
The supersonic motions may have been produced by an explosion   
about 3,500 years ago, similar to the event that occurred in 
Orion OMC1 about 550 years ago.   The explosion must have 
released more than $10^{47}$ ergs. 

2) Several models for the explosive motions are considered.  The explosion 
may have been produced by a period of common-envelope evolution while the
3 \Msol\ companion was engulfed by the bloated photosphere of S106~IR
as it was growing through a mass of 10 to 15 \Msol\ at a high accretion rate.
S106~IR may have experienced a dynamical interaction with another star or
binary in S106 such as S106~FIR $\sim$3,500 years ago.  Finally, S106~IR may have
experienced a major accretion burst $\sim$3,500 year ago, accreting 
$\sim$0.03 to 0.5 \Msol\ or more in an impulsive event.  It is unclear which, 
if any of these mechanisms occurred in S106.  

3) The South Bar at the southern end of the \Hii\ region is stationary
within the measurement errors.  It may trace the ionized surface of a 
clump of dense gas  that is too massive to respond to the outward 
flow of the \Hii\ region plasma.  The HST images show that nebular 
plasma is blowing past this obstruction on both the east and west sides. 

4) At radio wavelengths and in the \BrG\ emission line, the \Hii\ region
exhibits C-shaped symmetry.  The northern lobe of the S106 is
deflected towards the northwest by about 45$^o$ with respect to the
the axis of symmetry of the southern lobe, and the axis of symmetry of 
the mid- to far-infrared cavity in which the \Hii\ region is embedded.
The proper motion vector field also exhibits the same C-shaped symmetry.
In the less obscured \BrG\ hydrogen recombination line, the northern and 
southern lobes have remarkably similar morphologies.  Both lobes have bright 
bars at their ends, the South Bar and the Northwest Bar. 

5) S106~IR and the S106 \Hii\ region lobes are displaced from the 
center of the cylindrical infrared cavity surrounding the \Hii\ 
region.   S106~IR  may have  a proper motions towards the east with respect to 
the parent molecular cloud with a speed of $\sim 20$ km/s. 
The motion of the star may be responsible for the C-shaped symmetry of the 
ionized gas in S106.  Alternatively, photo-ablation from the dense gas 
located east of the nebula
and possibly falling towards the nebular core, may be responsible for
a side-wind which deflects the plasma flow towards the west.  
Such a side wind may be produced by photo-ablation of dense gas
from the cloud along the east-side of S106.

6) There are at least two Herbig-Haro (HH) objects in S106 outside the
\Hii\ region.  A bow shock is located along the axis of the outflow
emerging from the embedded Class~0 protostar, S106~FIR (not to be
confused with S106~IR).  No proper motions are detected because of its
diffuse nature. A second HH object located northwest of S106~FIR
exhibits proper motions of 174 \kms , the largest proper motion
in the S106 field.

7) Near-IR \Htwo\ imaging reveals additional knots suspected to
be `molecular hydrogen objects (MHOs) - shocks produced in the molecular
cloud surrounding S106.  
These objects are designated 
MHO~4079 through MHO~4085 in the catalog of Molecular Hydrogen Objects
\footnote{\url{http://astro.kent.ac.uk/~df/MHCat/}}

8) The \Htwo\ images  show that the PDR surrounding
S106 is asymmetric.  The PDR is farther from S106~IR towards the west than
towards the east.  

9) There is a $\sim$60\arcsec\ wide, conical cavity located due west of 
S106~IR between it and the western PDR.  The cavity harbors the Class 0
source, S106~FIR at its eastern tip.  The cavity may be shadowed by 
a clump of dust whose emission at 350 $\mu$m peaks between S106~IR and
S106~FIR in addition to the edge-on circumstellar matter surrounding
S106~IR.  

S106~IR joins the growing list of forming massive stars or young 
massive stars that have spawned explosive outflows.   These include
the highly obscured OMC1 in Orion \citep{Zapata2009,Bally2020}, 
the massive but distant G5.89 star forming region 
\citep{Zapata2019,Zapata2020}, and DR21 \citep{Zapata2013}.
Explosive events from embedded protostars may have a particularly powerful
feedback impact on their parent clouds. 

\section*{Acknowledgements}

J.B. and K.E.I.T. acknowledge support by National Science Foundation through grant No. 
AST-1910393.  A.G. acknowledges support from NSF grant No. AST-2008101.

K.E.I.T. acknowledges support by JSPS KAKENHI Grant Nos. JP19H05080, 
JP19K14760, JP21H00058, JP21H01145.

Based on observations made with the NASA/ESA Hubble Space Telescope and obtained 
from the Hubble Legacy Archive, which is a 
collaboration between the Space Telescope Science Institute (STScI/NASA), 
the European Space Agency (ST-ECF/ESAC/ESA), and the Canadian Astronomy 
Data Centre (CADC/NRC/CSA).  Some/all of the data presented in this paper 
were obtained from the Mikulski Archive for Space Telescopes (MAST). STScI 
is operated by the Association of Universities for Research in Astronomy, Inc., 
under NASA contract NAS5-26555.

Some of the work presented here is based on observations obtained with the Apache Point Observatory 3.5-meter telescope, which is owned and operated by the Astrophysical Research Consortium.    We thank the Apache Point Observatory Observing Specialists for their assistance during the observations. 

This work has made use of data from the European Space Agency (ESA) mission 
{\it Gaia} (\url{https://www.cosmos.esa.int/gaia}), processed by the {\it Gaia} 
Data Processing and Analysis Consortium (DPAC,
\url{https://www.cosmos.esa.int/web/gaia/dpac/consortium}). National 
institutions have provided funding for the DPAC, particularly the 
institutions participating in the {\it Gaia} Multilateral Agreement.

The MHO catalogue is hosted by the University of Kent
(See \url{http://astro.kent.ac.uk/~df/MHCat/}). 

This analysis used the  Python3 packages {\sf scipy} and 
{\sf image$\_$registration} packages
and the  {\sf DrizzlePac} software suite from STScI.

%%%%%%%%%%%%%%%%%%%%%%%%%%%%%%%%%%%%%%%%%%%%%%%%%%



%\bibliographystyle{mnras}


%%%%%%%%%%%%%%%%%%%%%%%%%%%%%%%%%%%%%%%%%%%%%%%%%%

%%%%%%%%%%%%%%%%% APPENDICES %%%%%%%%%%%%%%%%%%%%%

\appendix

\section{Proper Motions in the Four Sub-Fields}

\begin{figure*}
    \begin{center}
	\includegraphics[width=7.5in]{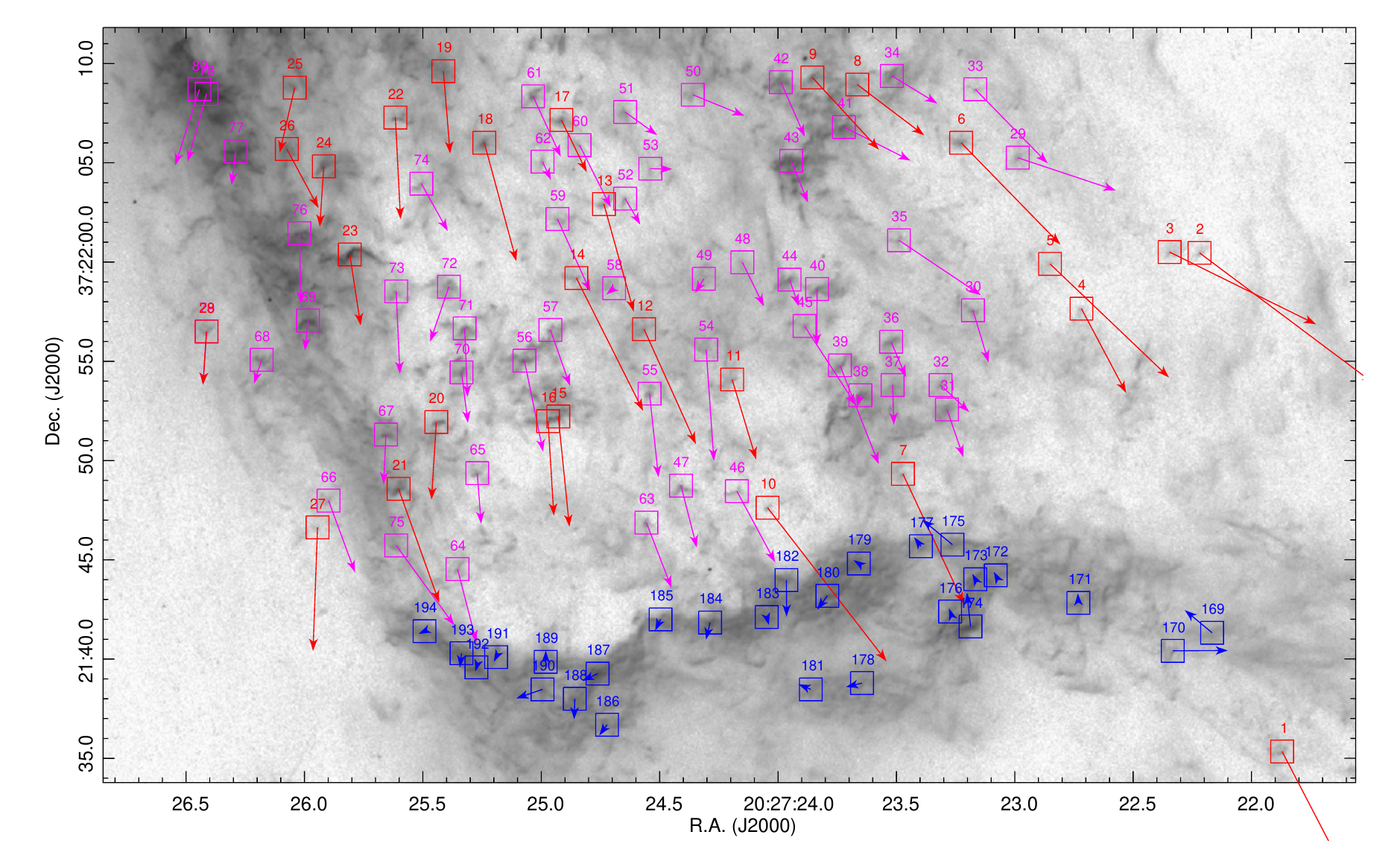}
    \caption{Proper motions in the S2 sub-field
    superimposed on the 2011 [\Nii ] image in the S106 reference frame, 
    measured using the cross-correlation
    methods described in the text.  The vector lengths correspond to the
    motions over the next 400 years.  
    Red dots indicate {\it fast} motions, magenta dots 
    indicate {\it slower} motions, and blue dots indicate 
    slow-moving knots in the South Bar at the South end of the
    South lobe of S106. 
    The ds9 region files are available as data behind the Figure.
    }
    \label{fig_S2_PMs}
    \end{center}
\end{figure*}

\begin{figure*}
    \begin{center}
	\includegraphics[width=7.5in]{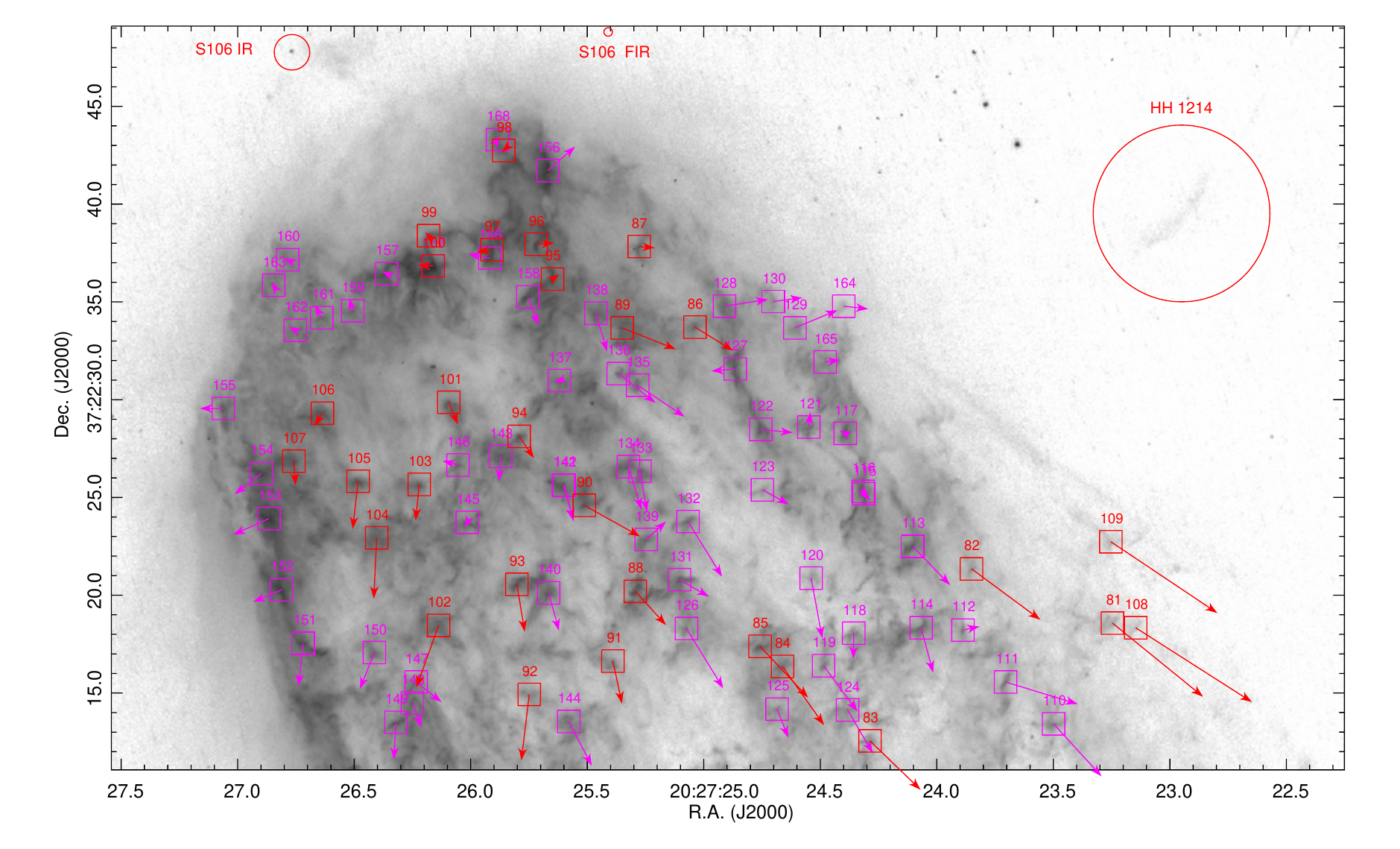}
    \caption{Proper motions in the S1 sub-field
    superimposed on the 2011 [\Nii ] image in the S106 reference frame, 
    measured using the cross-correlation
    methods described in the text.  The vector lengths correspond to the
    motions over the next 400 years.  
    The ds9 region files are available as data behind the Figure.
    }
    \label{fig_S1_PMs}
    \end{center}
\end{figure*}

\begin{figure*}
    \begin{center}
	\includegraphics[width=7.5in]{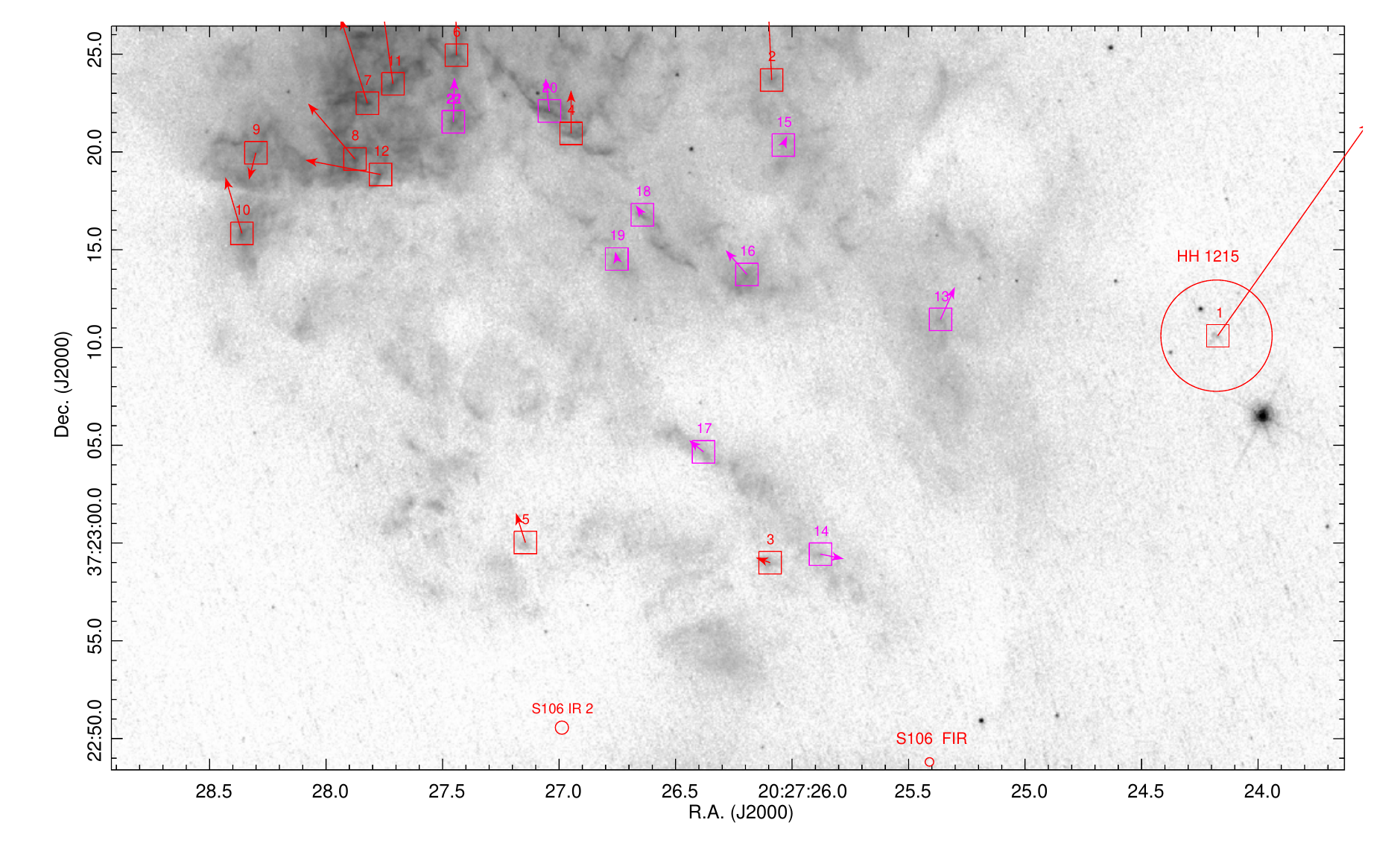}
    \caption{Proper motions in the N1 sub-field
    superimposed on the 2011 [\Nii ] image in the S106 reference frame, 
    measured using the cross-correlation
    methods described in the text.  The vector lengths correspond to the
    motions over the next 400 years.   
    The large arrow on the right
    marks the proper motion of HH 1215 (entry \#1 in Table 9).
    The ds9 region files are available as data behind the Figure.
    }
    \label{fig_N1_PMs}
    \end{center}
\end{figure*}

\begin{figure*}
    \begin{center}
	\includegraphics[width=7.5in]{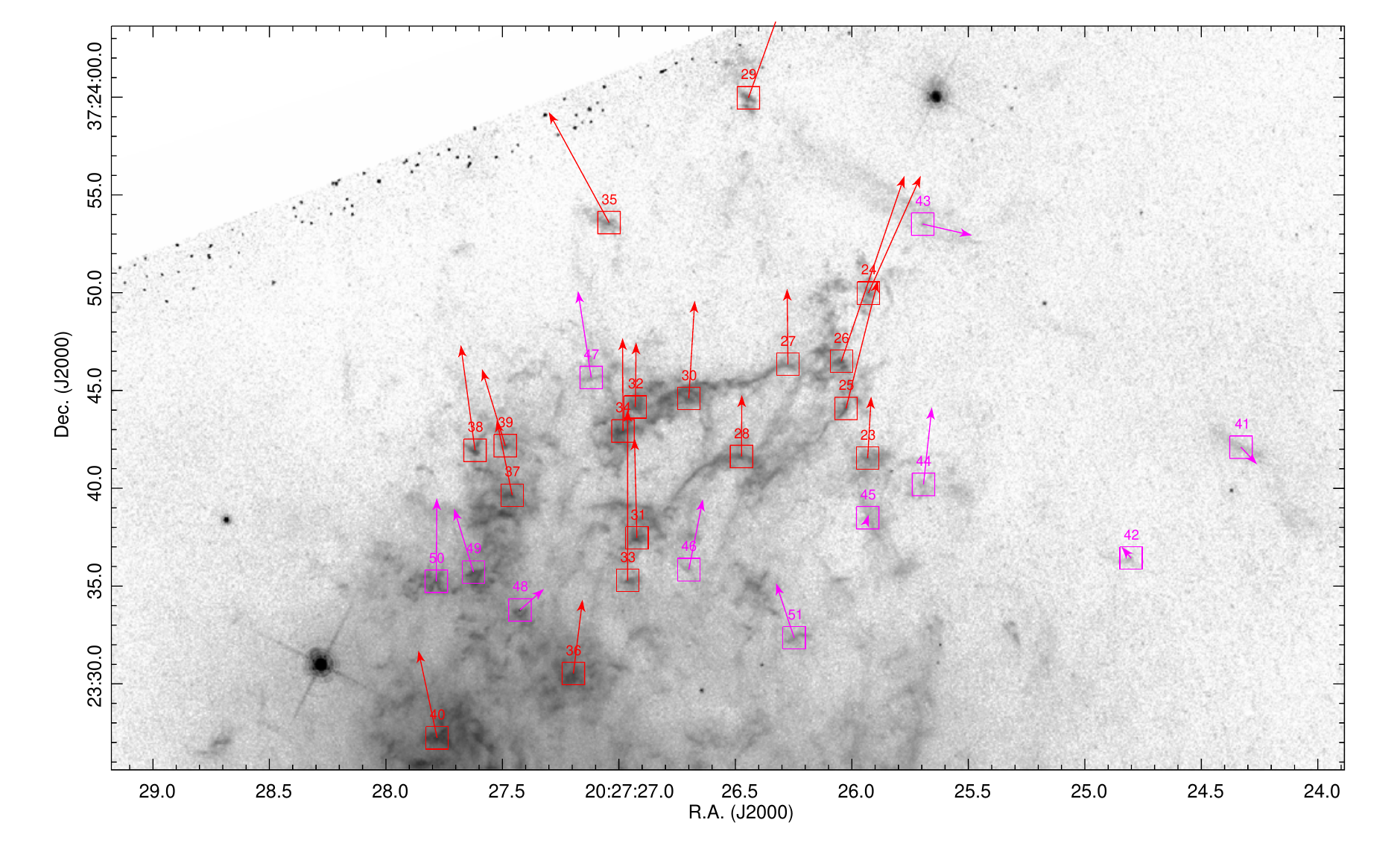}
    \caption{Proper motions in the N2 sub-field
    superimposed on the 2011 [\Nii ] image in the S106 reference frame, 
    measured using the cross-correlation
    methods described in the text.  The vector lengths correspond to the
    motions over the next 400 years.  
    The ds9 region files are available as data behind the Figure.
    }
    \label{fig_N2_PMs}
    \end{center}
\end{figure*}

Figures \ref{fig_S2_PMs} to \ref{fig_N2_PMs} show the proper motions in 
each of the four sub-fields extracted from the final, registered HST 
mosaics.   The figures show the spatial extents of the various
proper motion measurement boxes on the 2011 HST image.  The associated
vector lengths have been scaled to show the motions over the next 400 years.
The numbers above each box correspond to the entries in the  Tables
\ref{tab:table_south_lobe} and  Table \ref{tab:table_north_lobe}. 
These Tables list the proper motions in each of the four sub-fields.
The vectors shown in red mark the fastest motions
at a given projected distance from S106~IR, namely those that lie closest
to the dashed lines in the speed vs. projected distance plots.   
Vectors shown in magenta  mostly exhibit lower proper motions.  
However, in a few cases some of these speeds are as fast as the 
vectors shown in red.  Blue vectors
show the very slow motions in the South Bar.  Each category (fast, slow, and
South Bar) is listed separately in the Tables.
The Figures use the same color scheme for these categories
of fast and slow vectors.  The Northwest Bar is located outside of the 
field covered by HST.  Figure \ref{fig_PMs_All_VvsD} 
shows plots of the proper motions as a function of projected distance 
from S106~IR.

% Table 4
\begin{table}
	\centering
	\caption {South Lobe Proper Motions Between 1995 and 2011.} 
	\label{tab:table_south_lobe}
	\begin{tabular}{llllllccl} % columns, alignment for each
		\hline
\#		& $\alpha$(J2000) &  $\delta$(J2000) & D & PM &  V  & PA & Comments \\ 
\		& ($\arcdeg $) & ($\arcdeg $) & (pc) &  (mas  yr$^{-1}$) & (km s$^{-1}$) & ($\arcdeg$ ) & \\ 
\hline  \hline
     &  South     &  Lobe     & S2     &     Fast  \\
\hline  
1 & 306.8411 & 37.3598 & 0.545 & 32.3 & 167.1 & 207 \\ 
2 & 306.8426 & 37.3668 & 0.438 & 27.4 & 141.8 & 233 \\ 
3 & 306.8431 & 37.3668 & 0.43 & 20.5 & 106.2 & 243 \\ 
4 & 306.8447 & 37.366 & 0.416 & 12.0 & 62.1 & 207 \\ 
5 & 306.8452 & 37.3666 & 0.4 & 20.7 & 107.0 & 226 \\ 
6 & 306.8468 & 37.3683 & 0.357 & 17.9 & 92.5 & 224 \\ 
7 & 306.8478 & 37.3637 & 0.404 & 18.1 & 93.5 & 205 \\ 
8 & 306.8486 & 37.3692 & 0.32 & 10.6 & 54.6 & 232 \\ 
9 & 306.8494 & 37.3692 & 0.308 & 12.4 & 63.8 & 222 \\ 
10 & 306.8502 & 37.3632 & 0.384 & 24.5 & 126.4 & 217 \\ 
11 & 306.8508 & 37.365 & 0.35 & 10.5 & 54.2 & 196 \\ 
12 & 306.8524 & 37.3657 & 0.322 & 15.9 & 82.0 & 204 \\ 
13 & 306.8531 & 37.3675 & 0.287 & 14.0 & 72.6 & 195 \\ 
14 & 306.8535 & 37.3664 & 0.298 & 18.7 & 96.7 & 206 \\ 
15 & 306.8539 & 37.3645 & 0.328 & 14.0 & 72.1 & 185 \\ 
16 & 306.854 & 37.3644 & 0.328 & 11.9 & 61.5 & 183 \\ 
17 & 306.8538 & 37.3687 & 0.26 & 7.3 & 37.7 & 205 \\ 
18 & 306.8552 & 37.3683 & 0.252 & 15.5 & 79.9 & 195 \\ 
19 & 306.8559 & 37.3693 & 0.229 & 10.4 & 53.8 & 184 \\ 
20 & 306.856 & 37.3644 & 0.313 & 9.8 & 50.5 & 176 \\ 
21 & 306.8567 & 37.3635 & 0.327 & 15.2 & 78.5 & 199 \\ 
22 & 306.8567 & 37.3687 & 0.233 & 12.8 & 66.1 & 182 \\ 
23 & 306.8575 & 37.3668 & 0.262 & 9.0 & 46.7 & 188 \\ 
24 & 306.858 & 37.368 & 0.237 & 7.6 & 39.3 & 176 \\ 
25 & 306.8585 & 37.3691 & 0.214 & 8.4 & 43.4 & 167 \\ 
26 & 306.8586 & 37.3682 & 0.229 & 8.4 & 43.5 & 208 \\ 
27 & 306.8581 & 37.363 & 0.33 & 15.6 & 80.5 & 177 \\ 
28 & 306.8601 & 37.3657 & 0.273 & 6.7 & 34.7 & 176 \\
\hline
     &  South     &  Lobe     & S2     &     Slow  \\
\hline     
29 & 306.8458 & 37.3681 & 0.374 & 13.0 & 67.2 & 251 \\ 
30 & 306.8466 & 37.366 & 0.389 & 6.8 & 35.3 & 196 \\ 
31 & 306.847 & 37.3646 & 0.401 & 6.4 & 32.8 & 199 \\ 
32 & 306.8471 & 37.3649 & 0.395 & 4.9 & 25.5 & 226 \\ 
33 & 306.8465 & 37.3691 & 0.352 & 13.0 & 67.4 & 224 \\ 
34 & 306.848 & 37.3693 & 0.328 & 6.7 & 34.6 & 238 \\ 
35 & 306.8479 & 37.367 & 0.358 & 12.4 & 64.1 & 236 \\ 
36 & 306.848 & 37.3656 & 0.376 & 4.8 & 24.9 & 202 \\ 
37 & 306.848 & 37.3649 & 0.384 & 5.0 & 25.9 & 181 \\ 
38 & 306.8485 & 37.3648 & 0.379 & 1.5 & 7.8 & 162 \\ 
39 & 306.8489 & 37.3652 & 0.369 & 13.4 & 69.1 & 201 \\ 
40 & 306.8493 & 37.3663 & 0.349 & 7.1 & 36.9 & 179 \\ 
41 & 306.8488 & 37.3686 & 0.324 & 9.4 & 48.4 & 243 \\ 
42 & 306.8499 & 37.3692 & 0.301 & 7.5 & 38.6 & 203 \\ 
43 & 306.8498 & 37.3681 & 0.318 & 5.7 & 29.2 & 202 \\ 
44 & 306.8498 & 37.3664 & 0.341 & 3.5 & 18.2 & 198 \\ 
45 & 306.8495 & 37.3658 & 0.353 & 12.0 & 61.9 & 213 \\ 
\end{tabular}
\end{table}

% Table 4 continued
\begin{table}
	\centering
	\caption {South Lobe Proper Motions Between 1995 and 2011 (continued).} 
	\label{tab:table_south_lobe}
	\begin{tabular}{llllllccl} % columns, alignment for each
		\hline
\#		& $\alpha$(J2000) &  $\delta$(J2000) & D & PM &  V  & PA & Comments \\ 
\		& ($\arcdeg $) & ($\arcdeg $) & (pc) &  (mas  yr$^{-1}$) & (km s$^{-1}$) & ($\arcdeg$ ) & \\ 
\hline
46 & 306.8507 & 37.3635 & 0.375 & 10.2 & 52.5 & 208 \\ 
47 & 306.8517 & 37.3635 & 0.364 & 8.0 & 41.5 & 194 \\ 
48 & 306.8506 & 37.3667 & 0.327 & 6.1 & 31.8 & 206 \\ 
49 & 306.8513 & 37.3664 & 0.322 & 2.1 & 10.7 & 148 \\ 
50 & 306.8515 & 37.369 & 0.282 & 7.0 & 36.3 & 247 \\ 
51 & 306.8527 & 37.3688 & 0.271 & 5.0 & 25.6 & 234 \\ 
52 & 306.8527 & 37.3676 & 0.29 & 3.8 & 19.6 & 210 \\ 
53 & 306.8522 & 37.368 & 0.288 & 2.8 & 14.2 & 269 \\ 
54 & 306.8513 & 37.3654 & 0.338 & 14.0 & 72.6 & 184 \\ 
55 & 306.8523 & 37.3648 & 0.337 & 10.6 & 54.6 & 186 \\ 
56 & 306.8545 & 37.3653 & 0.31 & 11.7 & 60.4 & 191 \\ 
57 & 306.854 & 37.3657 & 0.306 & 7.5 & 38.6 & 199 \\ 
58 & 306.8529 & 37.3663 & 0.307 & 1.2 & 6.4 & 128 \\ 
59 & 306.8539 & 37.3673 & 0.282 & 10.0 & 51.9 & 204 \\ 
60 & 306.8535 & 37.3683 & 0.269 & 8.7 & 45.2 & 206 \\ 
61 & 306.8543 & 37.369 & 0.25 & 8.4 & 43.2 & 204 \\ 
62 & 306.8541 & 37.3681 & 0.266 & 2.5 & 13.1 & 204 \\ 
63 & 306.8523 & 37.363 & 0.367 & 8.8 & 45.4 & 201 \\ 
64 & 306.8556 & 37.3624 & 0.353 & 9.5 & 49.2 & 194 \\ 
65 & 306.8553 & 37.3637 & 0.331 & 6.4 & 33.0 & 184 \\ 
66 & 306.8579 & 37.3633 & 0.324 & 9.6 & 49.8 & 200 \\ 
67 & 306.8569 & 37.3643 & 0.311 & 6.2 & 32.3 & 177 \\ 
68 & 306.8591 & 37.3653 & 0.283 & 3.1 & 16.3 & 162 \\ 
69 & 306.8583 & 37.3658 & 0.276 & 3.8 & 19.6 & 174 \\ 
70 & 306.8556 & 37.3651 & 0.304 & 6.4 & 33.1 & 187 \\ 
71 & 306.8555 & 37.3657 & 0.294 & 8.6 & 44.5 & 182 \\ 
72 & 306.8558 & 37.3663 & 0.281 & 7.3 & 37.7 & 161 \\ 
73 & 306.8567 & 37.3663 & 0.276 & 10.4 & 53.7 & 182 \\ 
74 & 306.8563 & 37.3678 & 0.253 & 6.8 & 35.3 & 209 \\ 
75 & 306.8567 & 37.3627 & 0.341 & 12.4 & 64.3 & 216 \\ 
76 & 306.8584 & 37.3671 & 0.252 & 8.8 & 45.7 & 181 \\ 
77 & 306.8595 & 37.3682 & 0.227 & 4.2 & 22.0 & 173 \\ 
78 & 306.8601 & 37.369 & 0.21 & 8.8 & 45.7 & 164 \\ 
79 & 306.8601 & 37.3657 & 0.273 & 6.7 & 34.7 & 176 \\ 
80 & 306.8602 & 37.3691 & 0.208 & 10.0 & 51.9 & 162 \\
\hline
     &  South     &  Lobe     & S1     &     Fast  \\
\hline     
81 & 306.8469 & 37.3718 & 0.319 & 14.9 & 77.1 & 230 \\ 
82 & 306.8494 & 37.3726 & 0.27 & 10.9 & 56.6 & 233 \\ 
83 & 306.8512 & 37.3702 & 0.271 & 9.0 & 46.5 & 225 \\ 
84 & 306.8528 & 37.3712 & 0.235 & 9.2 & 47.5 & 215 \\ 
85 & 306.8532 & 37.3715 & 0.226 & 9.0 & 46.5 & 223 \\ 
86 & 306.8543 & 37.376 & 0.156 & 5.7 & 29.5 & 238 \\ 
87 & 306.8553 & 37.3772 & 0.129 & 2.0 & 10.2 & 265 \\ 
88 & 306.8554 & 37.3723 & 0.187 & 5.8 & 30.1 & 222 \\ 
89 & 306.8556 & 37.376 & 0.135 & 7.4 & 38.2 & 248 \\ 
90 & 306.8563 & 37.3735 & 0.158 & 8.2 & 42.1 & 240 \\
\end{tabular}
\end{table}

% Table 4 Continued
\begin{table}
	\centering
	\caption {South Lobe Proper Motions Between 1995 and 2011 (continued).} 
	\label{tab:table_south_lobe}
	\begin{tabular}{llllllccl} % columns, alignment for each
		\hline
\#		& $\alpha$(J2000) &  $\delta$(J2000) & D & PM &  V  & PA & Comments \\ 
\		& ($\arcdeg $) & ($\arcdeg $) & (pc) &  (mas  yr$^{-1}$) & (km s$^{-1}$) & ($\arcdeg$ ) & \\ 
\hline
91 & 306.8558 & 37.3713 & 0.198 & 5.6 & 28.8 & 192 \\ 
92 & 306.8573 & 37.3708 & 0.192 & 8.5 & 43.9 & 173 \\ 
93 & 306.8575 & 37.3724 & 0.163 & 6.1 & 31.8 & 189 \\ 
94 & 306.8575 & 37.3745 & 0.13 & 3.4 & 17.4 & 214 \\ 
95 & 306.8569 & 37.3767 & 0.108 & 1.0 & 5.4 & -53 \\ 
96 & 306.8572 & 37.3772 & 0.098 & 2.3 & 11.8 & -86 \\ 
97 & 306.858 & 37.3771 & 0.087 & 1.9 & 10.0 & 99 \\ 
98 & 306.8577 & 37.3785 & 0.077 & 0.3 & 1.6 & 133 \\ 
99 & 306.8591 & 37.3773 & 0.068 & 0.6 & 3.0 & 45 \\ 
100 & 306.859 & 37.3769 & 0.075 & 1.9 & 10.1 & 87 \\ 
101 & 306.8587 & 37.375 & 0.109 & 3.1 & 15.8 & 202 \\ 
102 & 306.8589 & 37.3718 & 0.163 & 8.4 & 43.5 & 160 \\ 
103 & 306.8593 & 37.3738 & 0.125 & 4.6 & 23.8 & 174 \\ 
104 & 306.86 & 37.373 & 0.135 & 7.8 & 40.1 & 177 \\ 
105 & 306.8603 & 37.3738 & 0.119 & 6.1 & 31.7 & 173 \\ 
106 & 306.861 & 37.3748 & 0.099 & 1.9 & 9.9 & 148 \\ 
107 & 306.8615 & 37.3741 & 0.111 & 3.1 & 16.0 & 184 \\ 
108 & 306.8464 & 37.3718 & 0.326 & 17.7 & 91.2 & 237 \\ 
109 & 306.8469 & 37.373 & 0.308 & 16.4 & 84.6 & 236 \\ 
\hline
     &  South     &  Lobe     & S1     &     Slow  \\
\hline     
110 & 306.8479 & 37.3704 & 0.316 & 9.1 & 47.0 & 222 \\ 
111 & 306.8488 & 37.371 & 0.296 & 9.6 & 49.8 & 253 \\ 
112 & 306.8495 & 37.3717 & 0.276 & 2.2 & 11.1 & -77 \\ 
113 & 306.8504 & 37.3729 & 0.25 & 6.9 & 35.6 & 224 \\ 
114 & 306.8503 & 37.3718 & 0.264 & 5.9 & 30.7 & 194 \\ 
115 & 306.8513 & 37.3737 & 0.228 & 1.2 & 6.2 & 225 \\ 
116 & 306.8513 & 37.3737 & 0.228 & 1.0 & 5.2 & 218 \\ 
117 & 306.8516 & 37.3745 & 0.214 & 0.8 & 4.3 & 63 \\ 
118 & 306.8515 & 37.3717 & 0.247 & 3.3 & 16.9 & 179 \\ 
119 & 306.852 & 37.3712 & 0.245 & 7.3 & 37.6 & 216 \\ 
120 & 306.8522 & 37.3725 & 0.227 & 7.8 & 40.4 & 190 \\ 
121 & 306.8523 & 37.3746 & 0.203 & 1.9 & 9.8 & -5 \\ 
122 & 306.8531 & 37.3746 & 0.189 & 4.1 & 21.1 & 264 \\ 
123 & 306.8531 & 37.3737 & 0.199 & 3.9 & 19.9 & 241 \\ 
124 & 306.8516 & 37.3706 & 0.259 & 6.4 & 33.1 & 210 \\ 
125 & 306.8529 & 37.3706 & 0.242 & 3.9 & 20.3 & 201 \\ 
126 & 306.8545 & 37.3717 & 0.206 & 9.0 & 46.4 & 211 \\ 
127 & 306.8536 & 37.3754 & 0.173 & 3.1 & 15.9 & 95 \\ 
128 & 306.8538 & 37.3763 & 0.162 & 5.6 & 29.1 & -81 \\ 
129 & 306.8525 & 37.376 & 0.186 & 5.9 & 30.5 & -67 \\ 
130 & 306.8529 & 37.3764 & 0.177 & 3.7 & 19.3 & -82 \\ 
131 & 306.8546 & 37.3724 & 0.194 & 4.4 & 22.5 & 240 \\ 
132 & 306.8544 & 37.3733 & 0.185 & 8.3 & 42.8 & 211 \\ 
133 & 306.8553 & 37.374 & 0.164 & 5.2 & 27.1 & 189 \\ 
134 & 306.8555 & 37.374 & 0.16 & 5.7 & 29.7 & 196 \\ 
135 & 306.8553 & 37.3752 & 0.148 & 7.2 & 37.3 & 236 \\
\end{tabular}
\end{table}

% Table 4 Continued
\begin{table}
	\centering
	\caption {South Lobe Proper Motions Between 1995 and 2011 (continued).} 
	\label{tab:table_south_lobe}
	\begin{tabular}{llllllccl} % columns, alignment for each
		\hline
\#		& $\alpha$(J2000) &  $\delta$(J2000) & D & PM &  V  & PA & Comments \\ 
\		& ($\arcdeg $) & ($\arcdeg $) & (pc) &  (mas  yr$^{-1}$) & (km s$^{-1}$) & ($\arcdeg$ ) & \\ 
\hline
136 & 306.8557 & 37.3754 & 0.141 & 6.0 & 31.3 & 231 \\ 
137 & 306.8567 & 37.3753 & 0.127 & 0.7 & 3.7 & 108 \\ 
138 & 306.8561 & 37.3762 & 0.125 & 5.1 & 26.5 & 196 \\ 
139 & 306.8552 & 37.373 & 0.179 & 3.4 & 17.7 & -48 \\ 
140 & 306.8569 & 37.3723 & 0.171 & 5.0 & 25.9 & 195 \\ 
141 & 306.8567 & 37.3738 & 0.149 & 4.6 & 23.9 & 196 \\ 
142 & 306.8567 & 37.3738 & 0.149 & 4.8 & 24.6 & 194 \\ 
143 & 306.8578 & 37.3742 & 0.131 & 3.2 & 16.3 & 175 \\ 
144 & 306.8566 & 37.3704 & 0.204 & 6.4 & 33.0 & 207 \\ 
145 & 306.8584 & 37.3733 & 0.141 & 0.7 & 3.8 & 158 \\ 
146 & 306.8586 & 37.3741 & 0.125 & 2.0 & 10.5 & 74 \\ 
147 & 306.8593 & 37.371 & 0.176 & 4.1 & 21.4 & 231 \\ 
148 & 306.8594 & 37.3707 & 0.181 & 3.4 & 17.4 & 200 \\ 
149 & 306.8597 & 37.3704 & 0.185 & 4.8 & 24.7 & 176 \\ 
150 & 306.8601 & 37.3714 & 0.165 & 5.0 & 25.9 & 158 \\ 
151 & 306.8613 & 37.3715 & 0.16 & 5.5 & 28.4 & 173 \\ 
152 & 306.8617 & 37.3723 & 0.146 & 3.9 & 20.3 & 113 \\ 
153 & 306.8619 & 37.3733 & 0.127 & 5.0 & 26.0 & 114 \\ 
154 & 306.8621 & 37.374 & 0.115 & 4.3 & 22.2 & 128 \\ 
155 & 306.8628 & 37.3749 & 0.1 & 3.0 & 15.4 & 89 \\ 
156 & 306.857 & 37.3783 & 0.093 & 4.6 & 23.9 & -48 \\ 
157 & 306.8598 & 37.3768 & 0.068 & 0.7 & 3.5 & 63 \\ 
158 & 306.8573 & 37.3765 & 0.104 & 3.9 & 20.1 & 201 \\ 
159 & 306.8604 & 37.3763 & 0.073 & 1.7 & 8.7 & 15 \\ 
160 & 306.8616 & 37.377 & 0.057 & 0.4 & 2.3 & 65 \\ 
161 & 306.861 & 37.3762 & 0.073 & 1.8 & 9.5 & 28 \\ 
162 & 306.8615 & 37.376 & 0.076 & 0.9 & 4.9 & 63 \\ 
163 & 306.8619 & 37.3766 & 0.064 & 0.5 & 2.4 & 28 \\ 
164 & 306.8517 & 37.3763 & 0.2 & 3.1 & 16.2 & 263 \\ 
165 & 306.852 & 37.3755 & 0.2 & 1.9 & 10.0 & -82 \\ 
166 & 306.858 & 37.377 & 0.088 & 2.8 & 14.5 & 79 \\ 
167 & 306.859 & 37.3769 & 0.075 & 1.9 & 10.1 & 87 \\ 
168 & 306.8579 & 37.3787 & 0.074 & 0.6 & 2.8 & -32 \\
\hline
   &  S2           & South        & Bar    &       \\
\hline   
169 & 306.8424 & 37.3615 & 0.506 & 4.5 & 23.0 & 50 \\ 
170 & 306.8431 & 37.3612 & 0.5 & 6.9 & 35.8 & -89 \\ 
171 & 306.8447 & 37.3619 & 0.469 & 1.2 & 6.2 & 2 \\ 
172 & 306.8462 & 37.3623 & 0.445 & 0.5 & 2.6 & 28 \\ 
173 & 306.8465 & 37.3622 & 0.441 & 0.8 & 4.0 & 28 \\ 
174 & 306.8466 & 37.3616 & 0.45 & 4.3 & 22.0 & 6 \\ 
175 & 306.8469 & 37.3627 & 0.43 & 4.9 & 25.5 & 50 \\ 
176 & 306.847 & 37.3618 & 0.443 & 0.4 & 2.1 & 19 \\ 
177 & 306.8475 & 37.3627 & 0.423 & 1.4 & 7.3 & 37 \\ 
178 & 306.8485 & 37.3608 & 0.441 & 2.1 & 10.7 & 103 \\ 
179 & 306.8486 & 37.3624 & 0.414 & 0.8 & 3.9 & 56 \\ 
180 & 306.8491 & 37.362 & 0.415 & 2.2 & 11.2 & 143 \\
\end{tabular}
\end{table}

% Table 4 Continued
\begin{table}
	\centering
	\caption {South Lobe Proper Motions Between 1995 and 2011 (continued).} 
	\label{tab:table_south_lobe}
	\begin{tabular}{llllllccl} % columns, alignment for each
		\hline
\#		& $\alpha$(J2000) &  $\delta$(J2000) & D & PM &  V  & PA & Comments \\ 
\		& ($\arcdeg $) & ($\arcdeg $) & (pc) &  (mas  yr$^{-1}$) & (km s$^{-1}$) & ($\arcdeg$ ) & \\ 
\hline
181 & 306.8494 & 37.3607 & 0.433 & 1.7 & 8.7 & 67 \\ 
182 & 306.8499 & 37.3622 & 0.404 & 4.6 & 23.8 & 180 \\ 
183 & 306.8502 & 37.3617 & 0.409 & 1.0 & 5.4 & 195 \\ 
184 & 306.8512 & 37.3616 & 0.4 & 2.2 & 11.4 & 166 \\ 
185 & 306.8521 & 37.3617 & 0.392 & 1.6 & 8.1 & 154 \\ 
186 & 306.853 & 37.3602 & 0.409 & 1.7 & 8.6 & 143 \\ 
187 & 306.8532 & 37.3609 & 0.396 & 2.1 & 11.0 & 115 \\ 
188 & 306.8536 & 37.3606 & 0.399 & 2.6 & 13.3 & 178 \\ 
189 & 306.8541 & 37.3611 & 0.386 & 1.4 & 7.2 & 0 \\ 
190 & 306.8541 & 37.3607 & 0.393 & 3.5 & 18.0 & 108 \\ 
191 & 306.855 & 37.3611 & 0.379 & 0.6 & 3.2 & 152 \\ 
192 & 306.8553 & 37.361 & 0.38 & 0.5 & 2.8 & 163 \\ 
193 & 306.8556 & 37.3612 & 0.375 & 1.8 & 9.2 & 173 \\ 
194 & 306.8562 & 37.3615 & 0.365 & 0.8 & 4.2 & 114 \\ 
\end{tabular}
\end{table}

% Table of PMs
\begin{table}
	\centering
	\caption {North Lobe Proper Motions Between 1995 and 2011.} 
	\label{tab:table_north_lobe}
	\begin{tabular}{llllllccl} % columns, alignment for each
	\hline
\#		& $\alpha$(J2000) &  $\delta$(J2000) & D & PM &  V  & PA & Comments \\ 
\		& ($\arcdeg $) & ($\arcdeg $) & (pc) &  (mas  yr$^{-1}$) & (km s$^{-1}$) & ($\arcdeg$ ) & \\ 
\hline \hline
     &  North      &  Lobe     & N1     &     Fast  \\
\hline     
1 & 306.8507 & 37.3863 & 0.238 & 33.6 & 173.8 & -35  & HH 1215 \\ 
2 & 306.8587 & 37.3899 & 0.197 & 9.4 & 48.6 & 2 \\ 
3 & 306.8587 & 37.3831 & 0.079 & 1.9 & 9.9 & 70 \\ 
4 & 306.8623 & 37.3892 & 0.175 & 5.5 & 28.5 & 0 \\ 
5 & 306.8631 & 37.3833 & 0.071 & 4.0 & 20.4 & 17 \\ 
6 & 306.8643 & 37.3903 & 0.203 & 12.7 & 65.6 & 0 \\ 
7 & 306.8659 & 37.3896 & 0.201 & 11.4 & 59.0 & 17 \\ 
8 & 306.8661 & 37.3888 & 0.189 & 9.3 & 48.1 & 40 \\ 
9 & 306.8679 & 37.3889 & 0.209 & 3.5 & 18.3 & 165 \\ 
10 & 306.8682 & 37.3877 & 0.194 & 7.5 & 38.6 & 16 \\ 
11 & 306.8655 & 37.3899 & 0.203 & 12.2 & 62.9 & 7 \\ 
12 & 306.8657 & 37.3886 & 0.182 & 9.8 & 50.6 & 79 \\
\hline
     &  North      &  Lobe     & N1     &     Slow  \\
\hline     
13 & 306.8557 & 37.3865 & 0.167 & 4.5 & 23.4 & -24 \\ 
14 & 306.8578 & 37.3832 & 0.093 & 3.0 & 15.7 & 258 \\ 
15 & 306.8585 & 37.389 & 0.181 & 1.3 & 6.5 & -22 \\ 
16 & 306.8591 & 37.3871 & 0.144 & 4.1 & 21.3 & 41 \\ 
17 & 306.8599 & 37.3846 & 0.094 & 2.4 & 12.3 & 51 \\ 
18 & 306.861 & 37.388 & 0.153 & 1.5 & 7.6 & 34 \\ 
19 & 306.8615 & 37.3874 & 0.141 & 1.0 & 5.0 & 12 \\ 
20 & 306.8627 & 37.3895 & 0.182 & 4.2 & 21.6 & 6 \\ 
21 & 306.8644 & 37.3893 & 0.186 & 5.6 & 29.1 & -1 \\ 
22 & 306.8644 & 37.3893 & 0.186 & 5.6 & 29.1 & -1 \\
\hline
    &  North      &  Lobe     & N2     &     Fast  \\
\hline    
23 & 306.858 & 37.3949 & 0.291 & 7.9 & 40.6 & -3 \\ 
24 & 306.858 & 37.3972 & 0.335 & 16.4 & 84.8 & -24 \\ 
25 & 306.8584 & 37.3956 & 0.303 & 16.9 & 87.2 & -14 \\ 
26 & 306.8585 & 37.3963 & 0.315 & 25.0 & 129.3 & -18 \\ 
27 & 306.8595 & 37.3962 & 0.311 & 9.6 & 49.9 & 0 \\ 
28 & 306.8603 & 37.3949 & 0.285 & 7.8 & 40.5 & 0 \\ 
29 & 306.8602 & 37.4 & 0.382 & 22.1 & 114.4 & -19 \\ 
30 & 306.8612 & 37.3957 & 0.3 & 12.5 & 64.5 & -3 \\ 
31 & 306.8622 & 37.3937 & 0.262 & 12.7 & 65.8 & 1 \\ 
32 & 306.8622 & 37.3956 & 0.298 & 8.2 & 42.6 & 0 \\ 
33 & 306.8623 & 37.3931 & 0.251 & 21.8 & 112.7 & 0 \\ 
34 & 306.8624 & 37.3953 & 0.292 & 11.9 & 61.3 & 0 \\ 
35 & 306.8627 & 37.3982 & 0.348 & 16.1 & 83.1 & 28 \\ 
36 & 306.8633 & 37.3918 & 0.228 & 9.5 & 49.1 & -7 \\ 
37 & 306.8644 & 37.3943 & 0.279 & 9.8 & 50.7 & 11 \\ 
38 & 306.8651 & 37.395 & 0.294 & 13.6 & 70.1 & 7 \\ 
39 & 306.8645 & 37.3951 & 0.293 & 10.2 & 52.9 & 16 \\ 
40 & 306.8658 & 37.3909 & 0.223 & 11.4 & 58.8 & 12 \\
\hline
   &  North      &  Lobe     & N2     &     Slow   \\
\hline
41 & 306.8514 & 37.395 & 0.345 & 3.0 & 15.4 & 223 \\ 
42 & 306.8533 & 37.3935 & 0.3 & 1.8 & 9.4 & 40 \\ 
43 & 306.8571 & 37.3982 & 0.357 & 6.4 & 33.3 & 257 \\ 
44 & 306.8571 & 37.3945 & 0.289 & 9.9 & 51.1 & -6 \\ 
45 & 306.858 & 37.394 & 0.276 & 0.4 & 1.9 & -17 \\
\end{tabular}
\end{table}

% Table of PMs
\begin{table}
	\centering
	\caption {North Lobe Proper Motions Between 1995 and 2011 (continued).} 
	\label{tab:table_north_lobe}
	\begin{tabular}{llllllccl} % columns, alignment for each
	\hline
\#		& $\alpha$(J2000) &  $\delta$(J2000) & D & PM &  V  & PA & Comments \\ 
\		& ($\arcdeg $) & ($\arcdeg $) & (pc) &  (mas  yr$^{-1}$) & (km s$^{-1}$) & ($\arcdeg$ ) & \\ 
\hline 
46 & 306.8612 & 37.3933 & 0.254 & 9.2 & 47.4 & -11 \\ 
47 & 306.863 & 37.396 & 0.307 & 11.2 & 57.7 & 8 \\ 
48 & 306.8643 & 37.3927 & 0.248 & 4.1 & 21.0 & -48 \\ 
49 & 306.8651 & 37.3933 & 0.262 & 8.5 & 43.8 & 17 \\ 
50 & 306.8658 & 37.3931 & 0.263 & 10.6 & 54.8 & 0 \\ 
51 & 306.8594 & 37.3923 & 0.239 & 7.4 & 38.0 & 18 \\ 

\end{tabular}
\end{table}

%%%%%  Proper Motion Tables (PM Tables)

\section{Difference Images}

\begin{figure*}
	\includegraphics[width=7.5in]{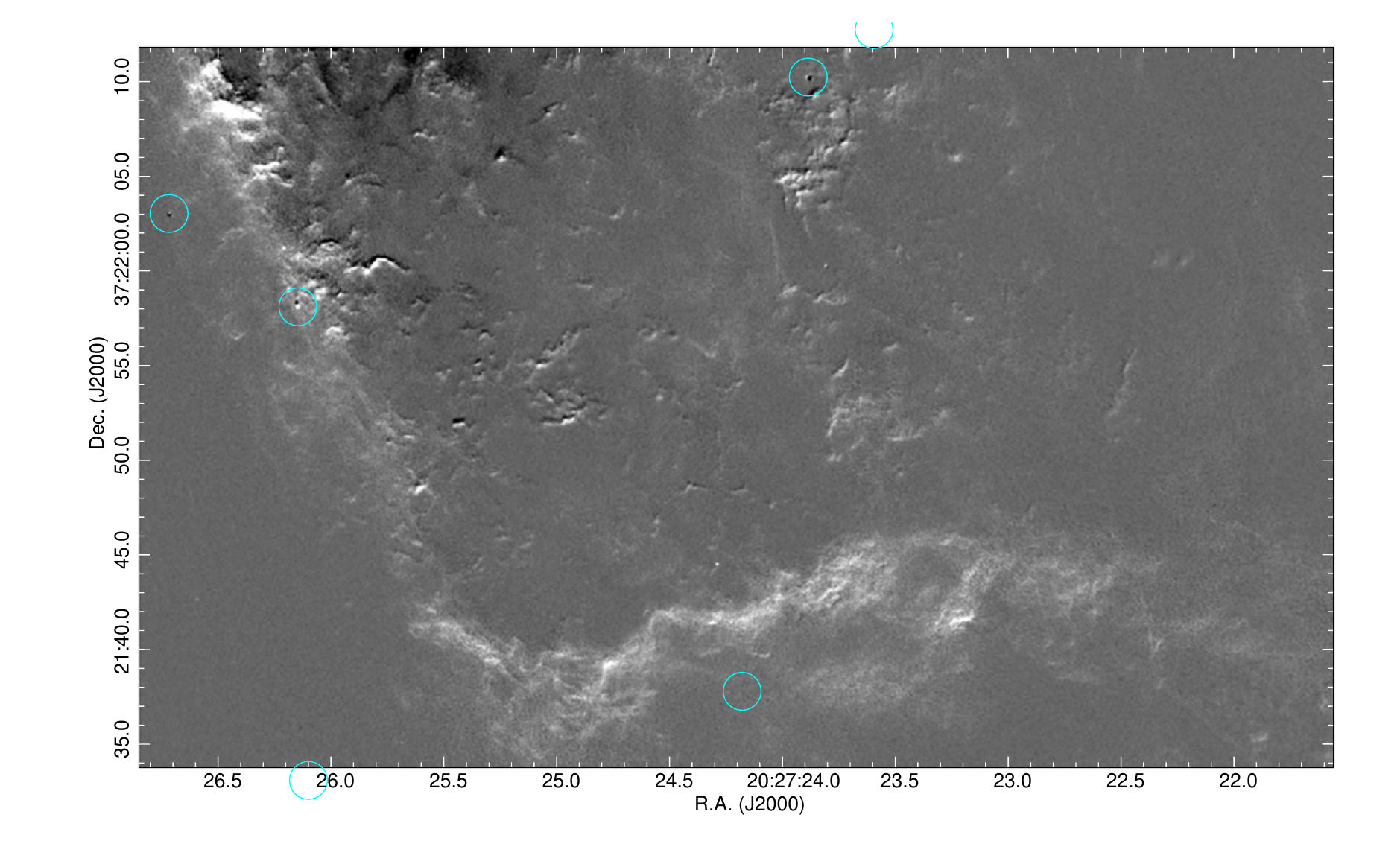}
    \caption{A difference image showing proper motions and changes 
    in the south 2 (S2) field between
    the 2011 and 1995 epoch HST images in the southern end-cap regions of S106. 
    An animated version of this difference image, blinking between the
    1995 and 2011 epochs, is available in the online Journal. The animation
    is 12 seconds in duration with a cadence of 1 epoch/second; it is not 
    annotated. The FITS files used to construct this difference image are  
    available as the data behind the figure. }
    \label{figA1diff}
\end{figure*}

\begin{figure*}
	\includegraphics[width=7.5in]{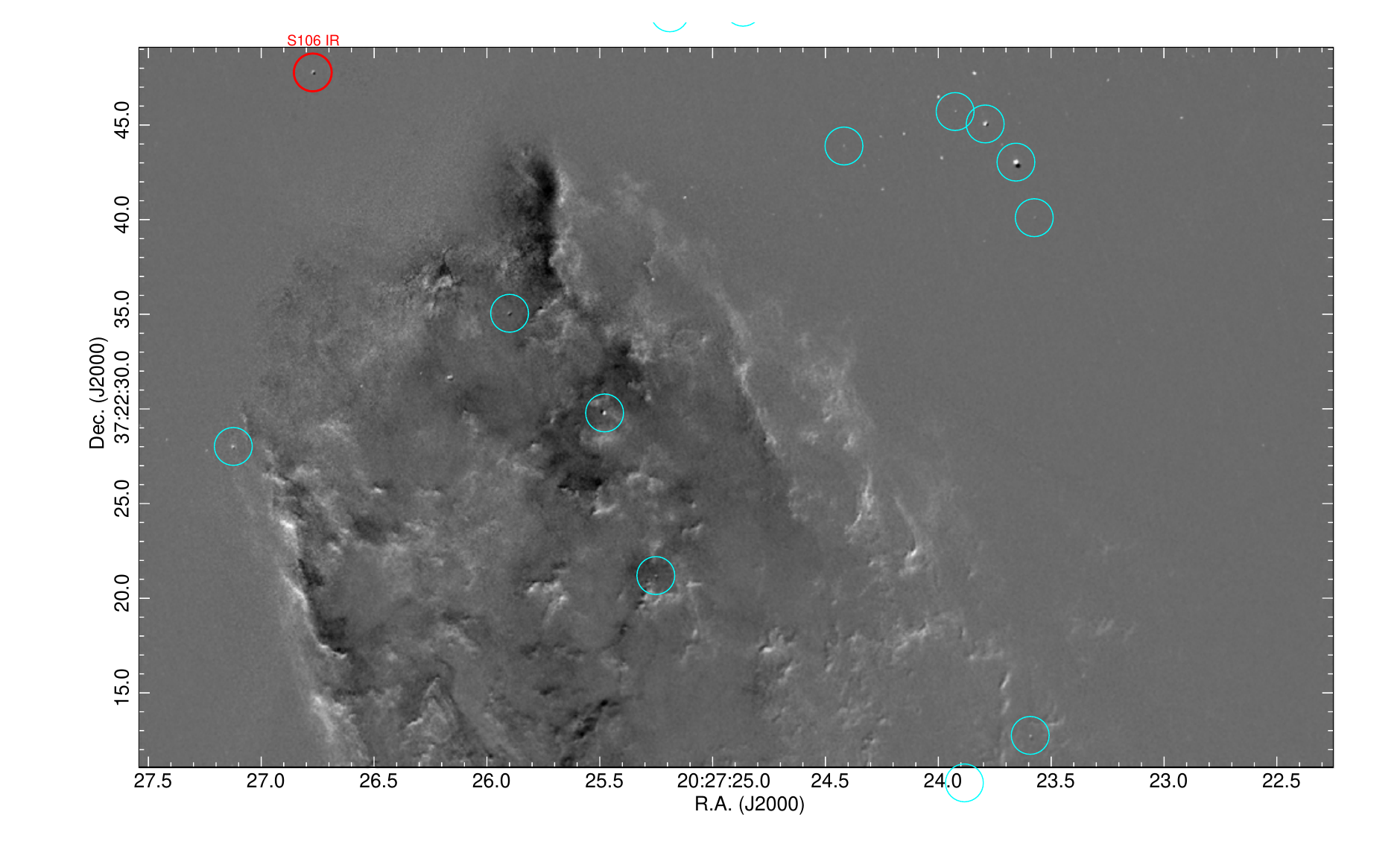}
    \caption{A difference image showing proper motions and changes 
    in the south 1 (S1) fields between the 2011 and 1995 epoch HST 
    images.   The time interval between these
    images is 15.6 years.   Cyan circles show the stars
    used to register this portion of the multi-epoch HST data.  The 2011 image is shown in 
    white.  The 1995 image is shown in black. This Figure shows the 
    field containing the inner part of the southern lobe of S106.
    An animated version of this difference image, blinking between the
    1995 and 2011 epochs, is available in the online Journal. The animation
    is 12 seconds in duration with a cadence of 1 epoch/second; it is not 
    annotated. The FITS files used to construct this difference image are  
    available as the data behind the figure. }
    \label{figA2diff}
\end{figure*}

\begin{figure*}
	\includegraphics[width=7.5in]{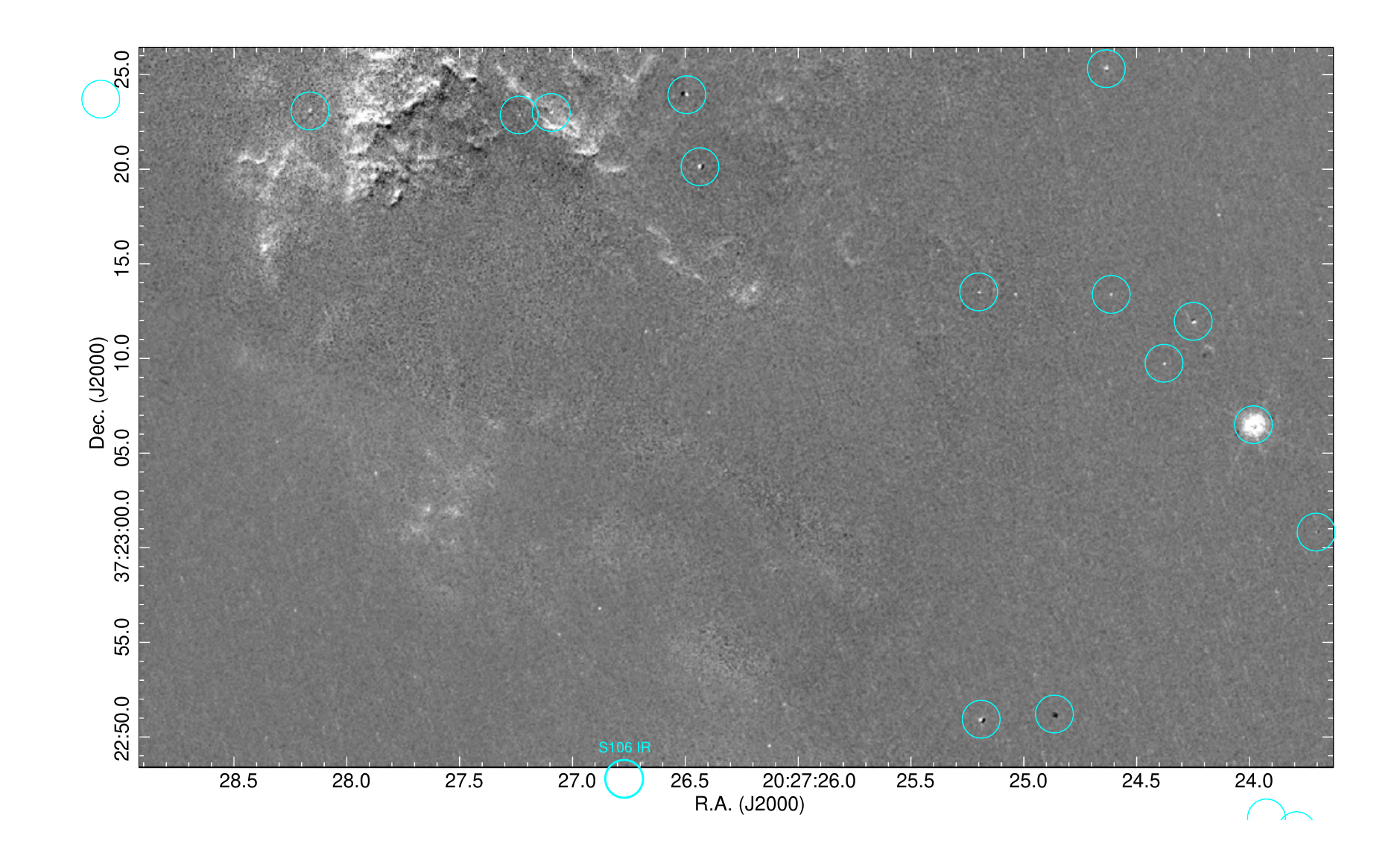}
    \caption{A difference image showing proper motions and changes 
    in the north 1 (N1) field between
    the 2011 and 1995 epoch HST images in inner part of the 
    northern lobe of S106. 
    An animated version of this difference image, blinking between the
    1995 and 2011 epochs, is available in the online Journal. The animation
    is 12 seconds in duration with a cadence of 1 epoch/second; it is not 
    annotated. The FITS files used to construct this difference image are  
    available as the data behind the figure. }
    \label{figA3diff}
\end{figure*}

\begin{figure*}
	\includegraphics[width=7.5in]{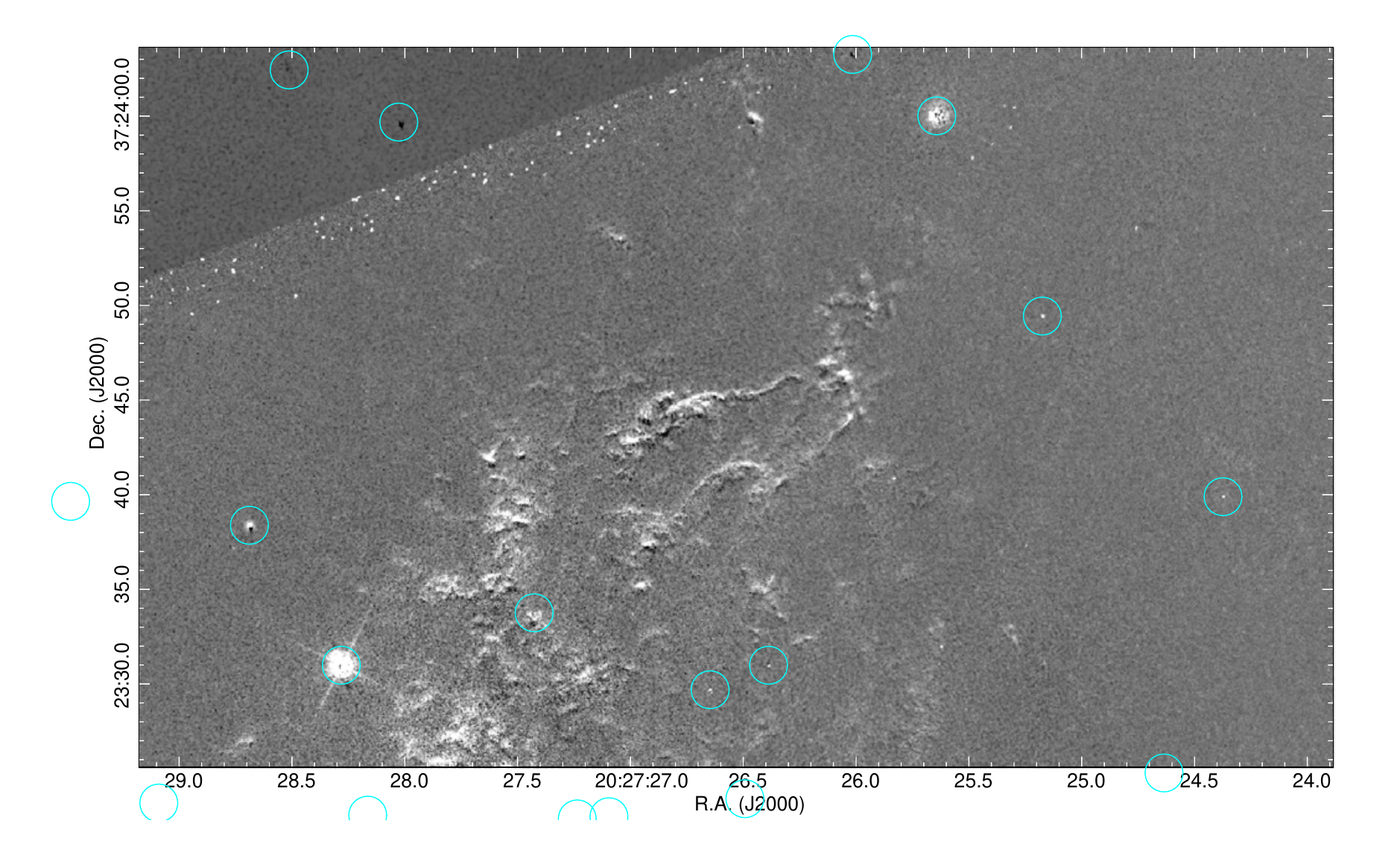}
    \caption{A difference image showing proper motions and changes 
    in the north 2 (21) field between
    the 2011 and 1995 epoch HST images in the portion of the northern
    lobe of S106.  The
    interval between images is 15.1 years.
    An animated version of this difference image, blinking between the
    1995 and 2011 epochs, is available in the online Journal. The animation
    is 12 seconds in duration with a cadence of 1 epoch/second; it is not 
    annotated. The FITS files used to construct this difference image are  
    available as the data behind the figure. }
    \label{figA4diff}
\end{figure*}

Images showing the difference between the 2011 and 1995 epoch data 
were formed from the four sub-fields shown in Figure \ref{fig_HST_Nii}.  
The images were normalized so that the peak nebular emission has a 
value of about 1.0.  The normalized 1995 image was then subtracted from
the normalized 2011 image.   The results are shown in Figures 
\ref{figA1diff} to \ref{figA4diff}.  
In the electronic version of this
manuscript, we present mp4 movies showing the proper motions and 
intensity changes in the four sub-fields between 1995 and 2011.
The  final mosaic assembled from the 1995 \Ha\ data
and the 2011 [\Nii ] image are made available in FITS format.
The registered and re-sampled 1995 and 2011 sub-frames 
used for proper motion measurements in the four sub-fields, 
S1, S2, N1, and N2, are made available in FITS format.

\section{Spitzer and Herschel mid-IR Images}

\begin{figure*}
\gridline{\fig {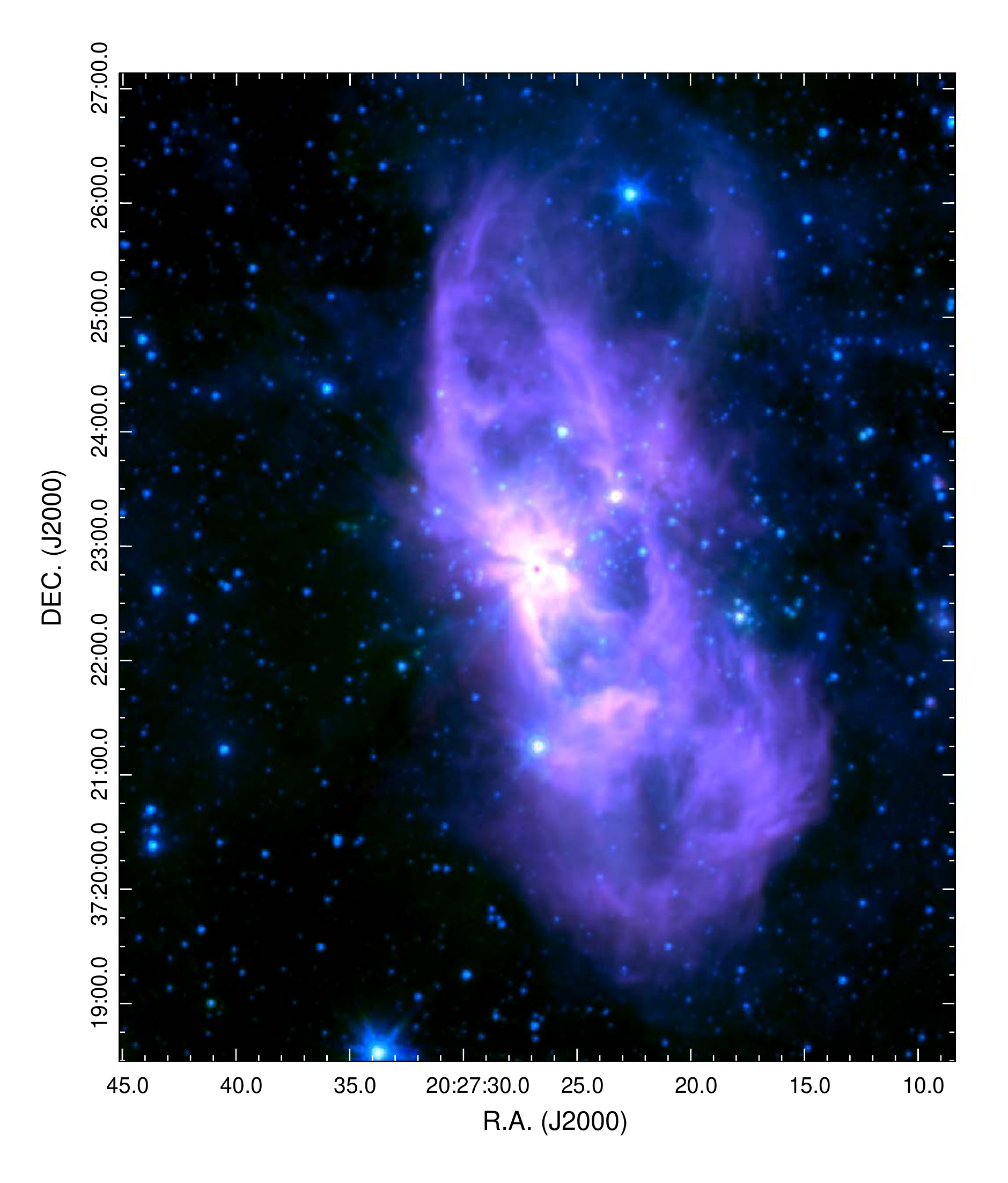} {0.5\textwidth} {(a)}
          \fig {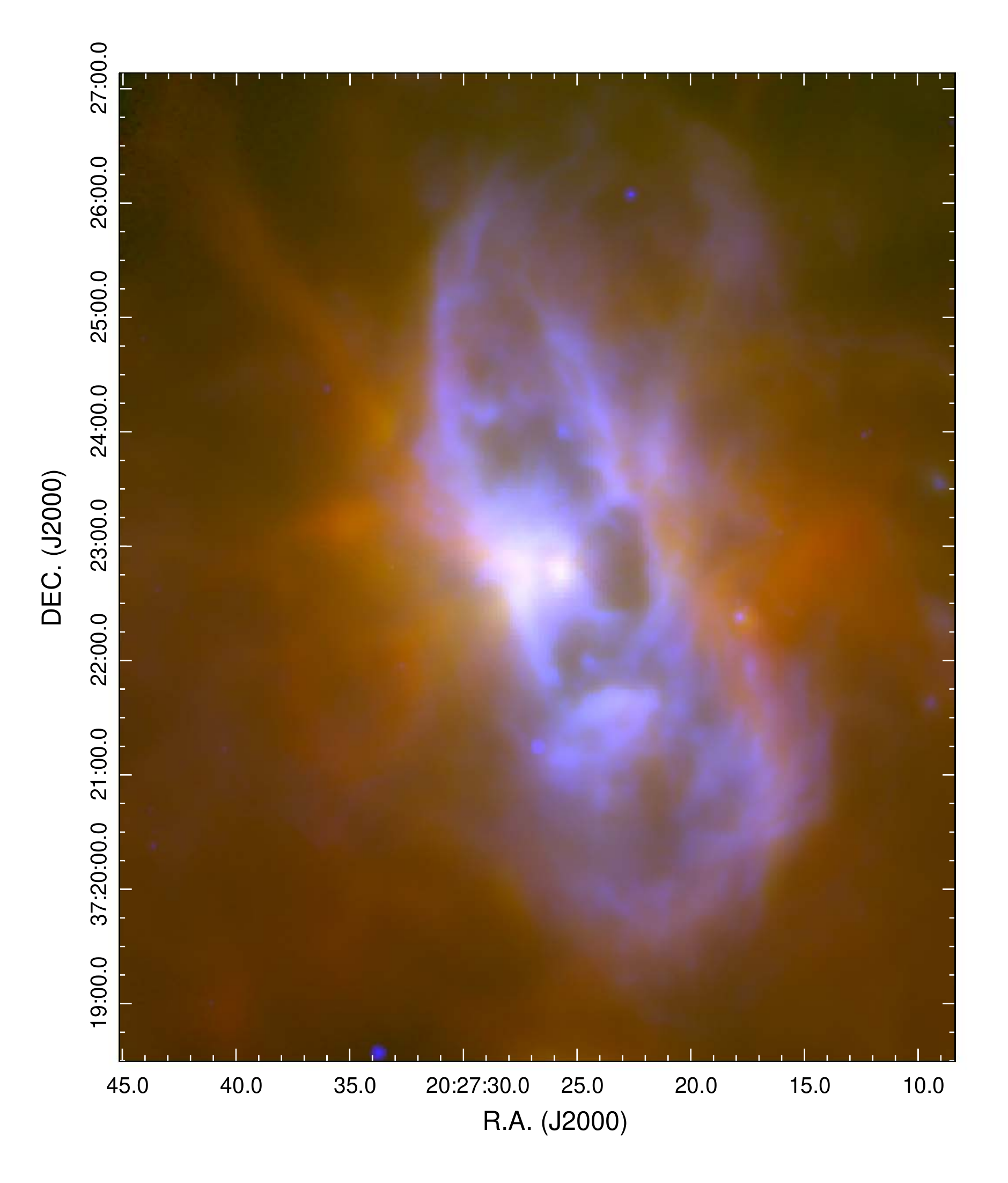} {0.5\textwidth} {(b)}
         }
\caption{{\bf Left:} 
    A wide-field view showing the cavities surrounding the S106 \Hii\ 
    region in the Spitzer/IRAC  3.6 $\mu$m (blue), 4.6 $\mu$m (green), and
    8.0 $\mu$m (red) images. The location of S106 IR is indicated by the 
    red dot in the brightest part of the image.
         {\bf Right:}
    A wide-field view showing the cavities surrounding the S106 \Hii\ 
    region in the Spitzer/IRAC  8.0 $\mu$m (blue),  Herschel/PACS 70 $\mu$m (green) and
    160.0 $\mu$m (red) images.
\label{Spitzer_FIR}}
\end{figure*}

Figure \ref{Spitzer_FIR} shows that the \Hii\ region sits inside
a roughly cylindrical, $\sim$6\arcmin\  (1.9 pc) long cavity bounded by warm dust 
and PAH emission at 3.6, to 70.0 $\mu$m.    This cavity is more than a factor of 
two longer and wider than the \Hii\ region at visual, near-IR, and radio wavelengths.   
S106~IR is displaced from the center of this cavity and located near its eastern edge.

Figure \ref{Spitzer_FIR} shows a color composite image showing 8, 70, and 160 $\mu$m
emission.   The cavity walls are clearly seen in the shorter wavelengths (blue
in Figure \ref{Spitzer_FIR}).   Warm dust in the surrounding molecular cloud
is traced by the 70 $\mu$m emission.  Note the V-shaped filaments
fanning away from the center of the \Hii\ region towards the north an northeast.  
Also note the bright bar located about 1.3\arcmin\ south of S106~IR.  This
feature is also seen at radio frequencies and marks the southern end of the 
\Hii\ region.   The dark feature located 30\arcsec\ to 60\arcsec\ west of S106~IR
remains dark at all wavelengths to 500 $\mu$m.  Thus, this feature may be
a cavity in the cloud surrounding S106 that is shielded from illumination by 
S106~IR by the dense molecular clump containing the Class~0 source, S106~FIR
\citep{Furuya1999,Furuya2000}.

\section{near-IR narrow-band images}

\begin{figure}
	\includegraphics[width=\columnwidth]{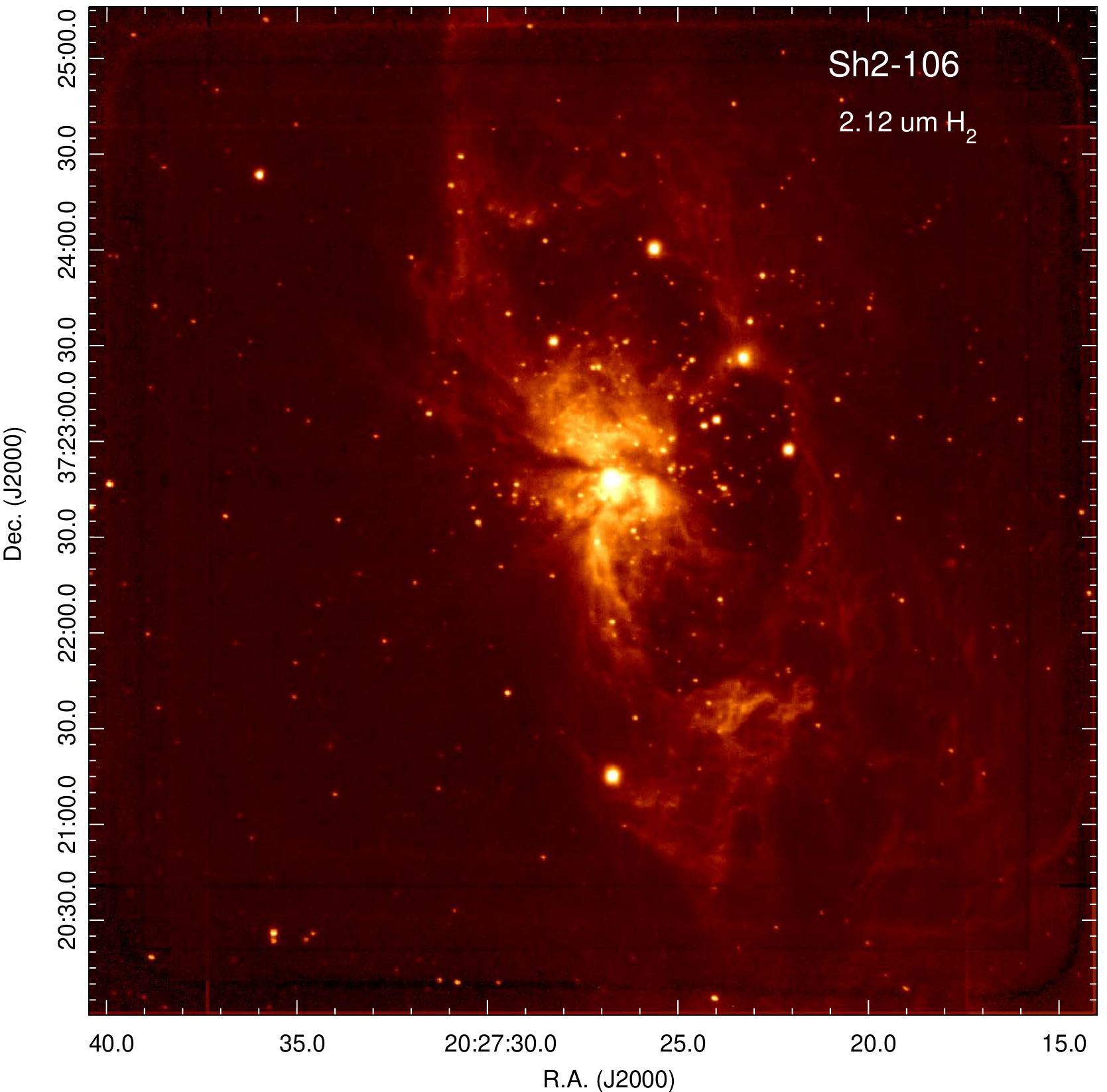}
    \caption{Near-infrared image showing S106 in a narrow-band
    filter centered on the $\lambda$=2.12 \um\ \Htwo\ emission line.}
        \label{fig_H2}
\end{figure}

\begin{figure}
	\includegraphics[width=\columnwidth]{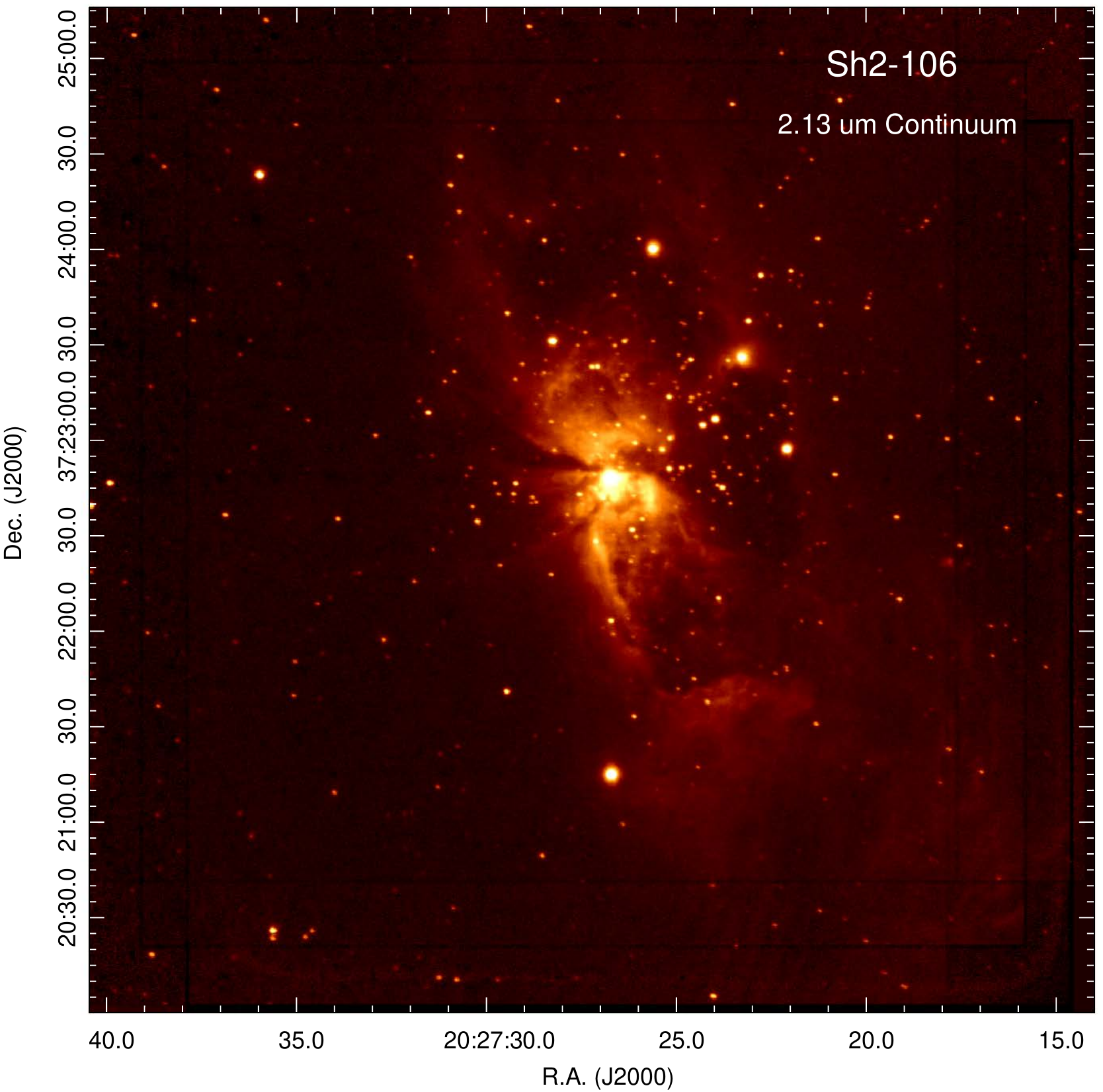}
    \caption{Near-infrared image showing S106 in a narrow-band
    filter centered at $\lambda$=2.13 \um\ off the 
    \Htwo\ emission line.}
        \label{fig_H2ref}
\end{figure}

Figure \ref{fig_H2} shows the narrow-band \Htwo\ image  taken
with the APO 3.5 meter telescope at a wavelength of 2.12 $\mu$m
with a filter pass-band of 0.5\%.
Figure \ref{fig_H2ref} shows the narrow-band \Htwo - off-line image  
taken with the APO 3.5 meter telescope at a wavelength of 2.13 $\mu$m
with a filter pass-band of 0.5\%.   These images show some of the
members of the S106 cluster.

\begin{figure}
	\includegraphics[width=\columnwidth]{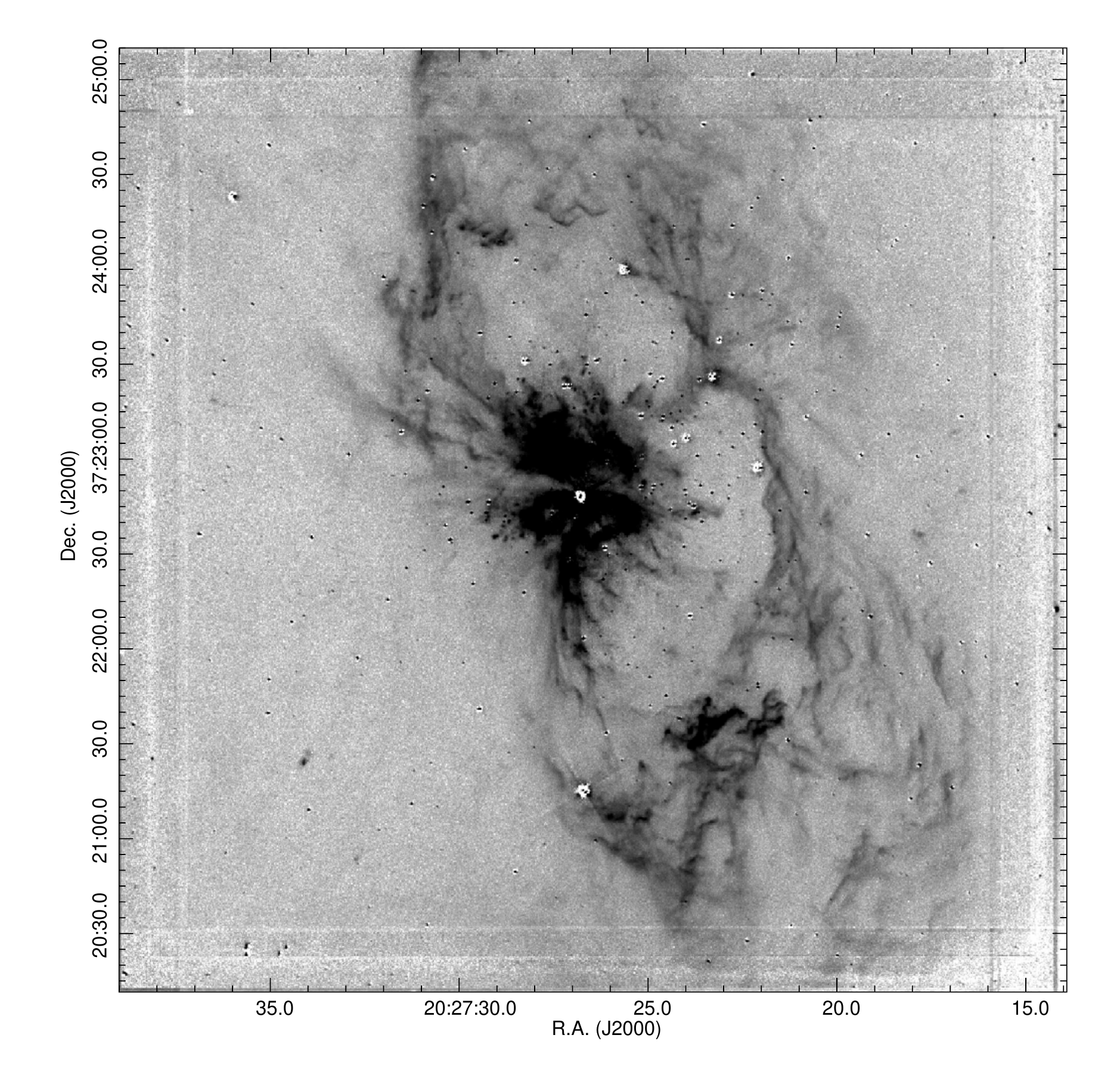}
    \caption{Continuum-subtracted near-infrared image showing 
    S106 in the $\lambda$=2.12 \um\ \Htwo\ emission line.  The
    intensity scale on this log display is set to emphasize faint
    features.}
        \label{fig_H2grey_deep_log}
\end{figure}

\begin{figure}
	\includegraphics[width=\columnwidth]{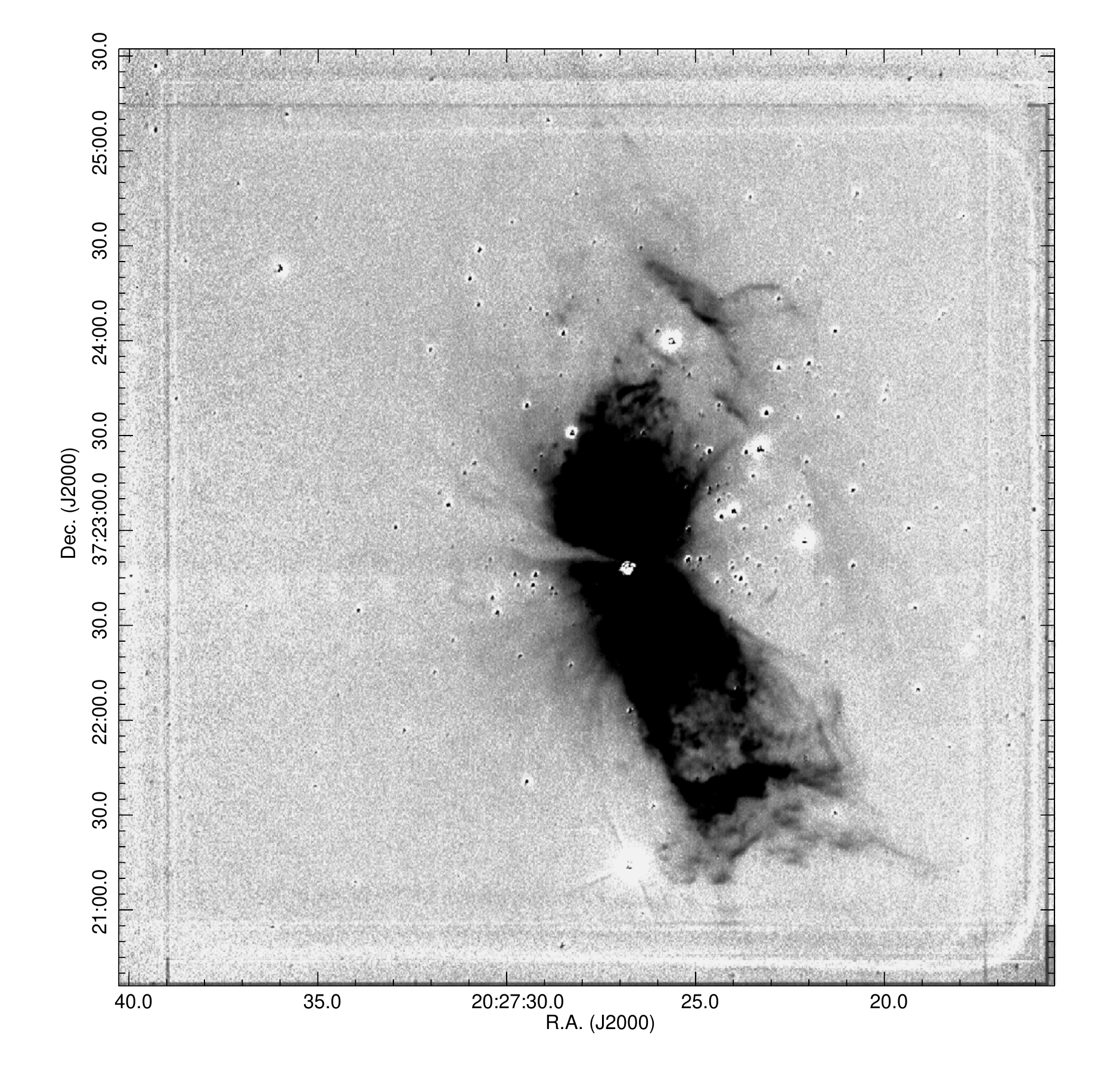}
    \caption{Continuum-subtracted near-infrared image showing 
    S106 in the $\lambda$=2.16 \um\ \BrG\ emission line.  The
    intensity scale on this log display is set to emphasize faint
    features.}
        \label{fig_BrGgrey_deep_log}
\end{figure}

Figure \ref{fig_H2grey_deep_log} shows a deep cut of Figure
\ref{fig_H2_subtracted} that emphasizes the faint feature of
\Htwo\ emission.  Figure \ref{fig_BrGgrey_deep_log} shows a 
deep cut of Figure
\ref{fig_BrGgrey} that emphasizes the faint feature of
\Htwo\ emission.

\end{document}